\documentclass[aps,twocolumn,amsmath,amssymb,preprintnumbers,superscriptaddress,floatfix]{revtex4}

\usepackage{physics}

\usepackage[utf8]{inputenc}
\usepackage[T1]{fontenc}
\usepackage{newtxtext}
\usepackage[upint]{newtxmath}
\usepackage{microtype}
\usepackage{textcomp}
\usepackage{dsfont}
\usepackage{eucal}
\usepackage{siunitx}
\usepackage{soul}

\usepackage{enumerate}
\usepackage{amsfonts}
\usepackage{color}
\usepackage{soul}

\usepackage{algorithm}
\usepackage{algpseudocode}
\usepackage{algorithmicx}

\usepackage{pgfplots}

\usepackage{graphicx}
\usepackage{todonotes}
\presetkeys%
    {todonotes}%
    {inline}{}
\usepackage[colorlinks,allcolors=blue]{hyperref}
\usepackage{cleveref}

\usetikzlibrary{decorations.text}
\usetikzlibrary{decorations.markings,arrows}
\usetikzlibrary{calc}

\newcommand{\me}[1]{\mathrm{e}^{#1}}                            % Differential scalar element
\newcommand{\transpose}[1]{\ensuremath{#1}^{T}}

\renewcommand{\vec}[1]{\boldsymbol{#1}}                           % Bold vector
\newcommand{\uv}[1]{\vec{e}_{#1}}       
\definecolor{DarkBlue}{rgb}{0,0,0.80}
\definecolor{DarkRed}{rgb}{0.80,0,0}
\definecolor{Purple}{rgb}{0.55,0,0.55}
\definecolor{Purple}{rgb}{0,0,0.8}

\DeclareMathOperator{\sgn}{sgn}
\DeclareMathOperator{\diag}{diag}

\newcommand*{\defeq}{\coloneqq}
\newcommand*{\eqdef}{\eqqcolon}

\newcommand{\up}{\uparrow}                                      % Spin up
\newcommand{\dn}{\downarrow}                                    % Spin down
\let\epsilon\varepsilon

\begin{document}
\title{Quasiclassical theory for antiferromagnetic metals}

\author{Eirik Holm Fyhn}
\affiliation{Center for Quantum Spintronics, Department of Physics, Norwegian \\ University of Science and Technology, NO-7491 Trondheim, Norway}
\author{Arne Brataas}
\affiliation{Center for Quantum Spintronics, Department of Physics, Norwegian \\ University of Science and Technology, NO-7491 Trondheim, Norway}
\author{Alireza Qaiumzadeh}
\affiliation{Center for Quantum Spintronics, Department of Physics, Norwegian \\ University of Science and Technology, NO-7491 Trondheim, Norway}
\author{Jacob Linder}
\affiliation{Center for Quantum Spintronics, Department of Physics, Norwegian \\ University of Science and Technology, NO-7491 Trondheim, Norway}

\date{\today}
\begin{abstract}
  \noindent Unlike ferromagnetism, antiferromagnetism cannot readily be included in the quasiclassical Keldysh theory because of the rapid spatial variation in the directions of of the magnetic moments.
  The quasiclassical framework is useful because it separates the quantum effects occurring at length scales comparable to the Fermi wavelength from other length scales, and has successfully been used to study a wide range of phenomena involving both superconductivity and ferromagnetism.
  Starting from a tight-binding Hamiltonian, we develop general quasiclassical equations of motion and boundary conditions which can, be used to describe two-sublattice metallic antiferromagnets in the dirty limit.
  The boundary conditions are applicable also for spin-active boundaries that can be either compensated or uncompensated.
  Additionally, we show how nonuniform or dynamic magnetic textures influence the equations and we derive a general expression for observables within this framework.
\end{abstract}
\maketitle
% Fakesection: Introduction
\section{Introduction}%
\label{sec:introduction}
The quasiclassical Keldysh Green's function technique~\cite{rammer1986,chandrasekhar2008,eilenberger1968,usadel1970,belzig1999} is a powerful tool to study mesoscopic structures~\cite{eschrig2003,buzdin2005,bergeret2005,houzet2008,heikkila2019,fermin2022,ojajarvi2022,belzig1999,fyhn2021_time,virtanen2010,fyhn2022,ali2017,hugdal2017,amundsen2018,amundsen2016,tancredi2022,wakamura2014,fyhn2020_zeeman,ali2018,fyhn2019}.
It is applicable to systems where the Fermi wavelength is much smaller than all other length scales and can be used to study a wide range of systems, including heterostructure with multiple competing types of order, such as superconductivity and ferromagnetism~\cite{eschrig2003,buzdin2005,bergeret2005,houzet2008,heikkila2019,fermin2022,ojajarvi2022}, both in and out of equilibrium.
In addition, the quasiclassical framework is versatile in regards to sample geometry~\cite{amundsen2016,amundsen2018,tancredi2022} and the details of external or intrinsic fields, such as applied magnetic fields~\cite{ali2018,fyhn2020_zeeman} or spin-orbit coupling~\cite{fyhn2022,wakamura2014}, whether they are time-dependent~\cite{houzet2008,virtanen2010,fyhn2021_time,fyhn2022} or spatially inhomogeneous~\cite{tancredi2022,fermin2022,fyhn2019}.
This makes the quasiclassical framework especially useful to the field of superconducting spintronics~\cite{linder2015}, which aims to utilize superconductivity in the field of spintronics.
In spintronics, spin is used as an information carrier rather than the electric charge used in conventional electronics~\cite{zutic2004,zabel2009}.
The combination of superconductivity and magnetism is therefore at the core of superconducting spintronics.

While the presence of a magnetic field typically suppresses superconductivity, the relationship between ferromagnets and superconductors (SC) can be synergistic~\cite{linder2015,bergeret2005}.
The interplay between magnetic and superconducting orders may give rise to spin-polarized superconductivity which can transport spin angular momentum with zero resistance~\cite{bergeret2005,keizer2006}, and the presence of superconductivity has also been shown to be beneficial for other central effects in spintronics, such as giving rise to infinite magnetoresistance~\cite{li2013}.

Antiferromagnets (AFs) have many important advantages over ferromagnets in the context of spintronics~\cite{baltz2018}.
The alternating magnetic moments mean that they are more robust and impervious to external magnetic fields while creating negligible magnetic stray fields of their own.
As a result, they are less intrusive to neighboring components.
Moreover, the resonance frequencies in AFs are on the order of terahertz~\cite{pimenov2009,baierl2016}, which allows for very fast information processing.
The fact that spin transport has been shown to be long-ranged in AFs~\cite{lebrun2018} also makes them promising and an active research topic in spintronics.

Superconductivity may coexist with antiferromagnetism~\cite{bulaevskii1985,lu2015,mukuda2012}, and AFs have a prominent role in the context of high-$T_c$ superconductivity~\cite{mukuda2012,zhang1997,orenstein2000}.
Despite this, AFs are much less studied in the field of superconducting spintronics compared to ferromagnets.
Heterostructures composed of superconductors and ferromagnets, including strongly polarized ferromagnets~\cite{eschrig2003}, has been studied theoretically in a wide range of systems~\cite{eschrig2003,buzdin2005,bergeret2005,houzet2008,heikkila2019,fermin2022,ojajarvi2022}, including in systems with complex geometries~\cite{amundsen2016,tancredi2022}.
On the other hand, while antiferromagnetic-superconductor junctions have been studied theoretically~\cite{jakobsen2020,zhou2019,bulaevskii2017,jakobsen2021,johnsen2021,zhen2019}, such studies are typically limited to simple geometries and clean systems.
This is because the rapid variation of the magnetic moments in AFs means that they, unlike ferromagnets, cannot readily be incorporated into the quasiclassical framework used for normal metals.
The quasiclassical Keldysh theory separates the short-range quantum effects from the long-range semiclassical dynamics, thereby allowing the inclusion of long-range spatial and temporal gradients.
As such, it is desirable with a quasiclassical framework that is applicable to systems with both superconductivity and AFs.

One approach, which has been used previously when studying the superconducting proximity effect in antiferromagnetic metals (AFMs)~\cite{hubener2002,hauser1966,cheng1990}, is to treat the AFM as a normal metal.
The reasoning is that the magnetic order is compensated on the length scale of the superconducting correlation length.
Using this framework, \textcite{hubener2002} studied AFM/SC/AFM structures and found an anomalous strong suppression of the proximity effect happening when the thickness of the AFM exceeded around $\SI{6}{\nano\meter}$.
They argued that the drop in superconducting critical temperature could possibly be associated with the onset of an incommensurate spin-density wave (SDW) state.
However, based on the theory presented in the present work, the observed suppression is expected even without the SDW state.
This is because \textcite{hubener2002} also reported a mean free path of $\SI{5.3}{\nano\meter}$ for their samples, and the theory presented here shows that even nonmagnetic impurities behave magnetically in the presence of antiferromagnetic order.
As such, conventional, spin-singlet superconductivity can be expected to be suppressed in antiferromagnetic systems when they enter the diffusive regime, and in particular more so than in diffusive normal metals.
This happens when the system size exceeds the mean free path, which was exactly the case in ref.~\cite{hubener2002}.

Quasiclassical equations of motion for AFMs, but without superconductivity, have been derived by \textcite{manchon2017}.
This was done by defining sublattice-resolved Green's function.
Such Green's functions can be treated quasiclassically because, while the magnetic order varies rapidly in the antiferromagnet, the Néel order varies slowly.
More recently, \textcite{bobkov2022} derived a sublattice-resolved quasiclassical theory for antiferromagnetic insulators with superconductivity.
Other related types of magnetically ordered systems that have been studied within quasiclassical theory are spiral ferromagnets~\cite{volkov2006,halasz2009} and SDW AFs~\cite{moor2011,dzero2015}.
Spiral ferromagnets have compensated magnetic order similar to AFs.
However, in order for these to be treated quasiclassically, the spatial modulation of the magnetic order must be slow compared to the Fermi wavelength.
SDW is also a state of matter with spatial modulation of the magnetic order, typically formed by itinerant particles with Fermi-surface nesting~\cite{gabovich2001,zhang2022}.
SDW can also coexist with SC~\cite{gabovich2001,vavilov2011}, and quasiclassical theory has been developed to model systems with both SDW and SC~\cite{moor2011,dzero2015}.
This is possible because the SDW state can be modeled using a mean-field approach with a slowly varying SDW order parameter.

\begin{figure}
  \centering
    \includegraphics[width=0.8\linewidth]{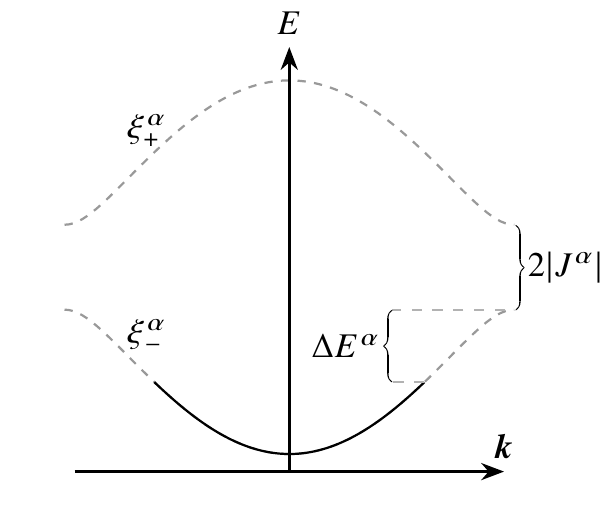}
    \caption{A sketch of the energy bands in an antiferromagnet, where $\xi^\alpha_\pm = -\mu^\alpha \pm \sqrt{(J^\alpha)^2 + (K^\alpha)^2}$. Here, $\alpha$ labels different materials, $\mu^\alpha$ is the chemical potential, $J^\alpha$ is the exchange coupling between itinerant electrons and localized magnetic moments and $K^\alpha$ is the kinetic energy and $\Delta E^\alpha$ is the smallest difference between the Fermi level and the edges of the conduction band. The gap between the energy bands is $2\lvert J^\alpha\rvert$. This gap can be arbitrary within the quasiclassical theory developed here, but $\Delta E^\alpha$ must be large compared to other energies in the system, not including the gap. }
\label{fig:sketch_energies}
\end{figure}

Here, we develop quasiclassical equations of motion for two-sublattice AFMs with superconductivity and impurities, as well as external fields and spin-orbit coupling, and where all the parameters, including the direction of the Néel vector, may be inhomogeneous in time and space, as long as it is not rapidly varying on the atomic length scale.
We also develop boundary conditions for the diffusive regime, which work also for spin-active interfaces that can be either uncompensated or compensated.
Because we consider antiferromagnetic metals, we assume that the Fermi level is deep within the conduction band compared to other energy scales except for the exchange energy between localized spins and itinerant electrons, as illustrated in \cref{fig:sketch_energies}.
This exchange energy may be either large or small compared to the distance between the Fermi level and the edges of the conduction band.
The quasiclassical theory can therefore not be used to model heavy-fermion antiferromagnetic superconductors, where the Fermi energy is comparable to the superconducting gap~\cite{petrovic2001}. 
On the other hand, it is well suited to study heterostructures or other systems in which the Fermi level can be assumed to lie deep within the conduction band.

Although our starting point is similar to that presented in refs.~\cite{manchon2017,bobkov2022}, except that we additionally consider the other effects mentioned above, there are a few important differences.
Instead of equations for sublattice-resolved Green's functions, we derive equations for the conduction band Green's functions.
This is possible because there is no rapidly varying magnetic order for these Green's functions, just as there is no rapidly varying magnetic order for sublattice-resolved Green's functions.
The reason why we project onto the conduction band is that only states close to the Fermi level contribute to the quasiclassical Green's function, and the Fermi level lies deep inside the conduction band.
As a result, we end up with fewer Green's functions to solve for.
More importantly, however, it means that the chemical potential drops out of the equations, similar to how it drops out in Keldysh theory for normal metals.
Therefore, we can consistently let it be much larger than other energies.
This procedure, leaving only the conduction band, means that the spin- and sublattice degrees of freedom are not independent.
An important consequence of this fact is that the effect of nonmagnetic impurities in AFMs is similar to the effect of magnetic impurities in normal metals.

We summarize the main results, outline how they are derived and describe the necessary assumptions in \cref{sec:outline}.
The derivations are presented in \cref{sec:hamiltonian,sec:green_s_functions,sec:impurity_averaging,sec:tunneling,sec:wigner_coordinates,sec:extracting_the_conduction_band,sec:quasiclassical_green_s_functions,sec:revisiting_impurity_self_energy,sec:the_dirty_limit,sec:boundary_condition,sec:nonuniform_magnetic_textures,sec:observables}.
This includes the derivation of quasiclassical equations of motion, boundary conditions for the diffusive regime and a general expression for computing observables.
Concluding remarks are given in \cref{sec:conclusion}.

\section{Outline}
\label{sec:outline}
The main results are equations for the isotropic part of the quasiclassical Green's function, $\check g_s^\alpha$, and the matrix current, $\check{\vec j}^\alpha$, where $\alpha$ labels the materials in the junction.
Under the assumptions that the quasiclassical Green's function is approximately spherically symmetric and that the energy difference between the Fermi level and the edges of the conduction band is larger than all other energy scales, except possibly the exchange energy $J^\alpha$, we find in \cref{sec:the_dirty_limit} that $\check g_s^\alpha$ and $\check{\vec j}^\alpha$ solve
\begin{subequations}
  \label{eq:outline}
\begin{align}
  i \tilde\nabla\circ\check{\vec j}^\alpha + \left[\tau_z\varepsilon - \check V_s^\alpha + \frac{i(J^\alpha)^2}{2\tau^\alpha_\text{imp}(\eta^\alpha)^2}\sigma_z\tau_z\check g_s^\alpha\sigma_z\tau_z,\,\check g_s^\alpha\right]_\circ
   = 0,
   \label{eq:usadel_outline}
   \\
  \check{\vec j}^\alpha = 
    -\check g^\alpha_s\circ\tilde\nabla\circ(D^\alpha\check g^\alpha_s)
    - \check g^\alpha_s\circ\left[\frac{(J^\alpha)^2}{2(\eta^\alpha)^2}\sigma_z\tau_z\check g_s^\alpha\sigma_z\tau_z,\,\check{\vec j}^\alpha\right]_\circ,
  \label{eq:gen_curr_outline}
\end{align}
\end{subequations}
where all the symbols are explained below.
In the absence of antiferromagnetism, $J^\alpha \to 0$, \cref{eq:outline} reduces to the well-known Usadel equation for normal metals~\cite{usadel1970}.
In the limit of very strong exchange coupling, such that $(J^\alpha/\eta^\alpha)^2 \to 1$, the short-range correlations become negligible in the diffusive limit, as we show in \cref{sec:the_dirty_limit}.

The itinerant electrons in an AFM are described by a Hamiltonian including kinetic energy $K^\alpha$, exchange energy to the magnetic lattice $J^\alpha$, chemical potential $\mu^\alpha$ as well as other additional terms coming from superconductivity, impurity scattering, external fields or spin-orbit coupling.
\Cref{eq:outline} is valid under the assumption that $K^\alpha$ at the Fermi level is large compared to all additional energies such as the impurity scattering rate and the superconducting gap.
Note that $K^\alpha$ need not be large compared to $J^\alpha$.
As a result, the fraction $(J^\alpha)^2/(\eta^\alpha)^2$, where $\eta^\alpha = \sqrt{(J^\alpha)^2 + (K^\alpha)^2}$, can take any value between $0$ and $1$.

The second assumption behind \cref{eq:outline} is that the system is in the dirty regime.
This means two things.
First, it means that the elastic impurity scattering rate, $1/\tau_\text{imp}$ is dominant out of all the additional energies in the system, not including $K^\alpha$, $J^\alpha$ and $\mu^\alpha$.
Second, it means that the matrix current $\check{\vec j}^\alpha$ is small compared to the Fermi velocity.
As we show in \cref{sec:the_dirty_limit}, this is the case if the variation in $\check g_s^\alpha$ is small compared to 1 over the length of the mean free path, either because the mean free path is short or because the proximity effect is weak.

To complete the theory for use in systems involving more than one material, we derive the boundary condition
\begin{equation}
  \uv n \cdot \check{\vec j}^\alpha = \left[\hat{\mathcal T}_l^{\alpha\beta} \circ \check g^\beta_{s}(\vec x_l^\beta) \circ\hat{\mathcal T}_l^{\beta\alpha} + i\transpose{(S_c^\alpha)}\hat R_l S_c^\alpha,\, \check g_s^\alpha\right]_\circ,
  \label{eq:bc_outline}
\end{equation}
which are valid when the quasiclassical Green's function is isotropic also close to the interface.
This is the case for instance when the tunneling is weak.
\Cref{eq:bc_outline} can be used to model interfaces that are compensated or uncompensated, magnetic or non-magnetic and conducting or isolating.
In the absence of antiferromagnetism, \cref{eq:bc_outline} reduces to the generalized Kupriyanov-Lukichev boundary condition for spin-active boundaries~\cite{KL1988,eschrig2015}.

In \cref{sec:observables}, we derive a general expression for computing observables which can be used to compute quantities such as densities and currents once $\check g_s^\alpha$ and $\check{\vec j}^\alpha$ have been found.
The expression, \cref{eq:observables}, contains not only the contribution from states captured by the quasiclassical Green's function but also a general expression for the contribution from states further away from the Fermi level.

We present a detailed, self-contained derivation of \cref{eq:outline,eq:bc_outline}, starting from a general tight-binding Hamiltonian with a tunneling contact, introduced in \cref{sec:hamiltonian}.
The full Green's functions and their equations of motion are presented in \cref{sec:green_s_functions}.
Impurity averaging is performed in \cref{sec:impurity_averaging}, where we derive the impurity self-energy to second order in the impurity potential.
This is valid as long as the impurity potential is weak, but since the self-energy depends only on the isotropic part of the Green's function, effects such as skew scattering~\cite{sinitsyn2007} would require going to third order.
In \cref{sec:tunneling}, we use the tunneling Hamiltonian to remove the intermaterial Green's functions from the equations of motion.
In \cref{sec:wigner_coordinates} we Fourier transform in relative coordinates, and it is taken into consideration both that the system is defined on a discrete lattice and, more importantly, different matrix elements correspond to different relative spatial positions because of the relative displacement between the two sublattices.
In \cref{sec:extracting_the_conduction_band} we transform the Green's functions into the basis of the antiferromagnetic energy bands, and thereby extract the conduction band.
From this, we carefully define the quasiclassical Green's functions in \cref{sec:quasiclassical_green_s_functions} and use them to remove higher-order spatial derivatives from the gradient expansion.
Next, in \cref{sec:revisiting_impurity_self_energy}, we derive the quasiclassical expression for the impurity scattering and show how it is modified by the antiferromagnetic order.
The main results are then derived in \cref{sec:the_dirty_limit} and \cref{sec:boundary_condition}.
Finally, in \cref{sec:nonuniform_magnetic_textures} we show how the equations are influenced by nonuniform magnetic textures.

\begin{figure}[htpb]
  \centering
  \includegraphics[width=0.6\linewidth]{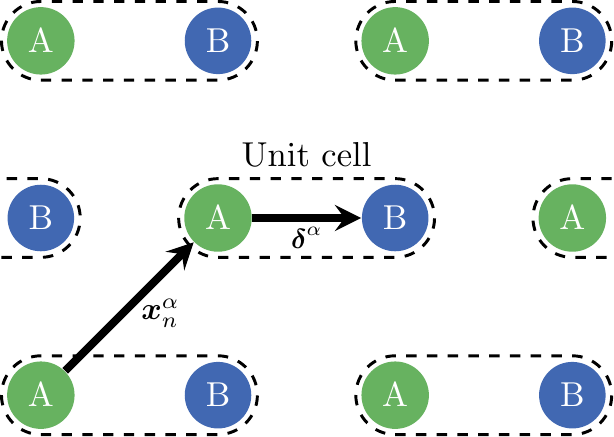}
  \caption{Sketch of a plane in material $\alpha$ for the case of a square lattice. Each unit cell contains two orbitals. One is located at sublattice A, $\vec x_n^\alpha$, and one is located at sublattice B, $\vec x_n^\alpha + \vec\delta^\alpha$.}%
  \label{fig:sublatticeSketch}
\end{figure}
\section{Hamiltonian}%
\label{sec:hamiltonian}
We consider a system composed of two materials, which we label material $L$ and material $R$, connected through a tunneling contact.
The Hamiltonian is
\begin{equation}
  \mathcal H(t) = \mathcal H_L(t) + \mathcal H_R(t) + \mathcal H_T.
\end{equation}
Here,
\begin{equation}
  \mathcal H_\alpha(t) = \sum_{n,m \in A_\alpha}c_n^{\alpha\dagger}\left[ H_0^\alpha(t)  + V^\alpha(t)\right]_{nm}c_m^\alpha,
  \label{eq:material_hamiltonian}
\end{equation}
where $\alpha \in \{L,R\}$ denotes material, $A_\alpha$ is the set of unit cells in material $\alpha$.
As sketched in \cref{fig:sublatticeSketch}, each unit cell, labeled by a 3-tuple $n$, contains one orbital associated with the $A$-sublattice at position $\vec x_n^\alpha$, and one orbital associated with the $B$-sublattice at position $\vec x_n^\alpha + \vec\delta^\alpha$.
We let the annihilation operators for the orbitals with spin $\sigma$ at unit cell $n$ in material $\alpha$ at the $A$- and $B$-sublattice be $c^{\alpha}_{nA\sigma}$ and  $c^{\alpha}_{nB\sigma}$, respectively, and define
\begin{equation}
  c_n^{\alpha\dagger} = \mqty(c^{\alpha\dagger}_{nA\up} & c^{\alpha\dagger}_{nA\dn} &c^{\alpha\dagger}_{nB\up} & c^{\alpha\dagger}_{nB\dn} &c^\alpha_{nA\dn} & -c^\alpha_{nA\up} &c^\alpha_{nB\dn} &-c^\alpha_{nB\up}).
\end{equation}

We include only nearest neighbor hopping and assume that this hopping is only between the two different sublattices.
The hopping parameter, $t^\alpha$, chemical potential, $\mu^\alpha$, and the exchange energy $J^\alpha$ between localized spins and conducting electrons are collected in $H_0^\alpha$.
The full electrochemical potential need not be constant.
However, we take $\mu^\alpha$ to be constant.
Any deviation in the electrochemical potential away from $\mu^\alpha$ is included in $V^\alpha$.
If $\vec \sigma$ is the vector of Pauli matrices in spin-space, $\vec \tau$ are the Pauli matrices in Nambu-space and $\vec \rho$ are the Pauli matrices in sublattice-space, then
\begin{multline}
  (H_0^\alpha)_{nm}(t) = -\frac 1 4 t^\alpha (\rho_x + i\rho_y)\tau_z\chi_\text{N.N.}(\vec x_n^\alpha - \vec\delta^\alpha - \vec x_m^\alpha) \\
  -\frac 1 4 t^\alpha (\rho_x - i\rho_y)\tau_z\chi_\text{N.N.}(\vec x_n^\alpha + \vec\delta^\alpha - \vec x_m^\alpha) - \frac 1 2 \delta_{nm}\mu^\alpha  \tau_z
  \\
  - \frac 1 2 \delta_{nm} J^\alpha\rho_z\vec\sigma \cdot 
  \left[\frac{1+\rho_z}{2}\vec n(\vec  x_n^\alpha, t) + \frac{1-\rho_z}{2}\vec n(\vec  x_n^\alpha + \vec \delta^\alpha, t) \right],
\end{multline}
where $\vec n = (\sin\theta\cos\phi, \sin\theta\sin\phi, \cos\theta)$ is the direction of the Néel vector, and $\chi_\text{N.N}(\vec x)$ is a nearest neighbor characteristic function which is 1 if $\vec x$ is a nearest neighbor vector between a $A$-lattice point and a $B$-lattice point and 0 otherwise.
Because the direction Néel vector generally is influenced by the dynamics of the itinerant electron, it should be solved for self-consistently.
This can be done with the Landau-Lifshitz-Gilbert equation~\cite{baltz2018}.

The term proportional to $V^\alpha$ in \cref{eq:material_hamiltonian} contains all additional effects that may be present in the model, such as superconductivity, external spin-splitting fields and corrections to the hopping term from the vector potential or spin-orbit coupling.
Additionally, $V^\alpha$ importantly also determines the spatial geometry of material $\alpha$ by a potential that is zero inside the material and very large outside the material.
We can therefore let the lattice, $A_\alpha$, run to infinity in all spatial directions, meaning that $A_\alpha = \mathbb{Z}^3$, where $\mathbb{Z}$ is the set of integers, while still having the system be confined to a finite region of space.
Note that the potential can also be spin-dependent, for instance if there is a spin-splitting field in the neighboring region.
This will influence the boundary condition we derive in \cref{sec:boundary_condition}.

Finally, the tunneling Hamiltonian is
\begin{equation}
  \mathcal H_T = \sum_{n,m \in \mathbb{Z}^3}c_n^{L\dagger} T^{LR}_{nm} c_m^R
  = \sum_{i,j \in \mathbb{Z}^3}c_n^{R\dagger} T^{RL}_{nm} c_m^L,
\end{equation}
where $T^{RL}$ and $T^{LR} = (T^{RL})^\dagger$ are matrices satisfying $T^{LR} = \diag(T, i\sigma_yT^*i\sigma_y)$ for some $4\times 4$ matrix $T$.

We rotate spin space such that the Néel vector is always parallel to the $z$-axis.
To do this we define the rotation matrix 
\begin{equation}
  R(\vec x, t) = \exp\left\{-i\frac{\theta[\vec n(\vec x, t) \times \uv z]\cdot \vec \sigma}{2\sin\theta}\right\},
  \label{eq:rotationMatrix}
\end{equation}
and
\begin{equation}
  \tilde c_n^\alpha(t) = \left[\frac{1+\rho_z}{2}R^\dagger(\vec x^\alpha_n, t) + \frac{1-\rho_z}{2}R^\dagger(\vec x_n^\alpha + \vec\delta^\alpha, t)\right]c_n^\alpha,
  \label{eq:rotc}
\end{equation}
such that
\begin{equation}
  \mathcal H_\alpha(t) = \sum_{n,m \in \mathbb{Z}^3}\tilde c_n^{\alpha\dagger}(t)\left[ \tilde H_0^\alpha(t)  + \tilde V^\alpha(t)\right]_{nm}\tilde c_m^\alpha(t),
\end{equation}
where, if we assume that $\vec n$ varies slowly in space over the distance of neighboring lattice points,
\begin{multline}
  (\tilde H_0^\alpha)_{nm}(t) = -\frac 1 2 \delta_{nm} \left[J^\alpha\rho_z\sigma_z + \mu  \tau_z\right] 
    + \frac 1 2 K^\alpha_{nm}\tau_z
    \\
    - \frac{\tau_z}{2} \left(K^\alpha_{nm}[\vec x_n^\alpha - \vec x_m^\alpha] + \left[\vec\delta^\alpha\rho_B,\, K_{nm}^\alpha\right]\right)\cdot \left(R^\dagger \nabla R\right)\left(\vec x_n^\alpha,t\right).
\end{multline}
where the kinetic term is
\begin{multline}
  K_{nm}^\alpha = -\frac{t^\alpha}{2} \bigl[(\rho_x + i\rho_y)\chi_\text{N.N.}(\vec x_n^\alpha - \vec\delta^\alpha - \vec x_m^\alpha) \\
  + (\rho_x - i\rho_y)\chi_\text{N.N.}(\vec x_n^\alpha + \vec\delta^\alpha - \vec x_m^\alpha)\bigr]
\end{multline}

Finally, we also define the projection operators in sublattice space,
\begin{align}
  \rho_A = \frac{1 + \rho_z}{2} \quad \text{and} \quad \rho_B = \frac{1 - \rho_z}{2},
\end{align}
for ease of notation.

\section{Green's functions and Equations of Motion}%
\label{sec:green_s_functions}
In this section, we define the full Green's functions.
These are the starting point of our derivation and will later be used to define the quasiclassical, impurity-averaged conduction band Green's functions which are the objects of the final equations.
To obtain the final equations we must first derive the equation of motion for the full Green's function.
These are called the Gor'kov equations and are derived in this section.

The retarded, advanced and Keldysh Green's functions are defined respectively as
\begin{subequations}
\begin{align}
  \hat G_{nm}^{R,\alpha\beta}(t_1, t_2)
  &= -i\tau_z\left\langle\acomm{\tilde c^\alpha_n(t_1)}{\tilde c^{\beta\dagger}_{m}(t_2)} \right\rangle\theta(t_1 - t_2)
  \\
  \hat G_{nm}^{A,\alpha\beta}(t_1, t_2)
  &= +i\tau_z\left\langle\acomm{\tilde c^\alpha_n(t_1)}{\tilde c^{\beta\dagger}_{m}(t_2)} \right\rangle\theta(t_2 - t_1)
  \\
  \hat G_{nm}^{K,\alpha\beta}(t_1, t_2)
  &= -i\tau_z\left\langle\comm{\tilde c^\alpha_n(t_1)}{\tilde c^{\beta\dagger}_{m}(t_2)} \right\rangle.
\end{align}
\end{subequations}
These are $8\times 8$ matrices, and are collected in larger $16\times 16$ matrices,
\begin{equation}
  \check G_{nm}^{\alpha\beta} = \mqty(\hat G_{nm}^{R,\alpha\beta} & \hat G_{nm}^{K,\alpha\beta} \\ & \hat G_{nm}^{A,\alpha\beta}),
\end{equation}
and even larger $32\times 32$ matrices,
\begin{equation}
  \breve G_{nm} = \mqty(\check G_{nm}^{LL} & \check G_{nm}^{LR} \\ \check G_{nm}^{RL} & \check G_{nm}^{RR}).
\end{equation}
We use the notation that $\hat{\cdot}$ indicates a nontrivial matrix structure in Nambu-space, $\check{\cdot}$ indicates a nontrivial structure in Keldysh-space and $\breve{\cdot}$ indicates a nontrivial structure in material-space.

In order to derive the equations of motion, we use that any operator, $A$, evolves in time according to
\begin{equation}
  \pdv t A = i \comm{\mathcal H}{A} + \left(\frac{\partial A}{\partial t}\right)_{\mathcal H}.
\end{equation}
From this, together with the relation $\comm{AB}{C} = A\acomm{B}{C} - \acomm{A}{C}B$, we find
\begin{multline}
  \frac{\partial \tilde c^\alpha_n}{\partial t} = -2i\sum_{m\in\mathbb{Z}^3}\left[ \tilde H_0^\alpha(t)  + \tilde V^\alpha(t)\right]_{nm}\tilde c^\alpha_{m}
  -i\sum_{m\in\mathbb{Z}^3} \tilde T^{\alpha\beta}_{nm}\tilde c^\beta_{m}
  \\
  -\left[\rho_A \left(R^\dagger \dot R\right)(\vec x_n, t)
  +\rho_B \left(R^\dagger \dot R\right)(\vec x_n + \vec\delta, t)\right]\tilde c^\alpha_n,
\end{multline}
where $\beta \neq \alpha$, and 
\begin{multline}
  \frac{\partial \tilde c^{\alpha\dagger}_n}{\partial t} = 2i\sum_{m\in\mathbb{Z}^3}\tilde c^{\alpha\dagger}_{m}\left[ \tilde H_0^\alpha(t)  + \tilde V^\alpha(t)\right]_{mn}
  +i\sum_{m\in\mathbb{Z}^3} \tilde T^{\alpha\beta}_{nm}\tilde c^\beta_{m}
  \\
  +\tilde c^{\alpha\dagger}_n\left[\rho_A \left(R^\dagger \dot R\right)(\vec x_n, t)
  +\rho_B\left(R^\dagger \dot R\right)(\vec x_n + \vec\delta, t)\right].
\end{multline}
From this, we derive the Gor'kov equations,
\begin{subequations}
  \label{eq:gorkov_breve}
\begin{align}
  i\tau_z\frac{\partial \breve G}{\partial t} - \breve\Sigma \bullet \breve G &= \delta(t_1 - t_2)\delta_{nm},
  \\
  \frac{\partial \breve G}{\partial t'}i\tau_z +  \breve G\bullet\breve\Sigma &= -\delta(t_1 - t_2)\delta_{nm},
\end{align}
\end{subequations}
where
\begin{equation}
  \breve\Sigma = \mqty(\hat H_0^L + \check V^L & \hat T^{LR} \\ \hat T^{RL} & \hat H_0^R + \check V^R),
\end{equation}
and
\begin{subequations}
  \begin{align}
    \label{eq:H0}
   (\hat H_0^\alpha)_{nm}(t_1,t_2) &= \left(K^\alpha_{nm} - \delta_{nm} \left[J^\alpha\rho_z\sigma_z\tau_z + \mu \right] 
    \right)\delta(t_1 - t_2),
    \\
     (\hat T^{\alpha\beta})_{nm}(t_1,t_2) &= \tilde T^{\alpha\beta}_{nm} \tau_z \delta(t_1-t_2),
      \\
      \check V^\alpha_{nm}(t_1, t_2) &= 
      \left(\check \Sigma^\alpha_\text{inel}\right)_{nm}(t_1,t_2)
      +
      \Biggl\{2\tilde V^\alpha_{nm}(t)\nonumber\\ 
                                     &-\tau_z\left(K^\alpha_{nm}[\vec x_n^\alpha - \vec x_m^\alpha]
                                     + \left[\vec\delta^\alpha\rho_B,\, K_{nm}^\alpha\right]\right)  \nonumber\\
     & \cdot \left(R^\dagger \nabla R\right)\left(\vec x_n^\alpha,t_1\right)
      -i\Bigl[\rho_A \left(R^\dagger \dot R\right)(\vec x_n, t_1)
        \nonumber\\
     &+\rho_B\left(R^\dagger \dot R\right)(\vec x_n + \vec\delta, t)\Bigr]\delta_{nm}\Biggr\}\tau_z\delta(t_1-t_2).
      \label{eq:V_beginning}
  \end{align}
\end{subequations}
We have added in $\check V_{nm}^\alpha$ a term which models inelastic processes, $\check \Sigma^\alpha_\text{inel}$.
The bullet product between two matrix-valued functions, $A$ and $B$, is defined as
\begin{equation}
  (A\bullet B)_{nm}(t_1,t_2) = \int_{-\infty}^{\infty}\dd{t} \sum_{l\in\mathbb{Z}^3} A_{nl}(t_1, t)B_{lm}(t, t_2).
\end{equation}
We also define the circle-product to be the integral over time,
\begin{equation}
     (A\circ B)(t_1,t_2) = \int_{-\infty}^{\infty}\dd{t} A(t_1, t)B(t, t_2).
\end{equation}

From \cref{eq:gorkov_breve} we also get the Dyson equations,
\begin{subequations}
  \label{eq:dyson}
 \begin{align}
  \breve G = \breve G_0 + \breve G_0\bullet\delta\breve \Sigma\bullet\breve G,
  \label{eq:dyson_a}
  \\
  \breve G = \breve G_0 + \breve G\bullet\delta\breve \Sigma\bullet\breve G_0,
  \label{eq:dyson_b}
 \end{align} 
\end{subequations}
if $\breve \Sigma = \breve\Sigma_0 + \delta\breve\Sigma$ and $\breve G_0$ solves
\begin{subequations}
  \label{eq:gorkov_breve_g0}
  \begin{align}
  i\tau_z\frac{\partial \breve G_0}{\partial t_1} - \breve\Sigma_0 \bullet \breve G_0 &= \delta(t_1 - t_2)\delta_{nm},
  \label{eq:gorkov_breve_g0_1}
  \\
  \frac{\partial \breve G_0}{\partial t_2}i\tau_z +  \breve G_0\bullet\breve\Sigma_0 &= -\delta(t_1 - t_2)\delta_{nm}.
  \label{eq:gorkov_breve_g0_2}
\end{align}
\end{subequations}
\Cref{eq:dyson} can be derived by taking bullet products of \cref{eq:gorkov_breve_g0_1,eq:gorkov_breve_g0_2} with $\breve G$ from the left and right, respectively, and using that $A\bullet(\partial B/\partial t_1) = -(\partial A/\partial t_2)\bullet B$ when $\lim_{t \to \pm \infty} A(t_1,t)B(t,t_2) = 0$.

\section{Impurity averaging}%
\label{sec:impurity_averaging}
In this section, we average over impurities and identify the self-energy which relates the impurity-averaged Green's function to the Green's function in the absence of impurities.
The impurity-averaged Green's function can then be found by replacing the impurity potential in the Gor'kov equations with this self-energy.
We determine this self-energy to second order in the impurity potential.
This is valid under the assumption that the impurity potentials are weak, although the number of impurities may be large.
By not going to third order, the self-energy depends only on the isotropic part of the Green's function and therefore does not capture effects such as skew scattering~\cite{sinitsyn2007}.

Let $m^{\alpha X}$ be the number of impurities in material $\alpha$ on sublattice $X \in \{A,B\}$.
Next, we assume that the impurity potentials are local and that the potential strength and position of the $i$'th impurity in material $\alpha$ on sublattice $X$ are $U^{\alpha X}_i$ and $r^{\alpha X}_i$, respectively.
The self-energy term from the impurity potential is then
\begin{multline}
  \breve V^\text{imp}_{nm} = \delta_{nm}\delta(t_1-t_2)\\
  \times\sum_{X \in\{A,B\}}\mqty(\sum_{i=1}^{m^{LX}}\rho_X U^{LX}_i \delta_{nr^{LX}_i} & \\ & \sum_{i=1}^{m^{RX}}\rho_X U^{RX}_i \delta_{nr^{RX}_i}).
\end{multline}

Next, we define the impurity average as the sum over all possible impurity locations and impurity potential strengths, weighted by some normalized distribution function $p_\text{imp}: \{U_i\},\{r_i\} \mapsto \mathbb{R}$, 
where $\{U_i\}$ and $\{r_i\}$ denote the set of potential strengths and locations, respectively.
That is,
\begin{multline}
\langle A \rangle_\text{imp} = \prod_{\alpha\in\{L,R\}}\prod_{X\in\{A,B\}}\prod_{i=1}^{m^{\alpha X}}\int_{-\infty}^{\infty}\dd{U^{\alpha X}_i}
\\
\times\sum_{r^{\alpha X}_i \in \mathbb{Z}^3}p_\text{imp}(\{U_i\},\{r_i\})A(\{U_i\},\{r_i\}).
\end{multline}
We do not specify $p_\text{imp}$, but we assume it is such that impurities are independently and uniformly distributed. 
By assuming that they are uniformly distributed in space, we have that $\langle\delta_{nr^{\alpha X}_j} \rangle_\text{imp} = 1/N^{\alpha} = n^{\alpha X}_\text{imp}/m^{\alpha X}$, where $N^\alpha$ is the number of unit cells in material $\alpha$ and $n^{\alpha X}_\text{imp} = m^{\alpha X}/N^\alpha$ is the impurity density on sublattice $X$ in material $\alpha$.
The assumption that impurities are independent means that $\langle U^{\alpha X}_i\delta_{nr^{\alpha X}_i} U^{\beta Y}_j\delta_{mr^{\beta Y}_j} \rangle_\text{imp} = \langle U^{\alpha X}_i \delta_{nr^{\alpha X}_i}\rangle_\text{imp}\langle U^{\beta Y}_j\delta_{mr^{\beta Y}_j} \rangle_\text{imp}$ if $i \neq j$, $\alpha\neq \beta$ or $X \neq Y$.
Finally, we also assume that the strengths and locations of impurities are uncorrelated, such that $\langle U^{\alpha X}_i \delta_{nr^{\alpha X}_i}\rangle_\text{imp} = \langle U^{\alpha X}_i\rangle_\text{imp} \langle\delta_{nr^{\alpha X}_i}\rangle_\text{imp}$, and that the impurities on each sublattice and material are identically distributed, such that $\langle U^{\alpha X}_i\rangle_\text{imp} = \langle U^{\alpha X}_j\rangle_\text{imp} \eqdef\langle U^{\alpha X}\rangle_\text{imp}$ for all $i$ and $j$.

To find how the impurity-averaged Green's function, $\breve G_\text{imp} \defeq \langle \breve G\rangle_\text{imp}$ is related to the Green's function in the absence of impurities, $\breve G_0$, we take the impurity average of \cref{eq:dyson_a} with $\delta\breve\Sigma = \breve V^\text{imp}$ to obtain
\begin{equation}
  \breve G_\text{imp} = \breve G_0 + \breve G_0\bullet\left\langle\breve V^\text{imp}\bullet\breve G\right\rangle_\text{imp}.
  \label{eq:Vimp_1}
\end{equation}
We want an equation on the form
\begin{equation}
\breve G_\text{imp} = \breve G_0 + \breve G_0\bullet \breve\Sigma_\text{imp}\bullet\breve G_\text{imp}.
\label{eq:Vimp_goal}
\end{equation}
That is, we want to remove $\breve G$, which depends on the specific realizations of the impurity configuration.
To find $\breve\Sigma_\text{imp}$ to second order in $\breve V^\text{imp}$, we again set $\delta\breve\Sigma = \breve V^\text{imp}$ and insert \cref{eq:dyson_a} twice into \cref{eq:Vimp_1} to obtain
\begin{multline}
  \breve G_\text{imp} = \breve G_0 + \breve G_0\bullet\left\langle\breve V^\text{imp}\right\rangle_\text{imp}\bullet\breve G_0 
  \\+ \breve G_0\bullet\left\langle\breve V^\text{imp}\bullet\breve G_0\bullet \breve V^\text{imp}\right\rangle_\text{imp}\bullet\breve G_0
  \\+ \breve G_0\bullet\left\langle\breve V^\text{imp}\bullet\breve G_0\bullet \breve V^\text{imp}\bullet\breve G_0\bullet \breve V^\text{imp}\bullet\breve G\right\rangle_\text{imp}.
  \label{eq:Vimp_2}
\end{multline}
We need $\breve G_0$ as a function of $\breve G_\text{imp}$ to get \cref{eq:Vimp_goal}.
This can be found to the appropriate order in $\breve V^\text{imp}$ in inserting \cref{eq:dyson_a} with $\delta\breve\Sigma = \breve V^\text{imp}$ once into \cref{eq:Vimp_1} and solving for $\breve G_0$, giving
\begin{multline}
  \breve G_0 = \breve G_\text{imp} - \breve G_0\bullet\left\langle\breve V^\text{imp}\right\rangle_\text{imp}\bullet\breve G_0 
  \\- \breve G_0\bullet\left\langle\breve V^\text{imp}\bullet\breve G_0\bullet \breve V^\text{imp}\bullet\breve G\right\rangle_\text{imp}.
\end{multline}
In order to find a self-consistent expression for the impurity self-energy $\breve \Sigma_\text{imp}$ as a function of $\breve V^\text{imp}$ and $\breve G_\text{imp}$, we insert the expression for $\breve G_0$ iteratively into \cref{eq:Vimp_2}.
By comparing the result to \cref{eq:Vimp_goal}, this gives that, to second order in $\breve V^\text{imp}$,
\begin{multline}
  \breve \Sigma_\text{imp} = \left\langle\breve V^\text{imp}\right\rangle_\text{imp}
  + \left\langle\breve V^\text{imp}\bullet\breve G_\text{imp}\bullet \breve V^\text{imp}\right\rangle_\text{imp}
  \\
  - \left\langle\breve V^\text{imp}\right\rangle_\text{imp}\bullet\breve G_\text{imp}\bullet \left\langle\breve V^\text{imp}\right\rangle_\text{imp}.
\end{multline}

Using the properties of $p_\text{imp}$, we see that the first order term,
\begin{multline}
  \left[\breve\Sigma_\text{imp}^{(1)}(t_1,t_2)\right]_{nm} = \left[\left\langle\breve V^\text{imp}\right\rangle_\text{imp}(t_1,t_2)\right]_{nm}
 =  \delta_{nm}\delta(t_1-t_2)\\
\times\sum_{X \in\{A,B\}}\mqty(n^{LX}_\text{imp}\rho_X\langle U^{LX}\rangle_\text{imp}  & \\ & n^{RX}_\text{imp}\rho_X\langle U^{RX}\rangle_\text{imp}),
\end{multline}
is an energy shift that may be sublattice-dependent if the number or strength of impurities is different on the two sublattices.
It may in general also be spin-dependent if the impurities are magnetic, meaning that $U^{\alpha X}_i$ has a nontrivial structure in spin-space.
Here we assume that the impurities are not magnetic.
Nevertheless, we shall see in \cref{sec:revisiting_impurity_self_energy} that they will have an effective magnetic component in the final equations.

To evaluate the second-order term,
\begin{equation}
\breve\Sigma_\text{imp}^{(2)} = 
\left\langle\breve V^\text{imp}\bullet\breve G_\text{imp}\bullet \breve V^\text{imp}\right\rangle_\text{imp}
  - \left\langle\breve V^\text{imp}\right\rangle_\text{imp}\bullet\breve G_\text{imp}\bullet \left\langle\breve V^\text{imp}\right\rangle_\text{imp}.
\end{equation}
note that the assumption that the impurities are independent means that the contributions with different impurities to the left and right of the Green's function cancel.
Hence,
\begin{multline}
  \left[\breve\Sigma_\text{imp}^{(2)}(t_1,t_2)\right]_{nm}^{\alpha\beta}
  =  \delta_{\alpha\beta}\sum_{X \in\{A,B\}}\sum_{i=1}^{m^{\alpha X}}
  \\\times\Biggl[
  \rho_X(\check G_\text{imp}^{\alpha\alpha})_{nm}\rho_X\left\langle U^{\alpha X}_i \delta_{nr^{\alpha X}_i}U^{\alpha X}_i \delta_{mr^{\alpha X}_i}\right\rangle_\text{imp}\\
- \rho_X(\check G_\text{imp}^{\alpha\alpha})_{nm}\rho_X
\left\langle U^{\alpha X}_i \delta_{nr^{\alpha X}_i}\right\rangle_\text{imp}\left\langle U^{\alpha X}_i \delta_{mr^{\alpha X}_i}\right\rangle_\text{imp}
\Biggr]
\\
=\delta_{\alpha\beta}\sum_{X \in\{A,B\}}\delta_{nm}n_\text{imp}^{\alpha X}\left\langle U^{X \alpha}U^{X \alpha}\right\rangle_\text{imp}\rho_X(\check G_\text{imp}^{\alpha\alpha})_{nn}\rho_X
\\
-\delta_{\alpha\beta}\sum_{X \in\{A,B\}}\frac{n_\text{imp}^{\alpha X}}{N^{\alpha}}\left\langle U^{X \alpha}\right\rangle_\text{imp}^2\rho_X(\check G_\text{imp}^{\alpha\alpha})_{nm}\rho_X.
\end{multline}
We can neglect the second term because $N^\alpha$ is large and the amplitude of the Green's function decreases as a function of relative distance in the presence of impurities, as will be shown later.
Thus, to second order the impurity self-energy is
\begin{multline}
  \left[\breve\Sigma_\text{imp}(t_1,t_2)\right]_{nm}^{\alpha\beta}
  =\delta_{\alpha\beta}\delta_{nm}
  \sum_{X \in\{A,B\}}n_\text{imp}^{\alpha X}
   \left( \rho_X\left\langle U^{X \alpha}\right\rangle_\text{imp}\right. \\
  + \left.\left\langle U^{X \alpha}U^{X \alpha}\right\rangle_\text{imp}\rho_X(\check G_\text{imp}^{\alpha\alpha})_{nn}(t_1,t_2)\rho_X\right).
  \label{eq:impsecondorder}
\end{multline}
From here on we drop the subscript on the impurity averaged Green's function, such that $\breve G_\text{imp} \to \breve G$.

\section{Tunneling}%
\label{sec:tunneling}
In order to get closed equations for $\check G^{LL}$ and $\check G^{RR}$, we must first remove $\check G^{LR}$ and $\check G^{RL}$.
In this section, we do this by treating the tunneling self-energy as the perturbation in the Dyson equation.
However, we note that the derived effective tunneling self-energy is still of infinite order in the tunneling amplitudes $\hat T^{LR}$.

Let
\begin{equation}
  \breve T = \mqty( & \hat T^{LR} \\ \hat T^{RL} & ),
\end{equation}
and let $\breve G_0$ be the Green's function with $\hat T^{RL} = \hat T^{LR} = 0$, meaning that it solves \cref{eq:gorkov_breve_g0} with $\delta\breve \Sigma = \breve\Sigma - \breve T = \diag(\check\Sigma^{LL}, \check\Sigma^{RR})$.
Here $\delta\breve\Sigma$ includes the impurity self-energy term obtained from the impurity average above.
Note that this means that $\check G_0^{RR}$ still depends on $\check G^{LL}$.
This is because $\check G_0^{RR}$ depend on $\check G^{RR}$ through the impurity self-energy found in \cref{sec:impurity_averaging}, and $\check G^{RR}$ depend on $\check G^{LL}$.
For the same reason $\check G_0^{LL}$ depends on $\check G^{RR}$.

From the Dyson equation, \eqref{eq:dyson}, we have that 
\begin{subequations}
  \begin{align}
    \label{eq:tunnel1_a}
  \breve G = \breve G_0 + \breve G_0\bullet \breve T \bullet \breve G,\\
    \label{eq:tunnel1_b}
  \breve G = \breve G_0 + \breve G\bullet \breve T \bullet \breve G_0.
  \end{align}
\end{subequations}
From the upper right block of \cref{eq:tunnel1_a} we have that
\begin{multline}
  \check G^{LR} = \check G^{LR}_0 + \check G^{LR}_0\bullet \hat T^{RL} \bullet \check G^{LR} + \check G^{LL}_0\bullet \hat T^{LR} \bullet \check G^{RR}
  \\
  = \check G^{LR}_0\bullet \left(i\tau_z\frac{\partial \check G^{RR}}{\partial t} - \check \Sigma^{RR}\bullet\check G^{RR} - \hat T^{RL}\bullet\check G^{LR}\right)
  \\
  + \check G^{LR}_0\bullet \hat T^{RL} \bullet \check G^{LR} + \check G^{LL}_0\bullet \hat T^{LR} \bullet \check G^{RR}
  \\
  = -\left(\frac{\partial \check G^{LR}_0}{\partial t'}i\tau_z + \check G^{LR}_0\bullet \check \Sigma^{RR}\right)\bullet \check G^{RR} + \check G^{LL}_0\bullet \hat T^{LR} \bullet \check G^{RR}
  \\
  = \check G^{LL}_0\bullet \hat T^{LR} \bullet \check G^{RR},
\end{multline}
where we used \cref{eq:gorkov_breve_g0} in the last equality.

Doing the same for $\check G^{LR}$, and from similar calculations using \cref{eq:tunnel1_b} we find that
\begin{subequations}
  \begin{align}
    \check G^{LR} = \check G_0^{LL}\bullet \hat T^{LR}\bullet\check G^{RR} 
      = \check G^{LL}\bullet \hat T^{LR}\bullet \check G^{RR}_0,
      \\
    \check G^{RL} = \check G_0^{RR}\bullet \hat T^{RL}\bullet\check G^{LL} 
      = \check G^{RR}\bullet \hat T^{RL}\bullet\check G^{LL}_0.
  \end{align}
\end{subequations}
Inserting this into the Gor'kov equation, we can remove $\check G^{RL}$ and $\check G^{LR}$ and get a block-diagonal self-energy,
\begin{equation}
  \label{eq:SigmaBreve}
  \breve \Sigma = \breve H_0 + \breve V + \breve\Sigma_\text{imp} + \breve \Sigma_T,
  %\mqty(\hat H_0^L + \check V^L + \hat T^{LR}\bullet\check G_0^{RR}\bullet \hat T^{RL} & \\ &\hat H_0^R + \check V^R + \hat T^{RL}\bullet\check G_0^{LL}\bullet \hat T^{LR}) + \breve\Sigma_\text{imp}
\end{equation}
where
\begin{subequations}
  \begin{align}
    \breve H_0 &= \mqty(\hat H_0^L & \\ & \hat H_0^R),\\
    \breve V &= \mqty(\check V^L & \\ & \check V^R),\\
    \breve \Sigma_T &= \mqty(\hat T^{LR}\bullet \check G_0^{RR}\bullet \hat T^{RL} & \\ & \hat T^{RL}\bullet \check G_0^{LL}\bullet \hat T^{LR}).
  \end{align}
\end{subequations}

\section{Fourier transform and Wigner Coordinates}%
\label{sec:wigner_coordinates}
In the quasiclassical framework, functions vary slowly with the center-of-mass (COM) coordinates, and quickly with the relative coordinates.
It is therefore useful to Fourier transform in the relative coordinates to obtain functions of momentum, energy, COM time and COM position, also known as Wigner coordinates.
The Fourier transform in relative time reads
\begin{equation}
  \mathcal F_t(A)(T, \varepsilon) = \int_{-\infty}^\infty \dd{t} A(T + t/2, T - t/2)\me{i\varepsilon t},
\end{equation}
and for the Fourier transform in relative position we use
\begin{equation}
  \mathcal F_r(A)(\vec k, \vec x_n^\alpha) = \sum_{m\in\mathbb{Z}^3}\me{-i\rho_B\vec k \cdot \vec \delta^\alpha} A_{(n+m)n}\me{i\rho_B\vec k \cdot \vec \delta^\alpha}\me{-i\vec k \cdot \vec x_m^\alpha}.
  \label{eq:FT}
\end{equation}
This is is a three-dimensional discrete-time Fourier transform (DTFT), and the inverse transform is given by
\begin{equation}
  \mathcal F^{-1}_r(A)_{(n+m)n} = V_e^\alpha\int_{\Diamond_\alpha}\frac{\dd[3]{k}}{(2\pi)^3} \me{i\rho_B\vec k \cdot \vec \delta^\alpha}A(\vec k, \vec x_n^\alpha)\me{-i\rho_B\vec k \cdot \vec \delta^\alpha}\me{i\vec k \cdot \vec x_m^\alpha},
\end{equation}
where $V_e^\alpha$ is the volume of the unit cell and $\Diamond_\alpha$ is the first Brillouin zone in material $\alpha$.
Note that $\vec x_n^\alpha$ in \cref{eq:FT} is not exactly the COM position, since the COM position for term $m$ on the right is $(\vec x_n^\alpha + \vec x^\alpha_m)/2$.

We use the same symbols as before to denote the bullet and circle products in the Wigner coordinates, meaning that they satisfy
\begin{equation}
  \mathcal F_r[\mathcal F_t(A)]\bullet\mathcal F_r[\mathcal F_t(B)]
  = \mathcal F_r[\mathcal F_t(A\bullet B)]
\end{equation}
and
\begin{equation}
  \mathcal F_t(A)\circ\mathcal F_t(B)
  = \mathcal F_t(A\circ B).
\end{equation}
Thus, the Gor'kov equations in the Wigner coordinates read
\begin{subequations}
  \label{eq:gorkov_wigner}
\begin{align}
  \tau_z\varepsilon\circ \breve G - \breve\Sigma \bullet \breve G &= 1,
  \\
  \breve G\circ \tau_z\varepsilon -  \breve G\bullet\breve\Sigma &= 1.
\end{align}
\end{subequations}   

The circle product in the Wigner product is the same as in continuous models for normal metals~\cite{larkin1986,belzig1999},
\begin{equation}
  A\circ B = \exp(\frac i 2 \partial_\varepsilon^A \partial_T^B - \frac i 2 \partial_T^A \partial_\varepsilon^B)AB,
  \label{eq:circWigner}
\end{equation}
where the superscripts on the differential operators denote which function they act on.
The spatial part of the bullet product, on the other hand, is different, and there  are three reasons for this.
First, since we are working on a discrete lattice, we cannot Taylor expand, which is how the series expansion in differential operators is achieved in \cref{eq:circWigner}.
Second, since we are working with two sublattices that are located differently in space, the COM positions and relative positions are different for different matrix elements.
Third, the COM position is not set constant in the way we have defined the Fourier transform in \cref{eq:FT}.
Nevertheless, the bullet product can still be written as a series of differential operators of increasing order.
To derive the explicit series expansion, one can use the Newton forward difference equation, which is the discrete analog to the Taylor series expansion.
The zeroth order term is the same, namely just the normal matrix product, and we will end up keeping only the zeroth order terms, except for the kinetic energy term, the tunneling term, and the potential which is large outside the material.
We will evaluate these terms explicitly when considering the boundary condition.
Note, however, that we cannot neglect the higher-order terms at this stage because the Green's function is strongly peaked in momentum space.

To evaluate $\hat H_0^\alpha \bullet \check G^{\alpha\alpha}$ and $\check G^{\alpha\alpha}\bullet\hat H_0^\alpha$ in Wigner coordinates, note that
\begin{multline}
  \label{eq:bullet_wigner}
  \mathcal F_r\left[A\bullet B\right](\vec k, \vec x_n^\alpha) = \sum_{m\in\mathbb{Z}^3}\mathcal F[A](\vec k, \vec x_m^\alpha + \vec x_n^\alpha)\\
  \circ\me{-i\rho_B \vec k \cdot \vec \delta^\alpha}B_{(n+m)n}\me{i\rho_B \vec k \cdot \vec \delta^\alpha}\me{-i\vec k\cdot\vec x_m^\alpha}.
\end{multline}
Hence, as $\hat H_0^\alpha$ does not depend on COM-position,
\begin{equation}
  \mathcal F_t\left\{\mathcal F_r\left[\hat H_0^\alpha\bullet \check G^{\alpha\alpha}\right]\right\}(\vec k, \vec x_n^\alpha) = \hat H_0^\alpha(\vec k) \check G^{\alpha\alpha}(\vec k, \vec x_n^\alpha).
\end{equation}
Here,
\begin{equation}
  \hat H_0^\alpha(\vec k) = \rho_x K^\alpha(\vec k) - (J^\alpha\rho_z\sigma_z\tau_z + \mu),
\end{equation}
and
\begin{equation}
  K^\alpha(\vec k) = -\sum_{\vec \delta_i \in \text{N.N.}}t^\alpha\cos(\vec k \cdot \vec \delta_i)
  \label{eq:kineticEnergy}
\end{equation}
where the sum goes over all the six nearest neighbor displacement vectors.

Interchanging the order, we find that
\begin{multline}
  \mathcal F_t\left\{\mathcal F_r\left[\check G^{\alpha\alpha}\bullet\hat H_0^\alpha \right]\right\} = \check G^{\alpha\alpha}\hat H_0^\alpha \\
  - \frac 1 2 \sum_{\vec\delta_i \in \text{N.N.}} \left(\Delta_R \check G^{\alpha\alpha}\right) \cdot (\vec\delta_i + \vec\delta^\alpha)(\rho_x + i\rho_y)t^\alpha \me{-i\vec k\cdot \vec\delta_i}
  \\
  - \frac 1 2 \sum_{\vec\delta_i \in \text{N.N.}} \left(\Delta_R \check G^{\alpha\alpha}\right) \cdot (\vec\delta_i - \vec\delta^\alpha)(\rho_x - i\rho_y)t^\alpha \me{-i\vec k\cdot \vec\delta_i}
  % + i\left(\Delta_R \check G^{\alpha\alpha}\right)\cdot\left(\rho_x \nabla_k K^\alpha + \vec\delta^\alpha \rho_y K^\alpha\right),
\end{multline}
where the symbols on the right-hand side denote functions of Wigner coordinates and the discrete finite difference operator is defined as
\begin{equation}
  \vec x_m^\alpha \cdot \Delta_R \check G^{\alpha\alpha}(\vec k, \vec x_n^\alpha)
  = \check G^{\alpha\alpha}(\vec k, \vec x_n^\alpha + \vec x_m^\alpha) - \check G^{\alpha\alpha}(\vec k, \vec x_n^\alpha).
\end{equation}
The finite difference is only well-defined when $\vec x^\alpha_m$ is a lattice vector, meaning that $\vec x^\alpha_m$ is a displacement vector from one unit cell to another.
However, we can define
\begin{equation}
  \vec \delta_i \cdot \Delta_R \check G^{\alpha\alpha}(\vec k, \vec x_n^\alpha)
  = [\check G^{\alpha\alpha}(\vec k, \vec x_n^\alpha + 2\vec \delta_i) - \check G^{\alpha\alpha}(\vec k, \vec x_n^\alpha)]/2.
\end{equation}
This is possible because we assume that $2\vec\delta_i$ is a lattice vector when $\vec\delta_i$ is a nearest neighbor displacement vector.
With this,
\begin{multline}
  \vec \delta_i \cdot \Delta_R \check G^{\alpha\alpha}(\vec k, \vec x_n^\alpha)
  + 
  \vec \delta^\alpha \cdot \Delta_R \check G^{\alpha\alpha}(\vec k, \vec x_n^\alpha)
  \\
  = (\vec \delta_i + \vec \delta^\alpha)\cdot\Delta_R \check G^{\alpha\alpha}(\vec k, \vec x_n^\alpha)
  + \frac 1 2 \check G^{\alpha\alpha}(\vec k, \vec x_n^\alpha + 2\vec\delta_i) \\
  + \frac 1 2 \check G^{\alpha\alpha}(\vec k, \vec x_n^\alpha + 2\vec\delta^\alpha)
  - \check G^{\alpha\alpha}(\vec k, \vec x_n^\alpha + \vec\delta_i + \vec\delta^\alpha).
  % +\abs{\delta_i + \delta^\alpha}^2\frac{\check G^{\alpha\alpha}(\vec k, \vec x_n^\alpha) - 2\check G^{\alpha\alpha}(\vec k, \vec x_n^\alpha) + \check G^{\alpha\alpha}(\vec k, \vec x_n^\alpha)}{\abs{\delta_i + \delta^\alpha}^2}
\end{multline}
The last three terms are equal to $\abs{\vec\delta_i - \vec\delta^\alpha}^2/2$ times the second order central difference of $\check G^{\alpha\alpha}$, so they are negligible under the assumption that the Green's function changes slowly as a function COM position compared to the interlattice spacing.
By the same reasoning we also approximate $(-\vec x_m^\alpha)\cdot\Delta_R \check G^{\alpha\alpha} = -\vec x_m^\alpha\cdot\Delta_R \check G^{\alpha\alpha}$, since the difference is equal to the $\abs{\vec x_m^\alpha}^2$ multiplied by the second order derivative of $\check G^{\alpha\alpha}$.
With this we have
\begin{multline}
  \mathcal F_t\left\{\mathcal F_r\left[\check G^{\alpha\alpha}\bullet\hat H_0^\alpha \right]\right\} = \check G^{\alpha\alpha}\hat H_0^\alpha \\
  % + \frac 1 2 \sum_{\vec\delta_i} \left(\Delta_R \check G^{\alpha\alpha}\right) \cdot (\vec\delta_i + \vec\delta^\alpha)(\rho_x + i\rho_y)t^\alpha \me{-i\vec k\cdot \vec\delta_i}
  \\
  % + \frac 1 2 \sum_{\vec\delta_i} \left(\Delta_R \check G^{\alpha\alpha}\right) \cdot (\vec\delta_i - \vec\delta^\alpha)(\rho_x - i\rho_y)t^\alpha \me{-i\vec k\cdot \vec\delta_i}
  + i\left(\Delta_R \check G^{\alpha\alpha}\right)\cdot\left(\rho_x \nabla_k K^\alpha + \vec\delta^\alpha \rho_y K^\alpha\right).
\end{multline}
The dot product in the last term must be interpreted in the following sense:
If $\nabla_k K^\alpha = A_1\vec \delta_1 + A_2\vec \delta_2 + A_3\vec \delta_3$, where $\vec \delta_1$, $\vec \delta_2$ and $\vec \delta_3$ are three different, linearly independent, nearest neighbor displacement vectors, then 
\begin{equation}
  \label{eq:dotprodinterp}
  \left(\Delta_R \check G^{\alpha\alpha}\right)\cdot \nabla_k K^\alpha
  = \sum_{i=1}^3 \vec\delta_i\cdot\left(\Delta_R \check G^{\alpha\alpha}\right) A_i.
\end{equation}

\section{Extracting the conduction band}%
\label{sec:extracting_the_conduction_band}
The main idea behind the quasiclassical theory is that most of the interesting physics happens close to the Fermi surface.
Therefore, it is of interest to isolate the contribution from states close to the Fermi surface.
In our model there are two energy bands that are not overlapping, so only one of these can pass through the Fermi surface.
In real materials, it is not always the case that the energy bands are not overlapping.
It is sufficient that the energy bands are not overlapping near the Fermi surface.

To separate the two bands, we must diagonalize $\hat H_0^\alpha$.
We find that
\begin{equation}
  \hat H_0^\alpha = S^\alpha D^\alpha \transpose{(S^\alpha)},
\end{equation}
where
\begin{equation}
  D^\alpha = \diag(\xi_-^\alpha,\xi_-^\alpha,\xi_-^\alpha,\xi_-^\alpha,\xi_+^\alpha,\xi_+^\alpha,\xi_+^\alpha,\xi_+^\alpha)
\end{equation}
and $\transpose{(S^\alpha)}$ denotes the transpose of
\begin{multline}
  S^\alpha = \frac{1}{\sqrt{2\eta^\alpha}}\left[\mqty(-\sigma_0 & 0 & \sigma_0 & 0 \\ \sigma_0 & 0 & \sigma_0 & 0 \\ 0 & -\sigma_0 & 0 & \sigma_0 \\ 0 & \sigma_0 & 0 & \sigma_0) \bar s^\alpha \right.
    \\
  - \left.\mqty(\sigma_z & 0 & \sigma_z & 0 \\ \sigma_z & 0 & -\sigma_z & 0 \\ 0 & -\sigma_z & 0 & -\sigma_z \\ 0 & -\sigma_z & 0 & \sigma_z)\Delta s^\alpha
  \right],
\end{multline}
where $\sigma_0$ is the $2\times 2$ identity matrix, $\eta^\alpha = \sqrt{(J^\alpha)^2 + (K^\alpha)^2}$, $\xi^\alpha_\pm = -\mu^\alpha \pm \eta^\alpha$, $\bar s^\alpha = (s_+^\alpha + s_-^\alpha)/2$ and $\Delta s^\alpha = (s_+^\alpha - s_-^\alpha)/2$, with $s_\pm^\alpha = \sqrt{\eta^\alpha \pm J^\alpha}$.

Next, we define
\begin{equation}
  \mqty(\check G_{--}^{\alpha\alpha} & \check G_{-+}^{\alpha\alpha} \\ \check G_{+-}^{\alpha\alpha} & \check G_{++}^{\alpha\alpha}) = \transpose{(S^\alpha)}\check G^{\alpha\alpha} S^\alpha.
\end{equation}
We want an equation for the Green's function associated with the energy band which crosses the Fermi surface.
This can be either $\check G_{--}^{\alpha\alpha}$ or $\check G_{++}^{\alpha\alpha}$.
Here we choose $\check G_{--}^{\alpha\alpha}$.
To derive this equation, we first find that
\begin{equation}
  \transpose{(S^\alpha)}\rho_x \nabla_k K^\alpha S^\alpha
  = \nabla_k D  + \frac{J^\alpha\nabla_k \eta^\alpha}{K^\alpha}\mqty(0 & 0 & \sigma_z & 0 \\ 0 & 0 & 0 & -\sigma_z \\ \sigma_z & 0 & 0 & 0 \\ 0 & -\sigma_z & 0 & 0)
\end{equation}
and
\begin{equation}
  \transpose{(S^\alpha)}i\rho_y  S^\alpha
  =\mqty(0 & 0 & -\sigma_0 & 0 \\ 0 & 0 & 0 & -\sigma_0 \\ \sigma_0 & 0 & 0 & 0 \\ 0 & \sigma_0 & 0 & 0).
\end{equation}
Additionally, we continue to use $\tau_z$ to denote the third Pauli matrix in Nambu-space after transforming to the band basis, which means that
\begin{equation}
  \transpose{(S^\alpha)}\tau_z  S^\alpha = \mqty(\sigma_0 & 0 & 0 & 0 \\ 0 & -\sigma_0 & 0 & 0 \\ 0 & 0 & \sigma_0 & 0 \\ 0 & 0 & 0 & -\sigma_0)
  = \mqty(\tau_z & 0 \\ 0 & \tau_z).
\end{equation}

Transforming the first Gor'kov equation to the AFM energy band basis and extracting the block corresponding to the conduction band, we get
\begin{equation}
  % i\tau_z \varepsilon \circ \check G_{--}^{\alpha\alpha} - \xi^\alpha_-\check G_{--}^{\alpha\alpha} - \left\{\transpose{(S^\alpha)} \left[(\check \Sigma^{\alpha}-\hat H_0^\alpha)\bullet \check G^{\alpha}\right] S^\alpha\right\}_{--} = 1,
  \tau_z \varepsilon \circ \check G_{--}^{\alpha\alpha} - \xi^\alpha_-\check G_{--}^{\alpha\alpha} - \left[(\check \Sigma^{\alpha}-\hat H_0^\alpha)\bullet \check G^{\alpha\alpha}\right]_{--} = 1,
  \label{eq:gorkov_cond_1}
\end{equation}
where $\check \Sigma^{\alpha}$ is the block of $\breve\Sigma$, given by \cref{eq:SigmaBreve}, corresponding to material $\alpha$ and the subscript on the last term on the left-hand side means that one should take the upper left block in the conduction band basis.
That is, for a general matrix $A$ in the sublattice basis,
\begin{equation}
  \mqty(A_{--} & A_{-+} \\ A_{+-} & A_{++}) = \transpose{(S^\alpha)}A S^\alpha.
\end{equation}
The second Gor'kov equation becomes
\begin{multline}
  \check G_{--}^{\alpha\alpha}\circ \tau_z\varepsilon -\xi^\alpha_-\check G_{--}^{\alpha\alpha} - i\nabla_k\xi^\alpha_{-}\cdot\Delta_R\check G_{--}^{\alpha\alpha} \\
  - \frac{iJ^\alpha\nabla_k\eta^\alpha}{K^\alpha}\cdot\Delta_R\check G_{-+}^{\alpha\alpha}\tau_z\sigma_z- K^\alpha\vec\delta^\alpha\cdot\Delta_R\check G_{-+}^{\alpha\alpha}
  \\ - \left[\check G^{\alpha\alpha}\bullet(\check \Sigma^{\alpha}-\hat H_0^\alpha) \right]_{--} = 1.
  \label{eq:gorkov_cond_2}
\end{multline}

\section{Quasiclassical Green's functions}%
\label{sec:quasiclassical_green_s_functions}
In this section, we derive the quasiclassical equations of motion.
To do so, we must integrate the Green's function over momenta.
Note that since we only want the contribution from states close to the Fermi surface, we cannot integrate over all momenta, but must instead integrate over a contour close to the Fermi surface.
While it is true that the Green's function will be strongly peaked around the Fermi surface, the contribution from far away from the Fermi surface is not negligible.
This is because the retarded and advanced Green's function goes as $1/\xi_-^\alpha$ far away from the Fermi surface.

\begin{figure}
  \centering
  \begin{tikzpicture}[scale=0.6]
    \draw [>->, black, thick, decoration={markings, 
      mark=at position 0.5 with {\arrow{>}}}, postaction={decorate}]
      (-5,0) -- (-3,0) node [above] {$\int \dd{\xi_-^\alpha}$} -- (-1,0) node [right] {$=$};
    \node at (-0.2,0) {$\frac 1 2$};
    \draw [<-<, black, thick, decoration={markings,
      mark=at position 0.25 with {\arrow{>}},
        mark=at position 0.75 with {\arrow{>}}}, postaction={decorate}]
      (2,-0.1)  arc (0:-180:1) -- cycle;
    \draw [<-<, black, thick, decoration={markings,
      mark=at position 0.25 with {\arrow{<}},
      mark=at position 0.75 with {\arrow{<}}}, postaction={decorate}]
      (0,0.1)  arc (180:90:1) node [above] {$\oint \dd{\xi_-^\alpha}$} -- (1,1.1) arc(90:0:1) -- cycle;

    \node at (2.5,0) {$ + \frac 1 2$};
    \draw [<-<, black, thick, decoration={markings,
      mark=at position 0.25 with {\arrow{<}},
        mark=at position 0.75 with {\arrow{<}}}, postaction={decorate}]
        (7,-0.1) -- (6,-0.1)  arc (0:-180:1) -- (3,-0.1);
    \draw [>->, black, thick, decoration={markings,
      mark=at position 0.25 with {\arrow{>}},
      mark=at position 0.75 with {\arrow{>}}}, postaction={decorate}]
      (3,0.1) -- (4,0.1)  arc (180:90:1) node [above] {$\fint \dd{\xi_-^\alpha}$} -- (5,1.1) arc(90:0:1) -- (7,0.1);
  % \draw [>->, rounded corners, thick, black, decoration={markings, 
  %   mark=at position 0.15 with {\arrow{>}},
  %   mark=at position 0.65 with {\arrow{>}}},
  %   postaction={decorate}] (-5,0.2) -- (3,0.2) node [above] {$\mathcal C$} -- (8,0.2) -- (8,-0.2) node [right, black] {$\tau$} -- (-5,-0.2);
\end{tikzpicture}
\caption{A sketch of the integration decomposition introduced by Eilenberger~\cite{eilenberger1968}.}
\label{fig:eilenbergContour}
\end{figure}
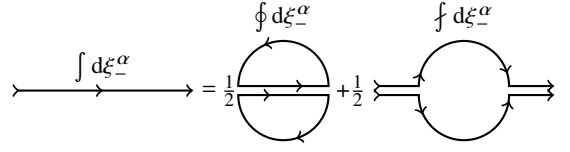

Observables are given as integrals over all momenta.
To extract the quasiclassical contribution, one must decompose this integral into one part which includes the contribution close to the Fermi surface and one part which includes the rest.
By using the Eilenberger decomposition~\cite{eilenberger1968}, as illustrated in \cref{fig:eilenbergContour}, the contribution from the Fermi surface is included as two closed contours in the complex plane, which simplifies the calculations.
We show how observables can be expressed as a quasiclassical contribution and a rest-term in \cref{sec:observables}.

To get the quasiclassical equations of motion, we must integrate the Gor'kov equations over the closed contours.
This allows us to simplify many of the bullet products when the self-energy varies slowly as a function of COM position, as we show in this section.
Note that the tunneling term and the potential which is large outside of the material change rapidly as a function of COM position.
These can therefore not be simplified in the same way.
However, these terms are only nonzero at the interface.
In this section, we consider only positions inside the material and therefore ignore these terms.
We will return to them when deriving the boundary conditions.
Hence, as long as we consider COM positions away from the boundaries,
\begin{equation}
  \check\Sigma^\alpha - \hat H_0^\alpha = \check V^\alpha + \check\Sigma_\text{imp}^\alpha.
\end{equation}

To simplify the bullet product, we can use the gradient expansion.
However, the gradient expansion is more complicated in our case compared to the continuous case.
This is because we are working with two discrete sublattices.
To derive the gradient expansion for discrete lattices, we can use the Newton forward difference equation.
If the basis vectors are $\{\vec v_1^\alpha, \vec v_2^\alpha, \vec v_3^\alpha\}$, such that $\vec x_m^\alpha = a_1^m\vec v_1^\alpha + a_2^m\vec v_2^\alpha + a_3^m\vec v_3^\alpha$, with integers $a_1^m$, $a_2^m$ and $a_3^m$, then
\begin{equation}
  A(\vec x_n^\alpha + \vec x^\alpha_m) = \sum_{j\in \mathbb{N}_0^3}
  \frac{(\vec x^\alpha_m)_j}{j!}\prod_{i=1}^3\left(\frac{[\sgn(a_i^m)\vec v_i^\alpha]\cdot \Delta_R}{\abs{\vec v_i^\alpha}}\right)^{j_i}A(\vec x_n^\alpha),
  %\frac{\Delta_{R\sgn(l)}^j A(\vec x^\alpha_n)}{j!}(\vec x^\alpha_m)_j
\end{equation}
where $j=(j_1,j_2,j_3)$ is a multi-index, $j! = j_1!j_2!j_3!$ and
\begin{subequations}
  \begin{align}
    (\vec x_m^\alpha)_j &= (a_1^m\vec v_1^\alpha)_{j_1}(a_2^m\vec v_2^\alpha)_{j_2}(a_3^m\vec v_3^\alpha)_{j_3},
  \\
    (a_i^m\vec v_i^\alpha)_{j_i} &= \sgn(a_i^m)^{j_i}\abs{\vec v_i^\alpha}\left[\abs{a_i^m} - (n-1)\right]\left(\abs{a_i^m\vec v_i^\alpha}\right)_{j_i-1},
  \end{align}
\end{subequations}
Hence, we see from \cref{eq:bullet_wigner} that the bullet product can be written as a series expansion in derivative operators,
\begin{equation}
  A \bullet B = A\circ B
  + [\Delta_R A] \circ (i\nabla_k B - [\vec\delta^\alpha \rho_B,\, B])
  + \dotsb,
\label{eq:gradientExpansion}
\end{equation}
where the circle product in the second term on the right-hand side includes a dot product, which must be interpreted according to \cref{eq:dotprodinterp}.
\Cref{eq:gradientExpansion} is the gradient expansion.
The gradient expansion is useful because the higher-order terms can be neglected after a proper integral over momenta.

We define the quasiclassical Green's function
\begin{equation}
  \check g^{\alpha} = \frac i \pi \oint \dd{\xi^\alpha_-}\check G^{\alpha\alpha}_{--},
  \label{eq:quasiclassical_definition}
\end{equation}
where the closed paths are illustrated in \cref{fig:eilenbergContour}.
They follow the real line from $\xi^\alpha_- = -E_c^\alpha$ to $\xi^\alpha_- = E_c^\alpha$ and then split into two semicircular paths to close the contours.
Here, $E_c^\alpha$ is some cutoff that is far larger than the other energies in the system, but smaller than $\lvert\mu^\alpha\rvert$.
Since the interval $(-E_c^\alpha,E_c^\alpha)$ must be inside the conduction band, $E_c^\alpha$ must also be smaller than, $\Delta E^\alpha$, which is the smallest energy difference between the Fermi level and the edges of the conduction band.

We can relate the kinetic energy $K^\alpha$ at the Fermi level to $\Delta E^\alpha$.
To do so, note that $\xi_-^\alpha + \Delta E^\alpha = -\mu^\alpha - \sqrt{(J^\alpha)^2+(K^\alpha)^2} + \Delta E^\alpha \le -\mu^\alpha - \lvert J^\alpha\rvert$ means that
\begin{equation}
  \left(\frac{J^\alpha}{K^\alpha}\right)^2 \le \frac{(J^\alpha)^2}{2\lvert J^\alpha\rvert \Delta E^\alpha + (\Delta E^\alpha)^2}.
  \label{eq:JoverK}
\end{equation}
It is possible that $\Delta E^\alpha \ll J^\alpha$.
For this reason, one can still consider $J^\alpha \gg K^\alpha$ within this framework, meaning that $J^\alpha/\eta^\alpha = J^\alpha/\sqrt{(J^\alpha)^2+(K^\alpha)^2} \to 1$.
The only requirement for the quasiclassical theory presented here to be valid is that $\Delta E^\alpha$ is large compared to all other energies except possibly the exchange energy $J^\alpha$.
We can have any ratio $J^\alpha/K^\alpha$, and the limit $J^\alpha/K^\alpha \to 0$ should reproduce the quasiclassical theory for normal metals.

Since the contours are closed in the complex plane and we assume that the functions are analytic in $\xi^\alpha_-$, we can use the residue theorem to evaluate
\begin{equation}
  \check g^\alpha = -\sum_{\xi_i}\sgn(\Im[\xi_i])\Res(\check G^{\alpha\alpha}_{--}, \xi_i),
\end{equation}
where the sum goes over all poles of $\check G^{\alpha\alpha}_{--}$ which are inside the contours and $\Res\big(\check G^{\alpha\alpha}_{--}, \xi_i)$ denote the corresponding residues.
To obtain an equation for the quasiclassical Green's function, we integrate the Gor'kov equations, \cref{eq:gorkov_cond_1,eq:gorkov_cond_2}, over momenta and use \cref{eq:gradientExpansion,eq:quasiclassical_definition}.

Consider first terms on the form $\check G_{--}^{\alpha\alpha}\bullet A$, for some $A$.
The zeroth order term in gradients is $\check G_{--}^{\alpha\alpha}\circ A$.
If $A(\xi_-^\alpha)$ has no poles inside the contour, we see that
\begin{multline}
  \frac i \pi \oint \dd{\xi^\alpha_-}\check G^{\alpha\alpha}_{--} \circ A
  = -\sum_{\xi_i}\sgn(\Im[\xi_i])\Res(\check G^{\alpha\alpha}_{--}, \xi_i)\circ A(\xi_-^\alpha)
  \\
  = \check g^\alpha \circ A(0)
  + \mathcal O(\check g^\alpha \circ a \partial_{\xi_-^\alpha}A),
  \label{eq:zerothapproxDerivation}
\end{multline}
where $a$ is the maximal distance from the poles of $\check G^{\alpha\alpha}_{--}$ to $\xi^\alpha_- = 0$.
It is therefore much smaller than $E_c^\alpha$.
We can neglect the second term when $A$ varies slowly as a function of $\xi_-^\alpha$, such that $\abs{a \partial_{\xi_-^\alpha}A} \ll \abs{A}$.
Note that this is not true when $A=\xi_-^\alpha$, which is the case for the second terms on the left-hand sides of \cref{eq:gorkov_cond_1,eq:gorkov_cond_2}.
We can therefore not evaluate $\oint \dd{\xi^\alpha_-}\xi_-^\alpha\check G^{\alpha\alpha}_{--}$ in terms of the quasiclassical Green's function.

From \cref{eq:impsecondorder} we see that $(\check\Sigma_\text{imp}^\alpha)_{--}$ only depend on momentum through $S$.
\Cref{eq:V_beginning} shows that $\check V^\alpha_{nm}$ depend on relative coordinates if there are magnetic textures or if the Hamiltonian includes terms other than the kinetic term which depend on relative position.
Corrections to the hopping term from the vector potential or spin-orbit coupling are included in $\check V^\alpha_{nm}$, and these terms will depend on relative position.
As a result, $(\check V^\alpha)_{--}$ depends on momentum, and therefore also on $\xi_-^\alpha$.
However, we assume that the dependence on momentum and $\xi_-^\alpha$ is sufficiently slow, such that the condition $\abs{a \partial_{\xi_-^\alpha}A} \ll \abs{A}$ is valid when $A = (\check V^\alpha + \check\Sigma_\text{imp}^\alpha)_{--}$.
As we now show, this assumption is reasonable as long as the Fermi level is far away from the bottom of the conduction band.

Fourier transforming the term in \cref{eq:V_beginning} coming from the magnetic texture, we get that
\begin{multline}
  \mathcal F_r\left\{\left(K^\alpha_{nm}[\vec x_n^\alpha - \vec x_m^\alpha]
                                     + \left[\vec\delta^\alpha\rho_B,\, K_{nm}^\alpha\right]\right)
     \cdot \left(R^\dagger \nabla R\right)\left(\vec x_n^\alpha,t_1\right)\right\}
     \\
     = i\nabla_k \mathcal F_r\{K^\alpha_{nm}\}(\vec k, \vec x_n^\alpha) \cdot \left(R^\dagger \nabla R\right)\left(\vec x_n^\alpha,t_1\right).
\end{multline}
As long as the Fermi level is sufficiently far away from the bottom of the conduction band, the gradient $\nabla_k \mathcal F_r\{K^\alpha_{nm}\} = \nabla_k K^\alpha$ will be approximately constant near the Fermi surface.
This can be seen from \cref{eq:kineticEnergy}, since
\begin{multline}
\left\lvert\frac{a\partial_{\xi^\alpha_-} \nabla_k K^\alpha}{\nabla_k K^\alpha}\right\rvert
  = \left\lvert \frac{a}{\vec v_F^\alpha \cdot \hat{\vec k}_F} \frac{\sum_{\vec \delta_i}\vec \delta_i (\vec \delta_i \cdot \hat{\vec k}_F)\cos(\vec k_F \cdot \vec \delta_i)}{\sum_{\vec\delta_i}(\vec \delta_i \cdot \hat{\vec k}_F)\sin(\vec k_F \cdot \vec \delta_i)}\right\rvert
  \\
  < 
   \left\lvert \frac{a}{\vec v_F^\alpha \cdot \hat{\vec k}_F} \frac{\sum_{\vec \delta_i}\vec \delta_i (\vec \delta_i \cdot \hat{\vec k}_F)\cos(\vec k_F \cdot \vec \delta_i)}{\sum_{\vec\delta_i}(\vec \delta_i \cdot \hat{\vec k}_F)(\vec k_F \cdot \vec \delta_i)}\right\rvert
   \lesssim \frac{a\lambda_F}{\lvert\vec v_F^\alpha \cdot \hat{\vec k}_F\rvert},
\end{multline}
where $\vec v_F^\alpha = \nabla_k\xi_-^\alpha$ is the Fermi velocity, $\lambda_F = 1/\lvert\vec k_F \rvert$ is the Fermi wavelength and $\hat{\vec k}_F$ is the unit vector in the direction of $\vec k_F$.
Hence, the variation in $\nabla_k K^\alpha$ is negligible provided that $\lambda_F \ll \lvert \vec v_F^\alpha\rvert/a$.

Physically, the condition can be understood in the following sense.
The inverse energy $1/a$ defines a time, so $\lvert \vec v_F^\alpha\rvert/a$ is the distance an electron with speed $\lvert \vec v_F^\alpha\rvert$ travels in this time.
For instance, when the dominant energy scale, other than $J^\alpha$ and $K^\alpha$, comes from the impurity scattering, then $a$ is at most the impurity scattering rate.
In this case, the condition $\lambda_F \ll \lvert \vec v_F^\alpha\rvert/a$ implies that the mean free path should be much greater than the Fermi wavelength.
This condition holds provided that the energy difference between the Fermi level and the bottom of the conduction band is sufficiently large.
Under this assumption, we can approximate 
\begin{equation}
  \frac i \pi \oint \dd{\xi^\alpha_-}\check G^{\alpha\alpha}_{--} \circ (\check V^\alpha + \check\Sigma_\text{imp}^\alpha)_{--}
  = \check g^\alpha \circ (\check V^\alpha + \check\Sigma_\text{imp}^\alpha)_{--}
\end{equation}
in the presence of an inhomogeneous magnetic texture.
A similar argument can be used to show that the same assumptions also imply that the condition $\abs{a \partial_{\xi_-^\alpha}A} \ll \abs{A}$ holds in the presence of corrections to the hopping amplitude, which can come from an external vector potential or spin-orbit coupling.
With these assumptions,
\begin{multline}
  \frac i \pi \oint \dd{\xi^\alpha_-}\left[(\check V^\alpha + \check\Sigma_\text{imp}^\alpha)\circ \check G^{\alpha\alpha}\right]_{--} \approx (\check V^\alpha + \check\Sigma_\text{imp}^\alpha)_{--} \circ \check g^\alpha 
  \\
  + \frac i \pi \oint \dd{\xi^\alpha_-}(\check V^\alpha + \check\Sigma_\text{imp}^\alpha)_{-+}\circ \check G^{\alpha\alpha}_{+-}.
  \label{eq:zerothorderapproxL}
\end{multline}
\Cref{eq:zerothapproxDerivation} works the same when reversing the order of $A$ and $\check G^{\alpha\alpha}_{--}$, so it is also true that
\begin{multline}
  \frac i \pi \oint \dd{\xi^\alpha_-}\left[\check G^{\alpha\alpha}\circ(\check V^\alpha + \check\Sigma_\text{imp}^\alpha)\right]_{--} \approx  \check g^\alpha\circ (\check V^\alpha + \check\Sigma_\text{imp}^\alpha)_{--} 
  \\
  + \frac i \pi \oint \dd{\xi^\alpha_-}\check G^{\alpha\alpha}_{-+}\circ (\check V^\alpha + \check\Sigma_\text{imp}^\alpha)_{+-}.
  \label{eq:zerothorderapproxR}
\end{multline}

If we are also sufficiently far away from the top of the conduction band, then the velocity $\vec v_F^\alpha = \nabla_k\xi_-^\alpha$ is also approximately constant at all the poles of the Green's function.
By approximately constant, we mean that the variation is small compared to $\vec v_F^\alpha$.
To see why, note that 
\begin{equation}
  \nabla_k\xi_-^\alpha = \frac{\sqrt{(\mu^\alpha - \xi_-^\alpha)^2 - (J^\alpha)^2}\nabla_k K^\alpha}{(\mu^\alpha - \xi_-^\alpha)}.
\end{equation}
Differentiating with respect to $\xi_-^\alpha$ gives
\begin{equation}
  \frac{\abs{a\partial_{\xi_-^\alpha}\nabla_k \xi_-^\alpha}}{\abs{\nabla_k \mathcal F_r\{K^\alpha_{nm}\}}} = \left\lvert \frac{a(J^\alpha)^2}{(\xi_-^\alpha - \mu^\alpha)(K^\alpha)^2} + \frac{a\partial_{\xi_-^\alpha}\nabla_k K^\alpha}{\nabla_k K^\alpha} \right\rvert.
  \label{eq:dkdxi}
\end{equation}
From \cref{eq:JoverK}, we know that $(J^\alpha/K^\alpha)^2 < \lvert J^\alpha\rvert/2\Delta E^\alpha$.
Since $\lvert J^\alpha/(\xi_-^\alpha -\mu^\alpha)\rvert \approx \lvert J^\alpha/\mu^\alpha\rvert < 1$ and $a/\Delta E^\alpha \ll 1$, the first term on the right-hand side of \cref{eq:dkdxi} is small.
We have shown that the second term on the right-hand side of \cref{eq:dkdxi} is also negligible.
As a result, integrating the third term on the left-hand side of \cref{eq:gorkov_cond_2} gives
\begin{equation}
  -\frac i \pi \oint \dd{\xi^\alpha_-}i\nabla_k \xi_-^\alpha \cdot \Delta_R\check G^{\alpha\alpha}_{--} = -i\vec v_F^\alpha \cdot \Delta_R\check g^\alpha.
\end{equation}

Next, consider the higher-order terms in the gradient expansion.
We will show that we can ignore these terms when the Hamiltonian, and therefore the Green's function, vary slowly in the center-of-mass (COM) spatial coordinate.
Assuming $\abs{a \partial_{\xi_-^\alpha}A} \ll \abs{A}$,
\begin{multline}
  \frac i \pi \oint \dd{\xi^\alpha_-}[\Delta_R\check G^{\alpha\alpha}_{--}] \circ (i\nabla_k A  - [\vec\delta^\alpha \rho_B,\, A])
  \\
  \approx [\Delta_R\check g^{\alpha}] \circ (i\nabla_k A  - [\vec\delta^\alpha \rho_B,\, A]),
  \label{eq:firstorderapprox}
\end{multline}
where we used \cref{eq:zerothapproxDerivation}.
The gradient $\nabla_k A$ is evaluated at the Fermi surface.
We define the characteristic COM length scale $L$ to be the smallest number satisfying
\begin{equation}
  \abs{\Delta_R \check g^\alpha} < \frac{\abs{\check g^\alpha}}{L},
  \label{eq:characteristicLength}
\end{equation}
everywhere and for all momentum directions,
where the norms can be understood using an appropriate matrix norm such as the Frobenius norm.
In the quasiclassical framework, $L$ is assumed to be much larger than the length of the nearest neighbor displacement vectors and the Fermi wavelength.
As a result,
\begin{equation}
  \abs{\Delta_R\check g^{\alpha} \circ [\vec\delta^\alpha \rho_B,\, A]} < \frac{\abs{\vec\delta^\alpha}}{L} \abs{\check g^{\alpha}} \circ \abs{A} \ll \abs{\check g^{\alpha}} \circ \abs{A},
\end{equation}
meaning that the second term in \cref{eq:firstorderapprox} is negligible compared to the zeroth order term, $\check g^\alpha \circ A$.
The magnitude of the first term is 
\begin{equation}
  \abs{[\Delta_R\check g^{\alpha}] \circ (i\nabla_k A)} < \frac{\abs{\vec v_F^\alpha}}{L}\abs{\check g^{\alpha}} \circ \abs{\partial_{\xi_-^\alpha}A}.
\end{equation}
Therefore, this term is negligible compared to the zeroth order term if $\abs{\abs{\vec v_F^\alpha}\partial_{\xi_-^\alpha}A/{L}} \ll \abs{A}$.
This is guaranteed to be the case if $L > \abs{\vec v_F^\alpha}/a$, since $\abs{a \partial_{\xi_-^\alpha}A} \ll \abs{A}$.
Physically, this criterion can again be understood by considering the time scale defined by $1/a$.
For instance, $1/a$ can be on the order of the elastic impurity scattering time.
The condition $L > \abs{\vec v_F^\alpha}/a$ then states that the variation is small over a distance equal to the mean free path.
However, we note that this condition is too strict.
It assumes only that $\abs{\partial_{\xi_-^\alpha}A}/\abs{A} \ll 1/a $, but if one can replace $1/a$ with a smaller number, then one can also loosen the condition on $L$.

With these assumptions, we neglect the first order terms in the gradient expansion of $\check G^{\alpha\alpha}_{--}\bullet A$ after integration over $\xi_-^\alpha$.
Since $L$ is large, higher order terms will be even smaller than the first order terms, so we neglect all terms except the zeroth order term in the gradient expansion of $\check G^{\alpha\alpha}_{--}\bullet A$.
Next, we must consider
\begin{equation}
  \frac i \pi \oint \dd{\xi^\alpha_-}[\Delta_R A] \circ (i\nabla_k \check G^{\alpha\alpha}_{--}  - [\vec\delta^\alpha \rho_B,\, \check G^{\alpha\alpha}_{--}]).
  \label{eq:qcagfo}
\end{equation}
We can use  \cref{eq:zerothapproxDerivation} one the second term on the right-hand side, which we see can be neglected since $\abs{\Delta_R A} < \abs{A}/L$ and $\abs{\vec\delta^\alpha}/L \ll 1$.
However, we cannot  use \cref{eq:zerothapproxDerivation} to evaluate the first term on the right-hand side of \cref{eq:qcagfo}.
This is because $\check G^{\alpha\alpha}_{--}$ varies rapidly as a function of $\vec k$ near its poles.
To proceed, we can use the contour integral of a total derivative is zero.
This implies that
\begin{equation}
  \oint d\xi A\frac{\partial B}{\partial k} = 
  \oint d\xi A\frac{\partial \xi_-^\alpha}{\partial k}\frac{\partial B}{\partial \xi_-^\alpha}
  = - \oint d\xi \frac{\partial}{\partial \xi_-^\alpha}\left(A\frac{\partial \xi_-^\alpha}{\partial k}\right)B,
\end{equation}
for any $A$ and $B$, where $\partial/\partial k$ is differentiation with respect to the amplitude of $\vec k$ in spherical coordinates.
This is not to be confused with the gradient operator $\nabla_k$.
We already assume that $\nabla_k\xi_-^\alpha$ is approximately constant on all the poles of $\check G^{\alpha\alpha}_{--}$.
Using this we find that
\begin{multline}
\frac i \pi \oint \dd{\xi} \Delta_R A \circ \nabla_k \check G^{\alpha\alpha}_{--}
= \frac{\Delta_R A}{k_F}  \circ \left[\uv\theta \frac{\partial}{\partial\theta} + \uv\phi \frac{1}{\sin\theta}\frac{\partial}{\partial\phi}\right]\check g^\alpha
  \\- \uv k \cdot \left(\partial_k \Delta_RA\right) \circ \check g^\alpha,
  \label{eq:firstord}
\end{multline}
where $k_F^\alpha$ is the Fermi momentum, satisfying $\xi_-^\alpha(k_F^\alpha) = 0$ and
$\theta$ and $\phi$ are the azimuthal and polar angles in momentum space, respectively.
As long as $\check g^\alpha$ does not vary rapidly as a function of $\theta$ and $\phi$, the right-hand side of \cref{eq:firstord} is negligible under the same assumptions as \cref{eq:firstorderapprox}.
Hence, we can also neglect the higher order terms in the gradient expansion of $A\bullet\check G^{\alpha\alpha}_{--}$.
Combining the above results,
\begin{multline}
  \frac i \pi \oint \dd{\xi^\alpha_-}\left[(\check \Sigma^{\alpha}-\hat H_0^\alpha)\bullet \check G^{\alpha\alpha}\right]_{--} = (\check \Sigma^{\alpha}-\hat H_0^\alpha)_{--} \circ \check g^\alpha 
  \\
  + \frac i \pi \oint \dd{\xi^\alpha_-}(\check \Sigma^{\alpha}-\hat H_0^\alpha)_{-+}\circ \check G^{\alpha\alpha}_{+-},
  \label{eq:bulletapproxL}
\end{multline}
and
\begin{multline}
  \frac i \pi \oint \dd{\xi^\alpha_-}\left[\check G^{\alpha\alpha}\bullet(\check \Sigma^{\alpha}-\hat H_0^\alpha)\right]_{--} =  \check g^\alpha\circ (\check \Sigma^{\alpha}-\hat H_0^\alpha)_{--} 
  \\
  + \frac i \pi \oint \dd{\xi^\alpha_-}\check G^{\alpha\alpha}_{-+}\circ (\check \Sigma^{\alpha}-\hat H_0^\alpha)_{+-}.
  \label{eq:bulletapproxR}
\end{multline}
The circle-products in the last terms on the right-hand side of \cref{eq:bulletapproxL,eq:bulletapproxR} comes from a truncation in the gradient expansion which is valid for the same reasons as the truncation in the gradient expansions involving $\check G^{\alpha\alpha}_{--}$.

To complete the derivation of the quasiclassical equations, we must remove the terms involving $\check G^{\alpha\alpha}_{-+}$ and $\check G^{\alpha\alpha}_{+-}$.
Physically, this can be done because the energy difference between the two bands is large for momenta close to the Fermi surface.
This means that there is negligible coupling between the electrons near the Fermi surface and the electrons in the other band.
In order to show 
\begin{equation}
  \abs{\frac i \pi \oint \dd{\xi^\alpha_-}\check G^{\alpha\alpha}_{-+}} \ll \abs{\check g^\alpha} \quad\text{and} \quad
  \abs{\frac i \pi \oint \dd{\xi^\alpha_-}\check G^{\alpha\alpha}_{+-}} \ll \abs{\check g^\alpha},
  \label{eq:gpm_negligible}
\end{equation}
we define
\begin{equation}
  \check g^\alpha_{\pm\mp} = \frac i \pi \oint \dd{\xi^\alpha_-}\check G^{\alpha\alpha}_{\pm\mp}.
\end{equation}
We get from the first Gor'kov equation that 
\begin{multline}
  \tau_z \varepsilon \circ \check g_{+-}^{\alpha} - \xi^\alpha_+\check g_{+-}^{\alpha} - (\check \Sigma^{\alpha}-\hat H_0^\alpha)_{++}\circ \check g^{\alpha}_{+-}\\
  -(\check \Sigma^{\alpha}-\hat H_0^\alpha)_{+-}\circ \check g^{\alpha} = 0,
\end{multline}
where $\xi^\alpha_+(\vec k_F) = -\mu^\alpha + \eta^\alpha(\vec k_F^\alpha)$ is evaluated at the Fermi surface defined by $\xi^\alpha_-(\vec k_F) = -\mu^\alpha - \eta^\alpha(\vec k_F^\alpha) = 0$.
As a result, $\abs{\xi_+^\alpha} = 2\abs{\mu^\alpha}$, which is much larger than $\abs{(\check \Sigma^{\alpha}-\hat H_0^\alpha)_{+-}}$ by assumption.
We will also assume $\abs{\varepsilon} \ll E_c^\alpha$, and consider larger $\abs{\varepsilon}$ separately when computing observables in \cref{sec:observables}.
Therefore, $\check g_{+-}^{\alpha} \approx (\check \Sigma^{\alpha}-\hat H_0^\alpha)_{+-}\circ\check g^\alpha/\xi_+^\alpha$ is negligible.
The same argument from the second Gor'kov equation shows that $\check g_{-+}^{\alpha}$ is negligible as well.

Finally, integrating the Gor'kov equations, \cref{eq:gorkov_cond_1,eq:gorkov_cond_2}, over the contours in  $\xi_-^\alpha$-space and using \cref{eq:bulletapproxL,eq:bulletapproxR,eq:gpm_negligible} we get
\begin{subequations}
  \begin{align}
    \tau_z \varepsilon \circ \check g^{\alpha} - (\check \Sigma^{\alpha}-\hat H_0^\alpha)_{--}\circ \check g^{\alpha} = \frac i \pi \oint \dd{\xi^\alpha_-}\xi^\alpha_-\check G_{--}^{\alpha\alpha},
  \label{eq:qc1_a}
    \\
  \check g^{\alpha}\circ \tau_z\varepsilon - i\vec v_F^\alpha\cdot\Delta_R\check g^{\alpha} - \check g^{\alpha}\circ(\check \Sigma^{\alpha}-\hat H_0^\alpha)_{--}
  \nonumber\\
  =\frac i \pi \oint \dd{\xi^\alpha_-}\xi^\alpha_-\check G_{--}^{\alpha\alpha}.
  \label{eq:qc1_b}
  \end{align}
\end{subequations}
We have no way to evaluate the right-hand sides because it would require first finding the poles of $\check G_{--}^{\alpha\alpha}$.
Instead, we can subtract \cref{eq:qc1_b} from \cref{eq:qc1_a} to obtain the Eilenberger equation,
\begin{equation}
  i\vec v_F^\alpha\cdot\Delta_R\check g^{\alpha} + \left[\tau_z\varepsilon - (\check \Sigma^{\alpha}-\hat H_0^\alpha)_{--},\, \check g^\alpha\right]_\circ = 0.
  \label{eq:eilenberger1}
\end{equation}

The distances between neighboring points are short compared to the characteristic COM length scale $L$, defined in \cref{eq:characteristicLength}, so we can approximate $\check g^{\alpha}$ by a continuous function in COM position and replace $\Delta_R$ by the gradient operator, $\nabla_R$.
One way to do this rigorously is to define the continuous function as a weighted average,
\begin{equation}
  \check g^\alpha_c(\vec R) = \sum_{n\in\mathbb{Z}^3} \check g^\alpha(\vec x_n^\alpha)\frac{1}{C(\vec R)} \me{-(\vec R - \vec x_n^\alpha)^2/l^2},
  \label{eq:cont_def}
\end{equation}
where $l \ll L$ and $C(\vec R) = \sum_{n\in\mathbb{Z}^3}\me{-(\vec R - \vec x_n^\alpha)^2/l^2}$.
From the fact that $l \ll L$, it is clear that $\check g^\alpha(\vec x_n^\alpha) \approx \check g^\alpha_c(\vec x_n^\alpha)$.
Moreover, if $\abs{\vec x_m^\alpha} \ll L$,
\begin{multline}
  (\vec x_m^\alpha \cdot \Delta_R\check g^\alpha)(\vec x_n^\alpha) \approx \sum_{n\in\mathbb{Z}^3} (\vec x_m^\alpha \cdot \Delta_R\check g^\alpha)(\vec x_n^\alpha)\frac{1}{C(\vec R)} \me{-(\vec R - \vec x_n^\alpha)^2/(2l)}
  \\ = \sum_{n\in\mathbb{Z}^3} \left[\check g^\alpha(\vec x_n^\alpha + \vec x_m^\alpha) - \check g^\alpha(\vec x_n^\alpha)\right]\frac{1}{C(\vec R)} \me{-(\vec R - \vec x_n^\alpha)^2/(2l)}
  \\ = \sum_{n\in\mathbb{Z}^3} \check g^\alpha(\vec x_n^\alpha)\left[\frac{\me{-(\vec R+\vec x_m^\alpha - \vec x_n^\alpha)^2/(2l)}}{C(\vec R+\vec x_m^\alpha)} - \frac{\me{-(\vec R - \vec x_n^\alpha)^2/(2l)}}{C(\vec R)}\right]
  \\
  \approx \vec x_m^\alpha \cdot \nabla_R \check g^\alpha_c(\vec x_n^\alpha).
  \label{eq:cont_grad}
\end{multline}
Inserting this into \cref{eq:eilenberger1} and relabeling $\check g^\alpha_c \to \check g^\alpha$, the Eilenberger equation now becomes, in terms of continuous COM coordinates,
\begin{equation}
  i\vec v_F^\alpha\cdot\nabla_R\check g^{\alpha} + \left[\tau_z\varepsilon - (\check \Sigma^{\alpha}-\hat H_0^\alpha)_{--},\, \check g^\alpha\right]_\circ = 0.
  \label{eq:eilenberger2}
\end{equation}

The Eilenberger equation does not have a unique steady-state solution.
This can be seen from the fact that any constant multiple of the identity matrix is a solution.
To compensate for this, one typically assumes a normalization condition.
In a spatially and temporally uniform system, we see from \cref{eq:gorkov_cond_2} that
\begin{multline}
  \check G^{\alpha\alpha}_{--} = \left(\tau_z\varepsilon - \xi_-^\alpha - \check V^\alpha - \check \Sigma^\alpha_\text{imp}\right)^{-1}
  \\
  = P(-\xi_-^\alpha + D)^{-1} P^{-1},
\end{multline}
where $\tau_z\varepsilon - \check V^\alpha - \check \Sigma^\alpha_\text{imp} = PDP^{-1}$ and $D$ is diagonal.
Since $D$ varies slowly as a function of $\xi_-^\alpha$ within the contour, we see that
\begin{equation}
  \frac i \pi \oint \dd{\xi_-^\alpha} (-\xi_-^\alpha + D)^{-1}_{ll} = -\sgn[\Im(D_{ll})],
\end{equation}
which implies that $\check g^\alpha \check g^\alpha = 1$.
More generally, we assume that $\check g^\alpha \circ \check g^\alpha = 1$.
This is consistent with the fact that $\check g^\alpha \circ \check g^\alpha = 1$ must also solve the Eilenberger equation, as can be seen by taking the circle product of the Eilenberger equation by $\check g^\alpha$ from the left and from the right, as well as the fact that the initial condition, if taken at $T\to -\infty$, should be a time-invariant state, such that $\check g^\alpha \circ \check g^\alpha = \check g^\alpha \check g^\alpha = 1$.
Moreover, it is possible to derive $\check g^\alpha \circ \check g^\alpha = 1$ if one defines the quasiclassical Green's function in terms of trajectory Green's function, as shown by Shelankov~\cite{shelankov1985}.

\section{Quasiclassical impurity self-energy}%
\label{sec:revisiting_impurity_self_energy}
Before deriving the dirty limit equation of motion for the quasiclassical Green's function, we must express the impurity self-energy in terms of the quasiclassical Green's function.
From \cref{sec:impurity_averaging} we have that
\begin{multline}
  \check \Sigma^\alpha_\text{imp}(\varepsilon, T, \vec k, \vec x_n^\alpha)
  =
  \sum_{X \in\{A,B\}}n_\text{imp}^{\alpha X}
  \left( \rho_X\left\langle U^{X \alpha}\right\rangle_\text{imp}\right. \\
  +\left. \left\langle U^{X \alpha}U^{X \alpha}\right\rangle_\text{imp}\rho_X(\check G^{\alpha\alpha})_{nn}(\varepsilon, T)\rho_X\right).
  \label{eq:impRe1}
\end{multline}
If on average there are an equal amount of impurities of equal average strength on both sublattices, and the impurities are not magnetic, then the first term is simply equivalent to a shift in the electrochemical potential.
It can therefore be absorbed into $\mu^\alpha$.

To evaluate the second term in \cref{eq:impRe1} we use the Eilenberger contour,
\begin{multline}
  (\check G^{\alpha\alpha})_{nn} = V_e^\alpha\int_{\Diamond_\alpha}\frac{\dd[3]{k}}{(2\pi)^3} \me{i\rho_B\vec k \cdot \vec \delta^\alpha}\check G^{\alpha\alpha}(\vec k, \vec x_n^\alpha)\me{-i\rho_B\vec k \cdot \vec \delta^\alpha}
  \\
  =V_e^\alpha\int \frac{\dd{\Omega}}{4\pi}\int_{\xi_\text{min}}^{\xi_\text{max}} \frac{p^2 \dd{\xi_-^\alpha}}{2\pi^2 (\xi_-^\alpha)'}\me{i\rho_B\vec k \cdot \vec \delta^\alpha}\check G^{\alpha\alpha}(\vec k, \vec x_n^\alpha)\me{-i\rho_B\vec k \cdot \vec \delta^\alpha}
  \\
  = V_e^\alpha\int \frac{\dd{\Omega}}{4\pi}\oint \frac{k^2\dd{\xi_-^\alpha}}{2\pi^2 v_F^\alpha} \me{i\rho_B\vec k \cdot \vec \delta^\alpha}\check G^{\alpha\alpha}(\vec k, \vec x_n^\alpha)\me{-i\rho_B\vec k \cdot \vec \delta^\alpha}
  \\
  + V_e^\alpha\int \frac{\dd{\Omega}}{4\pi}\fint \frac{k^2\dd{\xi_-^\alpha}}{2\pi^2 (\xi_-^\alpha)'} \me{i\rho_B\vec k \cdot \vec \delta^\alpha}\check G^{\alpha\alpha}(\vec k, \vec x_n^\alpha)\me{-i\rho_B\vec k \cdot \vec \delta^\alpha}.
  \label{eq:Gnn}
\end{multline}
Using that 
\begin{equation}
  \rho_X \me{i\rho_B\vec k \cdot \vec \delta^\alpha}\check G^{\alpha\alpha}(\vec k, \vec x_n^\alpha)\me{-i\rho_B\vec k \cdot \vec \delta^\alpha} \rho_X
  = 
  \rho_X \check G^{\alpha\alpha}(\vec k, \vec x_n^\alpha)\rho_X,
\end{equation}
where $X \in \{A,B\}$, we see that we can remove the exponentials in \cref{eq:Gnn}.
The first term on the right-hand side of \cref{eq:Gnn} is what gives us the quasiclassical Green's function.
To evaluate the second term, we can use the fact that we are far away from the Fermi surface, so, if we neglect spatial and temporal derivatives in the Gor'kov equations,
\begin{multline}
  \check G^{\alpha\alpha} \approx \left(\varepsilon\tau_z-\hat H_0^\alpha - \check V^\alpha - \check \Sigma_\text{imp}^\alpha\right)^{-1}
  = \left(-\hat H_0^\alpha\right)^{-1} \\
  - \left(\hat H_0^\alpha\right)^{-1}\left(\varepsilon\tau_z - \check V^\alpha - \check \Sigma_\text{imp}^\alpha\right)\left(\hat H_0^\alpha\right)^{-1} + \mathcal O([\xi_-^\alpha]^{-3})
\end{multline}
We can neglect the second term after integration for the following reason.
We can complete the contour in $\fint\dd{\xi_-^\alpha}$ with a semicircle of radius $(\abs{\xi_\text{min}} + \abs{\xi_\text{max}})/2$.
Since there are no poles inside the closed contour, the integral $\fint\dd{\xi_-^\alpha}$ must be equal to minus the integral over the semicircle arc.
The integral over this arc is negligible because it is less than $\pi(\abs{\xi_\text{min}} + \abs{\xi_\text{max}})/2 \times a \max(N_0^\alpha)/\min(\abs{\xi_\text{min}},\,\abs{\xi_\text{max}})^2$, which is $\mathcal O(N^\alpha_0(0) a/\Delta E^\alpha)$, where
$a$ is again an order of magnitude estimate of the elements of $(\varepsilon\tau_z - \check V^\alpha - \check \Sigma_\text{imp}^\alpha)$, and therefore much smaller than $\Delta E^\alpha $, and
\begin{equation}
  N_0^\alpha(\varepsilon) = \int \frac{\dd[3]{k}}{(2\pi)^3} \delta\left(\xi(\vec k) - \varepsilon\right) = \int \frac{\dd{\Omega}}{4\pi}\int\frac{k^2 \dd{\xi}}{2\pi^2 \xi'}\delta\left(\xi(\vec k) - \varepsilon\right)
\end{equation}
is the normal state density of states per spin.
For the same reason, the terms of higher order in $(\xi_-^{\alpha})^{-1}$ are also negligible.
The first term, however, is not negligible, as the same argument shows that this integral is $\mathcal O(N_0(0))$, which is the same as the quasiclassical term.

Evaluating the $\left(-\hat H_0^\alpha\right)^{-1}$ and applying the projection operators, we get
\begin{equation}
  \sum_{X\in{A, B}}\rho_X\left(-\hat H_0\right)^{-1}\rho_X
=  \frac{\mu^\alpha - J^\alpha\rho_z\sigma_z\tau_z}{\xi^\alpha_-\xi^\alpha_+}.
\end{equation}
Integrating out the momentum dependence, we see that we get constant matrices with the same matrix structure as a chemical potential and an antiferromagnetic spin-splitting.
We can therefore include this by renormalizing $\mu^\alpha$ and $J^\alpha$.

In order to evaluate the quasiclassical contribution, we define
\begin{equation}
  S^\alpha \mqty(1 \\ 0) = S_c^\alpha,
\end{equation}
where $1$ and $0$ are $4\times 4$ matrices,
such that
\begin{equation}
  A_{--} = \transpose{(S_c^\alpha)}A S_c^\alpha.
\end{equation}
Since only the contribution from the conduction band is non-negligible close to the Fermi surface, we have that
\begin{equation}
  \oint \frac{k^2\dd{\xi_-^\alpha}}{2\pi^2 v_F^\alpha} \check G^{\alpha\alpha}(\vec k, \vec x_n^\alpha) = -i\pi N^\alpha_0(0)S_c^\alpha \check g^\alpha \transpose{(S_c^\alpha)},
\end{equation}
where $S_c^\alpha$ is evaluated at the Fermi surface.

Hence, if we define
\begin{equation}
  \check g_s^\alpha \defeq \int \frac{\dd{\Omega}}{4\pi}\check g^\alpha = \left\langle\check g^\alpha\right\rangle,
\end{equation}
where in the last equality we also defined the angular average in momentum space as $\langle\cdot\rangle$, then
\begin{align}
  (\check \Sigma^\alpha_\text{imp})_{--} = -\frac{i}{\tau_\text{imp}} \sum_{X\in\{A,B\}}\transpose{(S_c^\alpha)}\rho_X S_c^\alpha \check g_s^\alpha\transpose{(S_c^\alpha)}\rho_X S_c^\alpha,
\end{align}
where
\begin{equation}
  \tau^\alpha_\text{imp} = \left(\pi N^\alpha_0(0) V_e^\alpha n_\text{imp}^{\alpha A}\left\langle U^{A\alpha}U^{A\alpha} \right\rangle_\text{imp}\right)^{-1}
\end{equation}
is the impurity scattering time.

Next, we find that
\begin{equation}
  \transpose{(S_c^\alpha)}\rho_{A/B} S_c^\alpha = \frac 1 2 \left(1 \pm \frac{J^\alpha}{\eta^\alpha} \sigma_z\tau_z\right),
  \label{eq:projectionOfProjection}
\end{equation}
such that
\begin{align}
  (\check \Sigma^\alpha_\text{imp})_{--} = -\frac{i}{2\tau^\alpha_\text{imp}} 
  \left(\check g_s^\alpha + \frac{(J^\alpha)^2}{(\eta^\alpha)^2}\sigma_z\tau_z\check g_s^\alpha\sigma_z\tau_z\right).
  % \left[\left(1 + \frac{J^2}{\eta^2}\right)\check g_s^\alpha + \frac{J^2}{\eta^2}\left[\sigma_z\tau_z,\, \check g_s^\alpha\right]\sigma_z\tau_z\right].
  \label{eq:imp_final}
\end{align}
This reduces to the normal state impurity self-energy in the absence of antiferromagnetism when $J^\alpha=0$.
However, when $J^\alpha\neq 0$ we get an additional term which is the same as one gets when adding magnetic impurities in the quasiclassical theory for normal metals.
This is an important result which means that impurities in the antiferromagnet behave as if they were magnetic.
This effect becomes important when the system size becomes larger than the mean free path, and this is why one should expect the critical temperature to decrease in superconducting proximity structures when the antiferromagnet becomes larger than its mean free path, which explains the findings of \textcite{hubener2002}, as alluded to in \cref{sec:introduction}.
Physical consequences of well \cref{eq:imp_final}, as well as a physical explanation for its existence is further discussed in ref.~\cite{prl_submission}.

The effective magnetic component of non-magnetic impurities is similar to how interfacial disorder in antiferromagnetic insulators has been shown to give rise to magnetic effects when the interface is uncompensated~\cite{kamra2018}, except that here it is a bulk effect.
As a result, it is present even though the magnetization is fully compensated.
Another type of material in which one can find effective ``magnetic'' coupling from non-magnetic impurities is in Rashba superconductors~\cite{houzet2015,illic2022}.
The strong coupling between spin and momentum degrees of freedom in Rashba superconductors means that non-magnetic impurities get a non-trivial matrix structure in the helical basis~\cite{houzet2015}.
However, the effective ``magnetic'' impurities in Rashba superconductors are different from what we see here.
They couple to the $p$-wave part of the Green's function and not the $s$-wave part.
They are ``magnetic'' in the sense that they couple different components in the helical basis, but not in the sense that it is \emph{as if} the system has magnetic impurities.
Here we find that non-magnetic impurities in AFMs are mathematically equivalent to having magnetic impurities in the original model.

\section{The dirty limit}%
\label{sec:the_dirty_limit}
In this section, we derive the equations of motion in the dirty limit, which are valid for diffusive systems.
There are two central assumptions in the dirty limit.
First, it is assumed that the quasiclassical Green's function is dominated by the $s$-wave and $p$-wave components.
Second, it is assumed that the elastic impurity scattering rate is large compared to the other energies in the system, except for the minimal distance between the Fermi level and the edges of the conduction band, $\Delta E^\alpha$, and possibly $J^\alpha$.
We show that the resulting equations are valid if the variation in $\check g^\alpha$ over the length scale of the mean free path is small compared to 1.
This is the case for instance if the system varies slowly in space or the proximity effect is small.
In the limit of very strong exchange coupling, such that $(J^\alpha)^2/(\eta^\alpha)^2 = \mathcal O(1)$, we show that the quasiclassical Green's function can be separated into short-range correlations and long-range components, where the former vanish in the diffusive limit.
Therefore, this regime can be solved by projecting the Green's function onto the set of long-range components.
The derivation is done by averaging the Eilenberger equation,
\begin{equation}
  i\vec v_F^\alpha \cdot \nabla_R \check g^\alpha  + [\tau_z\varepsilon - \check V^\alpha_{--} - (\check \Sigma_{\text{imp}}^\alpha)_{--},\,\check g^\alpha]_\circ = 0,
  \label{eq:eil_dirt}
\end{equation}
over momentum directions.
This will reduce the problem from having infinitely many coupled Green's functions, one for each momentum direction, to having only two coupled Green's functions.

Before proceeding, we first replace the gradient term with the covariant derivative. 
This is done by extracting the $p$-wave part of $\check V^\alpha_{--}$, meaning that we write
\begin{equation}
  \check V^\alpha_{--} = -\vec v_F^\alpha \cdot \hat{\vec A}^\alpha + \check V^\alpha_s
  + \Delta \check V^\alpha,
\end{equation}
where $\check V_s^\alpha = \langle \check V^\alpha_{--}\rangle$ is the $s$-wave part and $-\vec v_F^\alpha \cdot \hat{\vec A}^\alpha$ is the $p$-wave part of $\check V^\alpha_{--}$.
The $p$-wave contribution includes the vector gauge potential from the electromagnetic field as well as spin-orbit coupling and the spatial variation in the Néel vector.
The covariant derivative is then defined as
\begin{equation}
  \tilde\nabla\circ \check g = \nabla_R \check g - i \comm{\hat{\vec A}}{\check g}_\circ,
\end{equation}
such that
\begin{equation}
  i\vec v_F^\alpha \cdot\tilde\nabla\circ \check g^\alpha  + \left[\tau_z\varepsilon - \check V^\alpha_s - \Delta \check V^\alpha - (\check \Sigma_{\text{imp}}^\alpha)_{--},\,\check g^\alpha\right]_\circ = 0.
  \label{eq:eilenbergerCov}
\end{equation}

Doing an angular average of \cref{eq:eilenbergerCov}, we get
\begin{multline}
  i \tilde\nabla\circ\left\langle\vec v_F^\alpha\check g^\alpha\right\rangle + \left[\tau_z\varepsilon - \check V_s^\alpha + \frac{i(J^\alpha)^2}{2\tau^\alpha_\text{imp}(\eta^\alpha)^2}\sigma_z\tau_z\check g_s^\alpha\sigma_z\tau_z,\,\check g_s^\alpha\right]_\circ
   \\
   - \left\langle\left[\Delta \check V^\alpha,\,\check g^\alpha\right]_\circ\right\rangle
  = 0. 
\label{eq:momAv1}
\end{multline}
If we take the product with $\vec v_F^\alpha$ before averaging, we get
\begin{multline}
 i \tilde\nabla\circ\left\langle\vec v_F^\alpha \otimes \vec v_F^\alpha \check g^\alpha\right\rangle
 + \left[\tau_z\varepsilon - \check V_s^\alpha + \frac{i}{2\tau^\alpha_\text{imp}}\check g_s^\alpha,\,\left\langle\vec v_F^\alpha\check g^\alpha\right\rangle\right]_\circ
 \\
 + \left[\frac{i(J^\alpha)^2}{2\tau^\alpha_\text{imp}(\eta^\alpha)^2}\sigma_z\tau_z\check g_s^\alpha\sigma_z\tau_z,\,\left\langle\vec v_F^\alpha\check g^\alpha\right\rangle\right]_\circ
 \\
  - \left\langle\left[\Delta \check V^\alpha,\,\vec v_F^\alpha\check g^\alpha\right]_\circ\right\rangle
= 0,
\label{eq:momAv2}
\end{multline}
where $\otimes$ denotes the tensor product.
Next, we define the matrix current
\begin{equation}
  \check{\vec j}^\alpha \defeq \left\langle\vec v_F^\alpha\check g^\alpha\right\rangle.
\end{equation}
The aim is a set of equations for $\check{\vec j}^\alpha$ and $\check g^\alpha_s = \langle\check g^\alpha\rangle$.
This can be obtained from \cref{eq:momAv1,eq:momAv2} if we assume that $\Delta\check V^\alpha$ is negligible.
Neglecting the terms proportional to $\Delta\check V^\alpha$, multiplying \cref{eq:momAv2} by $\tau_\text{imp}^\alpha$, and defining the diffusion tensor,
\begin{equation}
  D^\alpha \defeq \tau^\alpha_\text{imp}\left\langle\vec v_F^\alpha \otimes \vec v_F^\alpha\right\rangle,
\end{equation}
\cref{eq:momAv1,eq:momAv2} become
\begin{multline}
  i \tilde\nabla\circ\check{\vec j}^\alpha + \left[\tau_z\varepsilon - \check V_s^\alpha + \frac{i(J^\alpha)^2}{2\tau^\alpha_\text{imp}(\eta^\alpha)^2}\sigma_z\tau_z\check g_s^\alpha\sigma_z\tau_z,\,\check g_s^\alpha\right]_\circ
 = 0. 
\label{eq:usadel}
\end{multline}
and
\begin{multline}
 \check g_s^\alpha\circ\check{\vec j}^\alpha
 =- \tilde\nabla\circ\left(D^\alpha\check g^\alpha_s\right)
 +i \tau^\alpha_\text{imp}\left[\tau_z\varepsilon - \check V_s^\alpha,\,\check{\vec j}^\alpha\right]_\circ
 \\
 - \left[\frac{(J^\alpha)^2}{2(\eta^\alpha)^2}\sigma_z\tau_z\check g_s^\alpha\sigma_z\tau_z,\,\check{\vec j}^\alpha\right]_\circ,
\label{eq:momAv2_2}
\end{multline}
respectively.
In \cref{eq:momAv2_2} we assumed that the higher order spherical harmonics in $\check g^\alpha$ are small, and used that $\{\check{\vec j}^\alpha,\,\check g_s^\alpha\}= 0$.
The latter follows from the former together with the $p$-wave component of the normalization condition, $\langle\vec v_F^\alpha\check g^\alpha\circ\check g^\alpha\rangle = \{\check{\vec j}^\alpha,\,\check g_s^\alpha\}= 0$.
The assumption that the $d$-wave component is negligible compared to 1 is consistent as long as $\check{\vec j}^\alpha$ is small compared to the Fermi velocity. 
To see why, note that the normalization condition,
\begin{equation}
  \check g^\alpha\circ\check g^\alpha = \check g^\alpha_s \circ\check g^\alpha_s + \{\check g_s^\alpha,\, \Delta\check g^\alpha\}_{\circ} + \Delta\check g^\alpha\circ\Delta\check g^\alpha = 1,
  \label{eq:norm_pre}
\end{equation}
must be satisfied for all momenta.
Hence, if $\Delta\check g^\alpha = \check g_p^\alpha + \check g_d^\alpha + \dotsb$, where $\check g_p^\alpha$ is the $p$-wave component and $\check g_d^\alpha$ is the $d$-wave component, the $d$-wave component resulting from $\check g_p^\alpha\circ\check g_p^\alpha$ must be cancelled by the $d$-wave term in $\{\check g_s^\alpha,\, \check g_d^\alpha\}_{\circ}$.
If $\check g_s = \mathcal O(1)$, then $\check g_d^\alpha$ will be $\mathcal O[(\check{\vec j}^\alpha\cdot\vec v_F^\alpha/(v_F^\alpha)^2)^2]$, which we assume is negligible compared to 1.
Hence,
\begin{multline}
  \tau^\alpha_\text{imp}\left\langle\vec v_F^\alpha \otimes \vec v_F^\alpha \check g^\alpha\right\rangle
  \approx D^\alpha\check g_s^\alpha + \tau^\alpha_\text{imp}\left\langle\frac{\vec v_F^\alpha \otimes \vec v_F^\alpha(\check{\vec j}^\alpha\cdot\vec v_F^\alpha) }{(v_F^\alpha)^2}\right\rangle
  \\
  \approx D^\alpha\check g_s^\alpha.
\end{multline}
If the Fermi surface is spherically symmetric, then $D^\alpha_{ij} = \delta_{ij}\tau^\alpha_\text{imp}(v_F^\alpha)^2/3$.

For a complete description in terms of $\check{\vec j}^\alpha$ and $\check g^\alpha_s$, we must also express the normalization condition, $\check g^\alpha\circ\check g^\alpha = 1$ in terms of $\check g^\alpha_s$ and $\check{\vec j}^\alpha$.
Taking the angular average of the normalization condition and using that $\langle \vec v_F^\alpha /(\vec v_F^\alpha)^2\rangle = 0$, we get that
\begin{equation}
\check g_s^\alpha\circ\check g_s^\alpha = 1 
 + \mathcal O(\lvert\check{\vec j}^\alpha/v_F^\alpha\rvert^2).
  \label{eq:norm_corr}
\end{equation}
We have already assumed that $(\check{\vec j}^\alpha\cdot\vec v_F^\alpha/(v_F^\alpha)^2)^2$ is negligible compared to 1, so
\begin{equation}
  \check g_s^\alpha\circ\check g_s^\alpha = 1.
  \label{eq:norm_corr_2}
\end{equation}
Using \cref{eq:norm_corr_2}, we can rewrite \cref{eq:momAv2_2} to
\begin{multline}
 \check{\vec j}^\alpha
 =- \check g^\alpha_s\circ\tilde\nabla\circ\left(D^\alpha\check g^\alpha_s\right)
 +i \tau^\alpha_\text{imp}\check g^\alpha_s\circ\left[\tau_z\varepsilon - \check V_s^\alpha,\,\check{\vec j}^\alpha\right]_\circ
 \\
 - \check g^\alpha_s\circ\left[\frac{(J^\alpha)^2}{2(\eta^\alpha)^2}\sigma_z\tau_z\check g_s^\alpha\sigma_z\tau_z,\,\check{\vec j}^\alpha\right]_\circ.
\label{eq:momAv2_3}
\end{multline}

\Cref{eq:norm_corr_2,eq:usadel,eq:momAv2_3} can be used to study systems with an arbitrary amount of disorder, provided that the matrix current squared, $\lvert\check{\vec j}^\alpha\rvert^2$, is small compared to the Fermi velocity squared, $\lvert\vec v_F^\alpha\rvert^2$.
To say that $\lvert\check{\vec j}^\alpha\rvert^2 \ll \lvert \vec v_F^\alpha\rvert^2$ is the same as saying that the quasiclassical Green's function is approximately isotropic in momentum space.
Physically, this is expected to be the case when the elastic scattering time, $\tau^\alpha_\text{imp}$, is small, but this it can also happen for example if the tunneling is weak.
In \cref{sec:boundary_condition} we show that the matrix current at the boundary is proportional to the square amplitude of the tunneling in the absence of spin-active boundaries.

We can also simplify \cref{eq:momAv2_3} a bit further if we assume that $\lvert \tau_\text{imp}^\alpha\check V_s^\alpha\rvert \ll 1$ and only consider energies $\lvert \varepsilon \rvert\ll 1/\tau_\text{imp}^\alpha$.
In this case, we can neglect the second term on the right-hand side of \cref{eq:momAv2_3}, since this term must be much smaller in magnitude than $\check{\vec j}^\alpha$.
Hence,
\begin{equation} 
  \check{\vec j}^\alpha = 
    -\check g^\alpha_s\circ\tilde\nabla\circ(D^\alpha\check g^\alpha_s)
    - \check g^\alpha_s\circ\left[\frac{(J^\alpha)^2}{2(\eta^\alpha)^2}\sigma_z\tau_z\check g_s^\alpha\sigma_z\tau_z,\,\check{\vec j}^\alpha\right]_\circ.
  \label{eq:gen_curr}
\end{equation}
At this point, it might be tempting to also assume that the last term in the commutator in \cref{eq:usadel} is dominant, but this is \emph{not} generally true.
Although $1/\tau_\text{imp}^\alpha \gg \lvert \check V_s^\alpha\rvert$, one can not say in general that
\begin{equation}
  \left\lvert\left[\frac{i(J^\alpha)^2}{2\tau_\text{imp}^\alpha(\eta^\alpha)^2}\sigma_z\tau_z\check g_s^\alpha\sigma_z\tau_z,\,\check g_s^\alpha\right]_\circ\right\rvert \gg \left\lvert\left[\check V_s^\alpha,\,\check g_s^\alpha\right]_\circ\right\rvert.
\end{equation}
This can be because the prefactor $(J^\alpha)^2/(\eta^\alpha)^2$ is small, or it can be because the matrices on the right-hand side commute.
even for very strong antiferromagnets with $(J^\alpha)^2/(\eta^\alpha)^2 = \mathcal O(1)$.
This is because, even though the prefactor can be large, the commutator can still be small.
Thus, one must in general keep all terms in \cref{eq:usadel}.

Next, consider the case of very strong exchange coupling, such that $(J^\alpha)^2/(\eta^\alpha)^2 = \mathcal O(1)$.
In this case the prefactor $(J^\alpha)^2/[2\tau_\text{imp}^\alpha(\eta^\alpha)^2]$ is large in the diffusive limit.
This will strongly suppress some components of the quasiclassical Green's function, making them negligible in the diffusive limit.
We can write the quasiclassical Green's functions in terms of Pauli matrices in spin-space and Nambu-space as
\begin{equation}
  \check g^\alpha = \sum_{i=0}^3\sum_{j=0}^3 c_{ij}\tau_i\sigma_j,
\end{equation}
where $\sigma_0$ and $\tau_0$ are identity matrices and $\{c_{ij}\}$ is a set of scalar functions.
We can separate these components into \emph{long-range} components, satisfying
\begin{equation}
  \sigma_z\tau_z c_{ij}\tau_i\sigma_j\sigma_z\tau_z = c_{ij}\tau_i\sigma_j,
\end{equation}
and \emph{short-range} components, satisfying
\begin{equation}
  \sigma_z\tau_z c_{ij}\tau_i\sigma_j\sigma_z\tau_z = -c_{ij}\tau_i\sigma_j.
\end{equation}
That is, long-range components have either $i\in\{0,3\}$ and $j\in\{0,3\}$ or $i\in\{1,2\}$ and $j\in\{1,2\}$, while the short-range components are the remaining components.
Note that the product of two long-ranged components or two short-ranged components is a long-range component, while the product of one long-range component and one short-range component is a short-range component.

Let the subscripts $\text{SR}$ and $\text{LR}$ denote the short-range and long-range components, respectively, such that $\check g^\alpha = \check g^\alpha_\text{SR} + \check g^\alpha_\text{LR}$.
Using the product properties of long-range and short-range components, the long-range component of \cref{eq:usadel} becomes
\begin{multline}
  i \tilde\nabla_\text{LR}\circ\check{\vec j}^\alpha_\text{LR} + \left[\tau_z\varepsilon - \check V_{\text{LR},s}^\alpha,\,\check g_{\text{LR},s}^\alpha\right]_\circ
  + \left[\hat{\boldsymbol A}_\text{SR},\,\check{\vec j}^\alpha_\text{SR}\right]_\circ
  \\
  - \left[\check V_{\text{SR},s}^\alpha,\,\check g_{\text{SR},s}^\alpha\right]_\circ
 = 0,
\label{eq:usadel_LR}
\end{multline}
where $\tilde{\nabla}_\text{LR}\circ\check{\vec j}^\alpha_\text{LR} = \nabla_R\cdot \check{\vec j}^\alpha_\text{LR} - i [\hat{\vec A}_\text{LR},\,\check{\vec j}^\alpha_\text{LR}]_\circ$.
We want to show that the short-range components vanish from the equations in the diffusive limit when $(J^\alpha)^2/(\eta^\alpha)^2 \to 1$.
This means that in this limit one can solve quasiclassical equations by simply setting the short-ranged components to zero.

Assuming that $(J^\alpha)^2/(\eta^\alpha)^2 \approx 1$, $\lvert \tau_\text{imp}^\alpha\check V_s^\alpha\rvert \ll 1$ and only considering energies $\lvert \varepsilon \rvert\ll 1/\tau_\text{imp}^\alpha$, the Eilenberger equation for the short-range components becomes
\begin{equation}
  \tilde\nabla\circ(\vec v_F^\alpha\check g^\alpha_\text{SR}) + \frac{1}{\tau^\alpha_\text{imp}}\left[\left\langle\check g^\alpha_\text{LR}\right\rangle,\,\check g_\text{SR}^\alpha\right]_\circ
  = 0. 
\label{eq:eilSR}
\end{equation}
The short-range correlations and the long-range correlations will generally not commute.
As a result, we see that the short-range correlations decay exponentially over a distance equal to the mean free path in this case.

Making no assumptions other than assuming that $\tau_\text{imp}$ is small and $\check g_s^\alpha\circ\check g_s^\alpha = 1$, which is valid even if the short-range components are not isotropic, provided they are small in magnitude, the short-range component of \cref{eq:momAv2} becomes
\begin{multline} 
  \check{\vec j}^\alpha_\text{SR} = 
    -\tau^\alpha_\text{imp}\left(\check g^\alpha_s\circ\tilde\nabla\circ\left\langle\vec v_F^\alpha \otimes \vec v_F^\alpha \check g^\alpha\right\rangle\right)_\text{SR}
    \\
    - \left(\frac{\check g^\alpha_s}{2}\circ\left[\left\langle\check g^\alpha_\text{LR}\right\rangle - \left\langle\check g^\alpha_\text{SR}\right\rangle,\,\check{\vec j}^\alpha\right]_\circ\right)_\text{SR}.
  \label{eq:curr_SR}
\end{multline}
Using that $\check g_\text{SR}^\alpha$ decays exponentially away from the interface with over a length-scale equal to the mean free path, \cref{eq:curr_SR} implies that, since $\check g^\alpha_\text{LR} = \mathcal O(1)$, 
\begin{equation}
  \check g_\text{SR}^\alpha = \mathcal O\left(\frac{l_\text{mfp}^\alpha\check{\vec j}^\alpha_\text{SR}}{\lvert D^\alpha \rvert}\right),
\end{equation}
where $l^\alpha_\text{mfp} = v_F^\alpha \tau^\alpha_\text{imp}$ is the mean free path.
The short-range component of the matrix current will be largest closest to the interface, where it will be determined by the boundary conditions.
Moreover, in the diffusive regime, the matrix current is small at the interface, as discussed in \cref{sec:boundary_condition}.
Hence, in the diffusive regime we see that $\check g_\text{SR}^\alpha = \mathcal O(\tau_\text{imp}^\alpha)$.
Hence, to zeroth order in $\tau_\text{imp}^\alpha$ the long-ranged components can be solved for consistently in the limit $(J^\alpha)^2/(\eta^\alpha)^2 \to 1$ by setting the short-ranged components to zero, effectively projecting out these components from the Green's function.

Very close to the interface the term $\left[\hat{\boldsymbol A}_\text{SR},\,\check{\vec j}^\alpha_\text{SR}\right]_\circ$ can give a contribution to \cref{eq:usadel_LR}.
This is not a problem if $\hat{\boldsymbol A}_\text{SR} = 0$, but in \cref{sec:nonuniform_magnetic_textures} we show that, similar to spin-orbit coupling, nonuniform magnetic textures can induce a non-zero $\hat{\boldsymbol A}_\text{SR}$.
This means that if there are domain walls very close to the interface to a spin-singlet superconductor, it can induce long-ranged superconducting correlations in the antiferromagnetic metal.
As long as $\hat{\boldsymbol A}_\text{SR} = 0$, the limit of very strong exchange coupling, $(J^\alpha)^2/(\eta^\alpha)^2 \to 1$, can be consistently captured by setting the short-range components to zero and solving 
\begin{equation}
  i \tilde\nabla_\text{LR}\circ\check{\vec j}^\alpha_\text{LR} + \left[\tau_z\varepsilon - \check V_{\text{LR},s}^\alpha,\,\check g_{\text{LR},s}^\alpha\right]_\circ
 = 0.
\label{eq:usadel_LR_2}
\end{equation}
The matrix current can be found by doing the same projection in \cref{eq:momAv2}, which in the limit $(J^\alpha)^2/(\eta^\alpha)^2 \to 1$ simply becomes
\begin{equation}
  \check{\vec j}^\alpha_\text{LR} = -\frac{\check g_{\text{LR},s}^\alpha\circ\tilde\nabla\circ\left(D^\alpha\check g_{\text{LR},s}^\alpha\right)}{2}.
\end{equation}

From \cref{eq:gen_curr} we see that $\check{\vec j}^\alpha\cdot\vec v_F^\alpha/(v_F^\alpha)^2 = \mathcal O(l^\alpha_\text{imp}\tilde\nabla\circ\check g^\alpha_s)$, where $l^\alpha_\text{imp} = v_F^\alpha \tau^\alpha_\text{imp}$ is the mean free path.
As a result, the assumption that $\check g_s^\alpha\circ\check g_s^\alpha = 1$ is consistent as long as the change in $\check g^\alpha_s$ over the length of the mean free path is small compared to 1.
In the limit of strong exchange coupling, the short-ranged components can decay over a length scale equal to the mean free path, but these components also become negligible, as shown above.
Therefore, although the short-ranged components are not necessarily isotropic in the limit $J^\alpha \to \infty$, one can still solve the diffusive equations as long as there is no strong spin-orbit coupling or sudden change in the Néel vector close to the interface.
To simplify the equations in this limit, one can project out the long-range components.
Spin-orbit coupling or non-uniform Néel vector close to the boundary can induce long-range components from the short-range components of the matrix current.
In this case, it is therefore not always consistent to simply set the short-range components to zero.
Instead, if the limit of very strong exchange coupling is necessary, one should solve the full Eilenberger equation for the short-ranged components.

\Cref{eq:gen_curr,eq:usadel} are our main results, together with the boundary condition derived in \cref{sec:boundary_condition}.
They provide general equations of motion which can be solved to obtain information about currents, densities, the local density of states and superconducting correlations in systems with antiferromagnetism and arbitrary geometry both in and out of equilibrium.
In the absence of antiferromagnetism, meaning that $J^\alpha \to 0$, \cref{eq:gen_curr,eq:usadel} reduce to the well-known Usadel equation for normal dirty metals~\cite{usadel1970}.
In the presence of antiferromagnetism, there are three important differences.
First, all self-energies must be projected onto the conduction band, which means that they must be transformed according to the $S_c^\alpha$ matrix.
Second, the coupling between spin and sublattice gives rise to effective magnetic impurities with scattering time $\tau_\text{imp}^\alpha (\eta^\alpha)^2/(J^\alpha)^2$.
Third, the magnetic impurities also modify the equation for the matrix current, which in the normal metal case is simply $\check{\vec j}^\alpha = -\check g^\alpha_s\circ\tilde\nabla\circ(D^\alpha\check g^\alpha_s)$.

One can solve \cref{eq:gen_curr} for $\vec j^\alpha$ in time-independent situations.
If we can diagonalize $(\check g_s^\alpha\sigma_z\tau_z\check g_s^\alpha\sigma_z\tau_z)_{ij} = \check U^{-1}_{ik} \lambda_k \check U_{kj}$, we find that
\begin{equation}
  \check{\vec j}^\alpha_{ij} = -\check U^{-1}_{ik}\frac{\check U_{km}[\check g^\alpha_s\tilde\nabla\cdot(D^\alpha \check g^\alpha_s)]_{mn}\check U^{-1}_{ml}}{1 + (J^\alpha)^2(\lambda_k + \lambda_l)/[2(\eta^\alpha)^2]}\check U_{lj},
\end{equation}
with summation over repeated indices.
Alternatively, since $(J^\alpha/\eta^\alpha)^2$ is smaller by 1 by definition, one can solve for $\check{\vec{j}}^\alpha$ by iteratively inserting into the right-hand side of \cref{eq:gen_curr}.
To get a series expansion with a faster convergence rate it can be beneficial to rewrite \cref{eq:gen_curr} as
\begin{multline}
  \check{\vec j}^\alpha = 
  -\left[1 + (J^\alpha/\eta^\alpha)^2\right]^{-1}\Biggl\{\check g^\alpha_s\circ\tilde\nabla\circ(D^\alpha\check g^\alpha_s)
  \\
+ \check g^\alpha_s\circ\left[\frac{(J^\alpha)^2}{2(\eta^\alpha)^2}\sigma_z\tau_z[\check g_s^\alpha,\,\sigma_z\tau_z],\,\check{\vec j}^\alpha\right]_\circ\Biggr\}.
\end{multline}
This is because the effective magnetic impurities in \cref{eq:usadel} will tend to suppress $[\check g_s^\alpha,\,\sigma_z\tau_z]$.
In the limit of small $J^\alpha/\eta^\alpha$ or vanishing $[\check g_s^\alpha,\,\sigma_z\tau_z]$, one can solve \cref{eq:usadel,eq:gen_curr} in the same way as the Usadel equation for normal metals, but with a renormalized diffusion coefficient, $D^\alpha \to D^\alpha/\left[1 + (J^\alpha/\eta^\alpha)^2\right]$, additional magnetic impurities and self-energies which are projected onto the conduction band of the antiferromagnet.
Otherwise, in the more general case, one can for instance solve \cref{eq:usadel,eq:gen_curr} numerically using the algorithm presented in \cref{sec:solAlg}.

\section{Boundary condition}
\label{sec:boundary_condition}
Next, we derive the boundary condition which is valid in the diffusive regime.
To do so, we must evaluate the two terms which we could neglect in the equation of motion inside the materials.
These are the tunneling terms and the potentials which are large only outside the materials.
Here we consider the interface between material $L$ and $R$.
To get the boundary condition at the interface to a vacuum or an insulator, one need only set the tunneling to zero.
As before, let $(\alpha,\beta)$ be either $(L,R)$ or $(R,L)$.
We assume that the Green's functions are approximately spherically symmetric also close to the interface.
This is the case as long as the matrix current at the interface is small compared to the Fermi velocity, which happens for instance when the tunneling amplitudes are small.

The way the boundary condition is derived here is that we sum the Gor'kov equations over a small set of unit cells which includes the interface.
We take this set to be the shape of a wide cylinder.
The width of this cylinder is much larger than its length but much smaller than the characteristic length scale $L$ of the bulk as defined in \cref{sec:quasiclassical_green_s_functions}.
Then we integrate over all momentum directions and integrate over the Eilenberger contour.
First, we consider the potential which is large only outside material $\alpha$,
\begin{equation}
  (\hat\Sigma_R^\alpha)_{nm}(t_1, t_2) = \hat R^\alpha_n(t_1) \delta_{nm}\delta(t_1-t_2),
\end{equation}
where $\hat R^\alpha_n$ is nonzero only at the boundary and outside of material $\alpha$.
Taking the bullet product with $\check G^{\alpha\alpha}$, we have
\begin{equation}
  (\check G^{\alpha\alpha}\bullet\hat\Sigma^\alpha_R)_{nm}(t_1,t_2) = \check G^{\alpha\alpha}_{nm}(t_1,t_2)\hat R^\alpha_m(t_2).
\end{equation}

Next, we sum this over a set of unit cells $V$ and define $I \subset V$ to be the subset of $V$ which is at the interface.
We get
\begin{equation}
  \left\langle \frac i \pi \oint \dd{\xi_-^\alpha} \sum_{n \in V}\check G^{\alpha\alpha}\bullet\hat\Sigma_R^\alpha\right\rangle
  = \sum_{n\in I}S_c^\alpha\check g_s^\alpha(\vec x^\alpha_n)\transpose{(S_c^\alpha)} \circ \hat R_n^\alpha.
  \label{eq:GR}
\end{equation}
Note that in our model $\hat\Sigma^\alpha_R$ is very large outside material $\alpha$, such that $\check G^{\alpha\alpha}(\vec k, \vec x)\hat\Sigma^\alpha_R(\vec x) \sim 1$ when $\vec x$ is outside material $\alpha$.
Nevertheless, only the points in $I$ contribute in \cref{eq:GR}.
This is because the poles of $\check G^{\alpha\alpha}$ are shifted outside of the Eilenberger contour when $\vec x$ is outside of material $\alpha$, rendering the quasiclassical Green's function exactly equal to zero.
The points at, or very close to, the interface are therefore the only points where both $\check g^\alpha$ and $\hat R^\alpha_n$ are different from 0.

Since the width of the cylinder is small compared to $L$, $\check g_s^\alpha$ is approximately constant on the points in $I$.
We further assume that $\hat R_n^\alpha$ is also approximately constant on the points in $I$.
This means that if $l \in I$ and $\Gamma$ is the number of unit cells in $I$, then 
\begin{equation}
  \left\langle \frac i \pi \oint \dd{\xi_-^\alpha} \sum_{n \in V}\check G^{\alpha\alpha}\bullet\hat\Sigma_R^\alpha\right\rangle
  = \Gamma S_c^\alpha\check g_s^\alpha(\vec x^\alpha_l)\transpose{(S_c^\alpha)} \circ \hat R_l^\alpha.
  \label{eq:ref_rev1}
\end{equation}

Next, we must evaluate  
\begin{multline}
  \left(\hat\Sigma_R^\alpha\bullet\check G^{\alpha\alpha}\right)(\vec k, \vec x_n^\alpha)
  =V_e^\alpha\sum_{m \in \mathbb{Z}^3}\int_{\Diamond_\alpha}\frac{\dd[3]{q}}{(2\pi)^3} \hat R_m^\alpha \\
  \circ \me{-i\rho_B (\vec k -\vec q) \cdot \vec\delta^\alpha}
  \check G^{\alpha\alpha}(\vec q, \vec x_n)\me{i\rho_B (\vec k - \vec q) \cdot \vec\delta^\alpha}
  \\ \times
  \me{-i(\vec k - \vec q) \cdot (\vec x_m^\alpha - \vec x_n^\alpha)}.
  \label{eq:RG_first}
\end{multline}
First, we evaluate the sum over $m$.
We use that $\hat R^\alpha_m = \hat R^\alpha_l$, where $\vec x_l^\alpha$ is a point on the interface close to $\vec x_n^\alpha$, whenever $\vec x_m^\alpha$ is on the interface.
Otherwise, $\hat R^\alpha_m = 0$.
We find that 
\begin{multline}
\left(\hat\Sigma_R^\alpha\bullet\check G^{\alpha\alpha}\right)_{ij}(\vec k, \vec x_n^\alpha)
  =\hat R^\alpha_l\circ\int_{\Diamond_\alpha}\frac{\dd[3]{q}}{(2\pi)^3}f_{ij}(\vec q)  \\
  \times
  \check G_{ij}^{\alpha\alpha}(\vec k + \vec q, \vec x_n),
  \label{eq:RG_second}
\end{multline}
where $f_{ij}$ is a normalized function which is peaked at $\vec q = 0$.
Next, integrating over the Eilenberger contour and averaging over momentum directions, we find that
\begin{equation}
  \left\langle \frac i \pi \oint \dd{\xi_-^\alpha} \sum_{n \in V}\hat\Sigma_R\bullet\check G^{\alpha\alpha}\right\rangle
  = \Gamma \hat R_l \circ S_c^\alpha\check g_s^\alpha(\vec x^\alpha_l)\transpose{(S_c^\alpha)}.
  \label{eq:ref_rev2}
\end{equation}

Next, we must evaluate the tunneling self-energy,
\begin{equation}
  \check \Sigma_T^\alpha = \hat T^{\alpha\beta}\bullet\check G_0^{\beta\beta}\bullet \hat T^{\beta\alpha}.
\end{equation}
To proceed, we must assume some properties of the tunneling term.
The tunneling should be short-ranged and only at lattice points at the interface between the two materials.
For each unit cell in material $\alpha$ at the interface we assume that there is exactly one connected unit cell in material $\beta$.
For simplicity, we label the connected unit cells the same.
This means that if $\vec x^\alpha_n$ is at the interface, then the connected unit cell in material $\beta$ is $\vec x^\beta_n$.
With this we have
\begin{equation}
  \hat T^{\alpha\beta}_{nm} = \sum_{l\in \text{int}} \hat t^{\alpha\beta}_{l} \delta_{ln}\delta_{lm},
\end{equation}
where the sum goes over all the points at the interface.
Hence, if $\chi_\text{int}$ is the characteristic function which is 1 if the argument is at the interface and 0 otherwise, then
\begin{multline}
  \left(\check \Sigma_T^\alpha\right)_{nm}(t_1,t_2) 
  \\
  = 
  \hat t^{\alpha\beta}_{n}(t_1)\left(\check G_0^{\beta\beta}\right)_{nm}(t_1,t_2) \hat t^{\beta\alpha}_{m}(t_2)\chi_\text{int}(n)\chi_\text{int}(m),
\end{multline}

In order to evaluate the bullet product
\begin{multline}
\left(\check G^{\alpha\alpha}\bullet\hat\Sigma_T^\alpha\right)(\vec k, \vec x_n^\alpha)
=\chi_\text{int}(n)\sum_{m \in \text{int}} \check G^{\alpha\alpha}(\vec k, \vec x_m^\alpha) \\
  \circ \me{-i\rho_B \vec k \cdot \vec\delta^\alpha}
  \hat t^{\alpha\beta}_{m}\circ\left(\check G_0^{\beta\beta}\right)_{mn}\circ \hat t^{\beta\alpha}_{n}\me{i\rho_B \vec k \cdot \vec\delta^\alpha}
  \me{-i\vec k \cdot (\vec x_m^\alpha - \vec x_n^\alpha)},
  \label{eq:tunn_bull_1}
\end{multline}
we write
\begin{multline}
  \left(\check G_0^{\beta\beta}\right)_{mn} = V_e^\beta \int_{\Diamond_\beta} \frac{\dd[3]{p}}{(2\pi)^3}\me{i\rho_B\vec p \cdot \vec \delta^\beta}
  \check G_0^{\beta\beta}(\vec x_n^\beta, \vec p)
  \me{-i\rho_B\vec p \cdot\vec \delta^\beta}
  \\\times
  \me{i\vec p\cdot (\vec x_m^\beta - \vec x_n^\beta)}.
\end{multline}
We can separate this integral into the quasiclassical contribution and a rest term, or high-energy contribution, according to the Eilenberger contour.
The high-energy contribution was not negligible when we calculated the impurity self-energy.
This was because we evaluated the Green's function at $m=n$.
The high-energy contribution to the term in \cref{eq:tunn_bull_1} with $m=n$ will only renormalize $\hat R_n$, because it only depends on $\hat H_0^\beta$, as we showed earlier.
When evaluated at $m \neq n$ the oscillating exponential suppresses the integral for the high-energy contribution.
For this reason, we neglect the high-energy contribution.

Next, we must evaluate the quasiclassical part.
Close to the Fermi surface we have $\xi^\beta_-(p) = 0 + (p - p_F^\beta)(\xi_-^\beta)'(p_F^\beta) = v_F^\beta(p - p_F^\beta)$.
Hence, if the poles are located at $\{\xi_i\}_i$,
\begin{multline}
 \frac i \pi  \oint \dd{\xi_-^\beta}\check G_0^{\beta\beta}\me{i\vec p \cdot \vec r}
  = -\sum_{\xi_i}\sgn(\Im[\xi_i])\Res(\check G_0^{\beta\beta}, \xi_i)
  \\\times\exp(i\vec r\cdot \uv p[p_F + \xi_i/v_F^\beta])
\end{multline}
From \cref{eq:imp_final} we know that impurity scattering gives rise to an imaginary shift in the pole location, such that $\abs{\Im(\xi_i)} \ge 1/2\tau_\text{imp}^\beta$.
Therefore,
\begin{equation}
  \abs{\me{ir\xi_i/v_F^\beta}} < \me {-r/2l^\beta_\text{mfp}},
\end{equation}
where $l^\beta_\text{mfp} = \tau^\beta_\text{imp} v_F^\beta$ is the mean free path.
The effective mean free path very close to the interface may additionally be lowered by interfacial disorder.

The exponential decay means that we need only consider relative distances up to around the mean free path in the sum over $m\in \text{int}$.
In the dirty limit, which is what we consider here, it is assumed that $1/2\tau^\beta_\text{imp}$ is much larger than all the other self-energy contributions, and therefore $\abs{\Im(\xi_i)}$ is much larger than the real part of $\xi_i$.
As a result, when $r < 2l_\text{mfp}^\beta$, $r\Re(\xi_i)/v_F^\beta < 2\tau^\beta_\text{imp}\Re(\xi_i) \ll 1$, which means that we can neglect $\Re\xi_i/v_F^\beta$ in the exponential function when $r < 2l_\text{mfp}^\beta$. Hence,
\begin{align}
 \frac i \pi  \oint \dd{\xi_-^\beta}\check G_0^{\beta\beta}\me{i\vec p \cdot \vec r}
 = S_c^\beta \check g^\beta_{0}\transpose{(S_c^\beta)}\me{i\vec r \cdot \vec p_F^\beta}f^\beta(\vec r).
\end{align}
where $f^\beta(\vec r)$ is an exponentially decaying function that gives rise to a soft cutoff as a function of relative distance at $\abs{\vec r} \approx 2l_\text{mfp}^\beta$.
Hence, we find that
\begin{multline}
 \left\langle \frac i \pi \oint \dd{\xi_-^\alpha} \sum_{n \in V}\check G^{\alpha\alpha}\bullet\hat\Sigma_T \right\rangle
 % = -i\pi V_e^\beta \sum_{n\in I}\sum_{\substack{m\in \text{int} \\ \abs{\vec x_m^\beta - \vec x_n^\beta} < 2l_\text{mfp}^\beta}}\int \frac{\dd{\Omega_p}}{4\pi} \\
 = -i\pi V_e^\beta \sum_{n\in I}\sum_{m\in \text{int}}\int \frac{\dd{\Omega_p}}{4\pi} \\
 \times\int \frac{\dd{\Omega_k}}{4\pi}
  \frac{f^\beta(\vec x_m^\beta - \vec x_n^\beta)(\vec p_F^\beta)^2}{2\pi \vec v_F^\beta}
  S_c^\alpha \check g^\alpha(\vec k_F^\alpha, \vec x_m^\alpha) \transpose{(S_c^\alpha)}
  \\
  \circ \me{-i\rho_B \vec k_F^\alpha \cdot \vec\delta^\alpha}
  \hat t^{\alpha\beta}_{n}
\me{i\rho_B \vec p_F^\beta \cdot \vec\delta^\beta}
S_c^\beta \check g^\beta_{0}(\vec p_F^\beta, \vec x_n^\beta) \transpose{(S_c^\beta)}
  \\\circ 
  \me{-i\rho_B \vec p_F^\beta \cdot \vec\delta^\beta}\hat t^{\beta\alpha}_{m}\me{i\rho_B \vec k_F^\alpha \cdot \vec\delta^\alpha}
 \me{-i\vec k_F^\alpha \cdot (\vec x_m^\alpha - \vec x_n^\alpha)}\\\times\me{i\vec p_F^\beta \cdot (\vec x_m^\beta - \vec x_n^\beta)}.
\end{multline}
Next, we assume that the averaging over all momentum directions for both $\vec p_F^\beta$ and $\vec k_F^\alpha$ gives the $s$-wave contribution from the Green's function together with a renormalization of the tunneling amplitudes.
This is the case because we assume that the Green's functions are approximately spherically symmetric also close to the interface.
As a result, we finally have
\begin{multline}
  \left\langle \frac i \pi \oint \dd{\xi_-^\alpha} \sum_{n \in V}\check G^{\alpha\alpha}\bullet\hat\Sigma_T \right\rangle_{--}
  \\
  = -i\sum_{n\in I} \check g_s^\alpha(\vec x_n^\alpha) \circ \hat{\mathcal T}_n^{\alpha\beta} \circ \check g^\beta_{0,s}(\vec x_n^\beta) \circ\hat{\mathcal T}_n^{\beta\alpha},
  \label{eq:tunn_rev1}
\end{multline}
where 
\begin{equation}
  \hat{\mathcal T}_n^{\alpha\beta} = \transpose{(S_c^\alpha)}
  \hat{\tilde t}^{\alpha\beta}_{n}
  S_c^\beta,
\end{equation}
and where $\hat{\tilde t}^{\alpha\beta}_{n}$ are the renormalized versions of $\hat{t}^{\alpha\beta}_{n}$ resulting from the average over momentum directions.
Similarly, $\hat{\mathcal T}_n^{\beta\alpha} = \transpose{(S_c^\beta)}\hat{\tilde t}^{\beta\alpha}_{n}S_c^\alpha$.
In a similar way, we find that
\begin{multline}
  \left\langle \frac i \pi \oint \dd{\xi_-^\alpha} \sum_{n \in V}\hat\Sigma_T\bullet\check G^{\alpha\alpha} \right\rangle_{--}
  \\
  = -i\sum_{n\in I} \hat{\mathcal T}_n^{\alpha\beta} \circ \check g^\beta_{0,s}(\vec x_n^\beta) \circ\hat{\mathcal T}_n^{\beta\alpha} \circ \check g_s^\alpha(\vec x_n^\alpha).
  \label{eq:tunn_rev2}
\end{multline}

We choose the volume defined by the unit cells in $V$ to be approximately the shape of a wide cylinder which includes the interface.
Let the discs at the ends of this cylinder have $\Gamma_2$ points and define a plane.
Let $\uv n$ be the unit vector that is orthogonal to this plane and points out of material $\alpha$.
We assume that the width of the cylinder is much larger than the length.
Inserting \cref{eq:tunn_rev1,eq:ref_rev1} into \cref{eq:gorkov_cond_2}, integrating over the Eilenberger contour and momentum directions and summing over the unit cells in $V$, we get that
\begin{multline}
  i\Gamma_2\uv n \cdot \check{\vec j}^\alpha(\vec x_l^\alpha)/\lvert\vec\delta\rvert + \sum_{n\in V}\check g^\alpha_s\circ\varepsilon \tau_z 
  \\
  - \left\langle \frac i \pi \oint \dd{\xi_-^\alpha} \sum_{n \in V}\check G^{\alpha\alpha}\bullet (\check \Sigma_\text{imp}^\alpha + \check V^\alpha) \right\rangle_{--}
\\
+i\Gamma\check g_s^\alpha(\vec x_l^\alpha) \circ \hat{\mathcal T}_l^{\alpha\beta} \circ \check g^\beta_{0,s}(\vec x_l^\beta) \circ\hat{\mathcal T}_l^{\beta\alpha} 
\\
-\Gamma \check g_s^\alpha(\vec x_l^\alpha)\circ \transpose{(S_c^\alpha)}\hat R_l S_c^\alpha
=\left\langle \frac i \pi \oint \dd{\xi_-^\alpha} \sum_{n \in V}\check G^{\alpha\alpha}\xi_-^\alpha \right\rangle,
\label{eq:bc_pre_1}
\end{multline}
where $l$ is again a unit cell in $I$ and $\lvert\vec\delta\rvert$ is the distance between nearest neighbors in the direction of $\uv n$.
We note that $\Gamma/\Gamma_2$ can in general be different from 1 because the interface need not lie in a perfect plane parallel to the ends of the cylinder.
We assume that the second and third terms on the left-hand side of \cref{eq:bc_pre_1} are negligible compared to the fourth and fifth terms because the width of the cylinder is much larger than its length and $\hat R_l$ and $\hat{\mathcal T}_l^{\alpha\beta}\hat{\mathcal T}_l^{\beta\alpha}$ are large compared to $\varepsilon$ and $(\check \Sigma_\text{imp}^\alpha + \check V^\alpha)$.
However, we cannot neglect the term on the right-hand side.
The way to remove this term is again to use the other Gor'kov equation.
From the other Gor'kov equation, \cref{eq:gorkov_cond_1}, we get, using \cref{eq:tunn_rev2,eq:ref_rev2}, that
\begin{multline}
  \sum_{n\in V}\varepsilon \tau_z \circ\check g^\alpha_s 
  - \left\langle \frac i \pi \oint \dd{\xi_-^\alpha} \sum_{n \in V}(\check \Sigma_\text{imp}^\alpha + \check V^\alpha)\bullet\check G^{\alpha\alpha} \right\rangle_{--}
\\
+i\Gamma\hat{\mathcal T}_l^{\alpha\beta} \circ \check g^\beta_{0,s}(\vec x_l^\beta) \circ\hat{\mathcal T}_l^{\beta\alpha} \circ \check g_s^\alpha(\vec x_l^\alpha)
\\
-\Gamma \transpose{(S_c^\alpha)}\hat R_l S_c^\alpha\circ\check g_s^\alpha(\vec x_l^\alpha)
=\left\langle \frac i \pi \oint \dd{\xi_-^\alpha} \sum_{n \in V}\check G^{\alpha\alpha}\xi_-^\alpha \right\rangle.
\label{eq:bc_pre_2}
\end{multline}
Here, we neglect the first two terms for the same reason as above.
Combining \cref{eq:bc_pre_1,eq:bc_pre_2} and absorbing the factor $\lvert\vec\delta\rvert\Gamma/\Gamma_2$ into the reflection and tunneling matrices, we finally get the boundary condition,
\begin{equation}
  \uv n \cdot \check{\vec j}^\alpha = \left[\hat{\mathcal T}_l^{\alpha\beta} \circ \check g^\beta_{0,s}(\vec x_l^\beta) \circ\hat{\mathcal T}_l^{\beta\alpha} + i\transpose{(S_c^\alpha)}\hat R_l S_c^\alpha,\, \check g_s^\alpha\right]_\circ.
  \label{eq:bc_final_m1}
\end{equation}

One can use the Dyson equation to write $\check g^\beta_{0,s}$ as a series expansion in $\check g^\beta_{s}$ and the tunneling matrix.
In principle, this should produce a generalization of the Nazarov boundary condition~\cite{nazarov1999,nazarov_blanter_2009}.
However, we are here interested in the diffusive regime, meaning that the matrix current is small compared to the Fermi velocity.
This is the case when the tunneling and reflection amplitudes are small.
For this reason, we need only consider \cref{eq:bc_final_m1} to the lowest order in the tunneling matrices, which are obtained by setting $\check g^\beta_{0,s} = \check g^\beta_{s}$, yielding
\begin{equation}
  \uv n \cdot \check{\vec j}^\alpha = \left[\hat{\mathcal T}_l^{\alpha\beta} \circ \check g^\beta_{s}(\vec x_l^\beta) \circ\hat{\mathcal T}_l^{\beta\alpha} + i\transpose{(S_c^\alpha)}\hat R_l S_c^\alpha,\, \check g_s^\alpha\right]_\circ.
  \label{eq:bc_final}
\end{equation}
We note that in the absence of antiferromagnetism, this exactly reproduces the generalized Kupriyanov-Lukichev boundary condition for spin-active boundaries in the quasiclassical theory for normal metals~\cite{KL1988,eschrig2015}.

\section{Nonuniform magnetic textures}%
\label{sec:nonuniform_magnetic_textures}
In this section, we derive the self-energy terms associated with nonuniform magnetic textures in antiferromagnets.
We find that a spatial gradient in the magnetic texture gives rise to a term in the covariant gradient, similar to spin-orbit coupling, and a temporal gradient gives rise to an effective magnetic field.

In both cases, we must evaluate $R^\dagger \partial R$, where $\partial$ can be either the time derivative or gradient operator and $R$ is given by \cref{eq:rotationMatrix}.
We find that
\begin{equation}
  R^\dagger \partial R = -\frac i 2 \partial (\theta\sin\phi \sigma_x - \theta\cos\phi\sigma_y),
\end{equation}
where the direction of the Néel vector is $\vec n = (\sin\theta\cos\phi,\,\sin\theta\sin\phi,\,\cos\theta)$.
From \cref{eq:V_beginning} we see that the spatial gradient of the Néel vector gives rise to a self-energy term equal to
\begin{multline}
  (\Sigma^\alpha_{s})_{nm}(t_1,t_2) = -\left(K^\alpha_{nm}\left[\vec x_n^\alpha - \vec x_m^\alpha\right] + \left[\vec\delta^\alpha\rho_B,\, K^\alpha_{nm}\right]\right)
  \\
  \cdot\left(R^\dagger\nabla R\right)(\vec x_n^\alpha, t_1)\delta(t_1-t_2).
\end{multline}
To get how it looks in the final equation, we must Fourier transform and project onto the conduction band by use of $S_c^\alpha$.
By doing this, we get
\begin{multline}
  \transpose{(S_c^\alpha)}\Sigma^\alpha_{s}(\vec k, \vec x_n^\alpha, T)S_c^\alpha
  = -\frac{\vec v_F^\alpha}{2}
   \cdot \nabla (\theta\sin\phi \sigma_x - \theta\cos\phi\sigma_y).
\end{multline}
Since this is a momentum-dependent self-energy, we see that this is supposed to go into the covariant derivative.
As a result, the covariant derivative looks like
\begin{multline}
  \tilde\nabla\circ \check g^\alpha = \nabla\check g^\alpha
  - i\Biggl[\frac{1}{2}\nabla (\theta\sin\phi \sigma_x
  - \theta\cos\phi\sigma_y),\,\check g^\alpha\Biggr]
  \\
  - i[\check{\vec A}_{\text{rest}},\,\check g^\alpha],
\end{multline}
where $\check{\vec A}_{\text{rest}}$ is the remaining $p$-wave contribution, which may come from the vector gauge field or spin-orbit coupling.

The temporal gradient gives rise to a term similar to a magnetic field.
From \cref{eq:V_beginning} we see that the temporal gradient of the Néel vector gives rise to a self-energy term equal to
\begin{align}
  (\Sigma^\alpha_{t})_{nm}(t_1,t_2) = -i\tau_z\left(R^\dagger\dot R\right)(\vec x_n^\alpha, t_1)\delta_{nm}\delta(t_1-t_2),
\end{align}
since
\begin{equation}
  \rho_A\left(R^\dagger\dot R\right)(\vec x_n^\alpha, t_1) + \rho_B\left(R^\dagger\dot R\right)(\vec x_n^\alpha+\vec\delta^\alpha, t_1)
  \approx\left(R^\dagger\dot R\right)(\vec x_n^\alpha, t_1).
\end{equation}
If we again Fourier transform in relative coordinates and transform using $S_c^\alpha$, we get
\begin{multline}
  \transpose{(S_c^\alpha)}\Sigma^\alpha_{t}(\vec k, \vec x_n^\alpha, T)S_c^\alpha
  = -\frac{1}{2}\sqrt{1 - (J^\alpha/\eta^\alpha)^2}
  \\\times
  \tau_z\partial_T (\theta\sin\phi \sigma_x - \theta\cos\phi\sigma_y).
\end{multline}
The factor $\sqrt{1 - (J^\alpha/\eta^\alpha)^2}$ comes from the projection of $\sigma_x$ and $\sigma_y$ onto the conduction band.
To understand the physical reason for this factor, consider a general electron state near the Fermi level.
An electron near the Fermi level will in general be in a superposition of spin-up and spin-down, but the spin-up component and the spin-down component will have different spatial distributions.
For the spin of this electron at a given lattice site to have a non-zero projection in a direction orthogonal to the Néel vector, it will need to be in a superposition of spin-up and spin-down.
At $J^\alpha/\eta^\alpha = 0$, an electron state near the Fermi level which is in an equal superposition of spin-up and spin-down will have spin everywhere orthogonal to the Néel vector.
However, as $J^\alpha/\eta^\alpha$ increases, the spin-up and spin-down component starts to separate in space, and in the limit $J^\alpha/\eta^\alpha \to 1$, any superposition of spin-up and spin-down has all of its spin-up component localized on one sublattice and all of its spin-down component localized on the other sublattice.
This means that it has spin along the Néel vector everywhere in space.
As a result, the effect of spin-splitting fields orthogonal to the Néel vector is suppressed as $J^\alpha/\eta^\alpha$ increases.

\section{Observables}%
\label{sec:observables}
Generally, observables such as densities or currents may be written
\begin{multline}
  \label{eq:observableDefinition}
  Q(\vec x_n^\alpha, T) = \bigl\langle c_n^{\alpha\dagger}(T)M(\vec x_n^\alpha, -i\Delta_R)c_n^\alpha(T) \\
  - \transpose{(c_n^{\alpha})}(T)\transpose M(\vec x_n^\alpha, -i\Delta_R)\transpose{(c_n^{\alpha\dagger})}(T) \bigr\rangle  + C,
\end{multline}
where $C$ is a constant and $M$ is a matrix that depends on the observable.
We can relate this to our Green's functions which are defined by the spin-rotated creation and annihilation operators $\tilde c_n^\alpha$, as defined by \cref{eq:rotc}, if we define
\begin{multline}
  \tilde M = \left[\rho_A R^\dagger(\vec x^\alpha_n, t) + \rho_B R^\dagger(\vec x_n^\alpha + \vec\delta^\alpha, t)\right] M
  \\
  \times\left[\rho_A R(\vec x^\alpha_n, t) + \rho_B R(\vec x_n^\alpha + \vec\delta^\alpha, t)\right].
\end{multline}
With this,
\begin{multline}
  Q(\vec x_n^\alpha, T) = C + i\int_{\Diamond^\alpha} \frac{\dd[3]{k}}{(2\pi)^3}\int_{-\infty}^{\infty}\frac{\dd{\varepsilon}}{2\pi}\\
  \times\Tr[\tilde M(\vec x_n^\alpha, \vec k)\tau_z\hat G^{K,\alpha\alpha}(\vec k, \vec x_n^\alpha, T, \varepsilon)].
\end{multline}

The quasiclassical treatment is only valid for $\varepsilon \ll E_c^\alpha$.
As a result, we should split the energy integral,
\begin{multline}
  \int_{\Diamond^\alpha} \frac{\dd[3]{k}}{(2\pi)^3}\int_{-\infty}^{\infty}\frac{\dd{\varepsilon}}{2\pi}\Tr[\tilde M(\vec x_n^\alpha, \vec k)\tau_z\hat G^{K,\alpha\alpha}(\vec k, \vec x_n^\alpha, T, \varepsilon)]
  \\
  = \int_{-a}^{a}\frac{\dd{\varepsilon}}{2\pi}\int_{\Diamond^\alpha} \frac{\dd[3]{k}}{(2\pi)^3}\Tr[\tilde M(\vec x_n^\alpha, \vec k)\tau_z\hat G^{K,\alpha\alpha}(\vec k, \vec x_n^\alpha, T, \varepsilon)]
  \\
  + \left(\int_{-\infty}^{-a}\frac{\dd{\varepsilon}}{2\pi}+\int_{a}^{\infty}\frac{\dd{\varepsilon}}{2\pi}\right)\int\frac{\dd[3]{k}}{(2\pi)^3}
  \\\times
  \Tr[\tilde M(\vec x_n^\alpha, \vec k)\tau_z\hat G^{K,\alpha\alpha}(\vec k, \vec x_n^\alpha, T, \varepsilon)],
  \label{eq:obs_gen}
\end{multline}
where $a$ is much smaller than $E_c^\alpha$.
In the diffusive regime, $a$ should also be much smaller than the elastic impurity scattering rate.
We can rewrite the first term on the right-hand side by again using the Eilenberger contour.
The Keldysh Green's function is $\sim 1/(\xi_-^\alpha)^2$ for large $(\xi_-^\alpha)^2$, so we can neglect the high energy contribution, $\fint\dd{\xi_-^\alpha}$.
Hence,
\begin{multline}
  \int_{-a}^{a}\frac{\dd{\varepsilon}}{2\pi}\int_{\Diamond^\alpha} \frac{\dd[3]{k}}{(2\pi)^3}\Tr[\tilde M(\vec x_n^\alpha, \vec k)\tau_z\hat G^{K,\alpha\alpha}(\vec k, \vec x_n^\alpha, T, \varepsilon)]
  \\
  = -i\pi N_0^\alpha\Biggl\langle\int_{-a}^{a}\frac{\dd{\varepsilon}}{2\pi} \Tr\big[\transpose{(S_c^\alpha)}\tilde M(\vec x_n^\alpha, \vec k_F)S_c^\alpha
   \\ \times \tau_z\hat g^{K,\alpha}(\vec k_F, \vec x_n^\alpha, T, \varepsilon)\big]\Biggr\rangle.
\end{multline}

Next, we must evaluate the second term on the right-hand side of \cref{eq:obs_gen}.
Generally, we can write
\begin{equation}
  \hat G^{R,\alpha\alpha} = \left(\tau_z\varepsilon - \hat H_0^\alpha - \hat\Sigma^{R,\alpha}\right)^{-1} + \delta\hat G^{R,\alpha\alpha}.
\end{equation}
Inserting this into the equation
\begin{equation}
  \tau_z\varepsilon \circ \hat G^{R,\alpha\alpha} - \hat H_0^\alpha\hat G^{R,\alpha\alpha} - \hat\Sigma^{R,\alpha} \bullet \hat G^{R,\alpha\alpha} = 1,
\end{equation}
one gets an equation for $\delta\hat G^{R,\alpha\alpha}$.
We find that the contribution to the expression for the observable from $\delta\hat G^{R,\alpha\alpha}$ is negligible, so we neglect it in the following.
We assume that $a$ is sufficiently large such that states at $\abs{\varepsilon} \ge a $ are either completely occupied or completely unoccupied.
Moreover, $a$ is much larger than the superconducting gap, so the density of states at energies above $a$ should not be affected by superconductivity.
For this reason, we assume that we can neglect superconductivity when considering the high-energy contribution.
When this is the case,
\begin{equation}
  \hat G^{K,\alpha\alpha} = \sgn(\varepsilon)\left[\hat G^{R,\alpha\alpha} - (\hat G^{R,\alpha\alpha})^\dagger\right].
\end{equation}

By neglecting $\delta \hat G^{R,\alpha\alpha}$, we find that
\begin{align}
  \hat G^{R,\alpha\alpha} = S^\alpha \left[\tau_z \varepsilon - \mqty(\dmat{\xi_-^\alpha,\xi_+^\alpha}) - \transpose{(S^\alpha)}\hat\Sigma^{R,\alpha}S^\alpha\right]^{-1} \transpose{(S^\alpha)}.
\end{align}
Let $\transpose{(S^\alpha)}\hat\Sigma^{R,\alpha}S^\alpha = A$, then
\begin{subequations}
  \label{eq:gk_comps_obs}
\begin{align}
  [\transpose{(S^\alpha)}\hat G^{R,\alpha\alpha}S^\alpha]_{--} &= \big[\tau_z\varepsilon - \xi_-^\alpha - A_{--}
    \nonumber\\
  - A&_{-+}(\tau_z\varepsilon - \xi_+^\alpha - A_{++})^{-1}A_{+-}\big]^{-1},
\\
  [\transpose{(S^\alpha)}\hat G^{R,\alpha\alpha}S^\alpha]_{++} &= \big[\tau_z\varepsilon - \xi_+^\alpha - A_{++}
    \nonumber\\
  -A&_{+-}(\tau_z\varepsilon - \xi_-^\alpha - A_{--})^{-1}A_{-+}\big]^{-1},
  \\
[\transpose{(S^\alpha)}\hat G^{R,\alpha\alpha}S^\alpha]_{-+}
          &= -[\transpose{(S^\alpha)}\hat G^{R,\alpha\alpha}S^\alpha]_{--}A_{-+}
  \nonumber \\&\quad\quad\quad\times(\tau_z\varepsilon - \xi_+^\alpha - A_{++})^{-1},
  \\
[\transpose{(S^\alpha)}\hat G^{R,\alpha\alpha}S^\alpha]_{+-}
          &= -[\transpose{(S^\alpha)}\hat G^{R,\alpha\alpha}S^\alpha]_{++}A_{+-}
  \nonumber \\&\quad\quad\quad\times(\tau_z\varepsilon - \xi_-^\alpha - A_{--})^{-1}.
\end{align}
\end{subequations}
If
\begin{subequations}
\begin{align}
  A_{--}- A_{-+}(\tau_z\varepsilon - \xi_+^\alpha - A_{++})^{-1}A_{+-}
  = P_- J_- P_-^\dagger,
  \\
  A_{++}- A_{+-}(\tau_z\varepsilon - \xi_-^\alpha - A_{--})^{-1}A_{-+}
  = P_+ J_+ P_+^\dagger,
\end{align}
\end{subequations}
where $J_-$ and  $J_+$ are diagonal, we find that
\begin{multline}
  [\transpose{(S^\alpha)}\hat G^{k,\alpha\alpha}S^\alpha]_{\pm\pm,ij} = 
  2\pi i\sgn(\varepsilon)\sum_{l}P_{\pm,il}
  \\ \times \frac{\Im(J_{\pm,ll})/\pi}{[\varepsilon\tau_{z,ll} - \xi_\pm^\alpha - \Re(J_{\pm,ll})]^2 + [\Im(J_{\pm,ll})]^2}
  P_{\pm,lj}^\dagger.
  \label{eq:lorentzians}
\end{multline}
If not for the fact that $J_\pm$ depends on $\varepsilon$, this would be a sum of Lorentz distribution as functions of $\varepsilon$.
However, the dependence of $J_\pm$ on $\varepsilon$ is very weak close to the peak of the distribution.
For this reason, we neglect the dependence of both $P_\pm$ and $J_{\pm}$ on $\varepsilon$.
From \cref{eq:gk_comps_obs} we can see that $[\transpose{(S^\alpha)}\hat G^{R,\alpha\alpha}S^\alpha]_{\pm\mp}$ are products of two functions with peaks at distantly separated values of $\varepsilon$.
One peak is close to $\xi_-^\alpha$ and the other is close to $\xi_+^\alpha$.
As a result, we neglect these terms.

To proceed, we must evaluate terms that look like
\begin{multline}
  I_\pm \defeq \left(\int_{-\infty}^{-a}\frac{\dd{\varepsilon}}{2\pi}+\int_{a}^{\infty}\frac{\dd{\varepsilon}}{2\pi}\right)\int_{\xi_\text{min}}^{\xi_\text{max}}\dd{\xi_-^\alpha}2\pi\sgn(\varepsilon)g(\xi_-^\alpha)
  \\
  \times\frac{\Im(J_{\pm,ll})/\pi}{[\varepsilon\tau_{z,ll} - \xi_\pm^\alpha - \Re(J_{\pm,ll})]^2 + [\Im(J_{\pm,ll})]^2},
\end{multline}
where the function $g$ can be identified from \cref{eq:obs_gen,eq:lorentzians}.

From the fact that the retarded Green's function should be nonzero only for positive relative times, we have that $\Im(J_{-,ll}) = \tau_{z,ll}\abs{\Im(J_{-,ll})}$.
If we define
\begin{equation}
  f(y) = \int_{-\infty}^{y}\dd{x} \frac{\abs{\Im(J_{\pm,ll})}/\pi}{x^2 + [\Im(J_{\pm,ll})]^2},
\end{equation}
we find that
\begin{multline}
  I_- = \int_{a-\Re(J_{\pm,ll})+C}^{\xi_\text{max}}\dd{\xi_-^\alpha}g(\xi_-^\alpha)
  - \int^{-a-\Re(J_{\pm,ll})-C}_{\xi_\text{min}}\dd{\xi_-^\alpha}g(\xi_-^\alpha)
  \\
  + \int_{-C}^C\dd{\xi_-^\alpha} f(\xi_-^\alpha)\big\{g[\xi_-^\alpha + a -\Re(J_{\pm,ll} )] \\
  - g[-\xi_-^\alpha - a -\Re(J_{\pm,ll})]\big\},
  \label{eq:obs1}
\end{multline}
where $C$ is a number which is on the order of $\Im(J_{\pm,ll})$, and sufficiently large such that $f(y) \approx 0$ for $y \le -C$ and $f(y) = 1$ for $y \ge C$.
From \cref{eq:obs_gen,eq:lorentzians}, one can see that $g$ is a slowly varying function.
For this reason, we can neglect the last integral in \cref{eq:obs1}.
Next, we rewrite $I_-$ as one term which depend on $J_{\pm,ll}$ and one which does not, as
\begin{multline}
  I_- \approx \int_{a+C}^{\xi_\text{max}}\dd{\xi_-^\alpha}g(\xi_-^\alpha)
  - \int^{-a-C}_{\xi_\text{min}}\dd{\xi_-^\alpha}g(\xi_-^\alpha)
  \\
  + \Re(J_{\pm,ll})[g(a)+ g(-a)],
\end{multline}
so that, since $g(\pm a) \approx g(0)$,
\begin{multline}
  \left(\int_{-\infty}^{-a}\frac{\dd{\varepsilon}}{2\pi}+\int_{a}^{\infty}\frac{\dd{\varepsilon}}{2\pi}\right)\int_{\xi_\text{min}}^{\xi_\text{max}}\dd{\xi_-^\alpha} g(\xi_-^\alpha)[\transpose{(S^\alpha)}\hat G^{k,\alpha\alpha}S^\alpha]_{--,ij}
  \\
  = i\delta_{ij}\int_{a+C}^{\xi_\text{max}}\dd{\xi_-^\alpha}g(\xi_-^\alpha)
  - i\delta_{ij}\int^{-a-C}_{\xi_\text{min}}\dd{\xi_-^\alpha}g(\xi_-^\alpha)
  \\
  + ig(0)[A_{--} + A_{--}^\dagger]_{ij},
\end{multline}
where we have used that for $\xi_-^\alpha \approx 0$,
\begin{equation}
  2P_- \Re(J_-)P_-^\dagger = P_- J_-P_-^\dagger + \left(P_- J_-P_-^\dagger\right)^\dagger
  \approx A_{--} + A_{--}^\dagger.
\end{equation}

Evaluating $I_+$ is less difficult because $\xi_+^\alpha \gg a$ for all $\vec k$. Hence,
\begin{equation}
  I_+ = \int_{\xi_\text{min}}^{\xi_\text{max}}\dd{\xi_-^\alpha} g(\xi_-^\alpha).
\end{equation}
Inserting this into the expression for the high-$\varepsilon$ contribution to the observable, we find
\begin{multline}
  \left(\int_{-\infty}^{-a}\frac{\dd{\varepsilon}}{2\pi}+\int_{a}^{\infty}\frac{\dd{\varepsilon}}{2\pi}\right)\int\frac{\dd[3]{k}}{(2\pi)^3}
  \Tr[\tilde M\tau_z\hat G^{K,\alpha\alpha}]
  \\
  = i\left\langle \int_{\xi_\text{min}}^{\xi_\text{max}}\dd{\xi_-^\alpha}N_0^\alpha(\xi_-^\alpha)\Tr\left\{[\transpose{(S^\alpha)}\tilde MS^\alpha]_{++}\tau_z\right\}\right\rangle
  \\
  +
  i\left\langle \int_{a+C}^{\xi_\text{max}}\dd{\xi_-^\alpha}N_0^\alpha(\xi_-^\alpha)\Tr\left\{[\transpose{(S^\alpha)}\tilde MS^\alpha]_{--}\tau_z\right\}\right\rangle
  \\
  -
  i\left\langle \int_{\xi_\text{min}}^{-a-C}\dd{\xi_-^\alpha}N_0^\alpha(\xi_-^\alpha)\Tr\left\{[\transpose{(S^\alpha)}\tilde MS^\alpha]_{--}\tau_z\right\}\right\rangle
  \\
  +
  i\left\langle N_0^\alpha\Tr\left\{[\transpose{(S^\alpha)}\tilde MS^\alpha]_{--}\tau_z[A_{--} + A_{--}^\dagger]\right\}\right\rangle.
\end{multline}
The first three terms on the right-hand side are just constants and can be absorbed into the constant $C$ in the expression for the observable.
By doing this, we get that the observable can finally be written
\begin{multline}
  Q = C + \frac{N_0^\alpha}{2}\Biggl\langle\int_{-a}^{a}\dd{\varepsilon} \Tr\big[\transpose{(S_c^\alpha)}\tilde MS_c^\alpha \tau_z\hat g^{K,\alpha}\big]\Biggr\rangle
   \\
 -  N_0^\alpha\Big\langle\Tr\Big\{[\transpose{(S^\alpha_c)}\tilde MS^\alpha_c\tau_z\\
 \times\transpose{(S^\alpha_c)}\left[\hat\Sigma^{R,\alpha} + (\hat\Sigma^{R,\alpha})^\dagger\right]S^\alpha_c\Big\}\Big\rangle.
 \label{eq:observables}
\end{multline}
To compute observables from the quasiclassical Green's functions, one therefore generally also need to take into account the contribution from the self-energy term.
Note that since the quasiclassical Green's function is not gauge-invariant, the second term in \cref{eq:observables} is required to make the observables gauge-invariant.

For concrete examples of observables, consider the electric charge density in material $\alpha$, $n_e^\alpha$, and the spin densities in material $\alpha$, $\vec s^\alpha = (s_{x}^\alpha, s_{y}^\alpha, s_{z}^\alpha)$.
For the electric charge density $\tilde M = e\tau_z/4$, which can be confirmed by inserting this into \cref{eq:observableDefinition}.
The denominator $4$ comes from the fact that we count each electron 4 times in \cref{eq:observableDefinition}.
To derive the formula for electric charge density, we can insert this into \cref{eq:observables}, giving
\begin{equation}
  n_e^\alpha = \frac{N_0^\alpha e}{8}\int_{-a}^a \dd{\varepsilon} \Tr(\hat g_s^{K,\alpha}) - 2N_0^\alpha e \phi_e^\alpha,
  \label{eq:obs_charge}
\end{equation}
where we dropped the constant and $\phi_e^\alpha$ is the deviation in the electrochemical potential away from $\mu^\alpha$, and may therefore vary in both time and space.
In other words, $\phi_e^\alpha$ is the real, diagonal part of the self-energy.
\Cref{eq:obs_charge} reproduces earlier results for charge density in the quasiclassical regime~\cite{eliashberg1972,rammer1986}.
We can see that the second term in \cref{eq:obs_charge} is necessary to retain gauge invariance.
Take for example a non-superconducting stationary system in equilibrium with an electrochemical potential $\phi_e^\alpha$.
The symmetric part of the quasiclassical Keldysh function is then $\hat g_s^{K,\alpha} = 2\diag(\operatorname{tanh}[\beta(\varepsilon + \phi_e^\alpha)/2],\operatorname{tanh}[\beta(\varepsilon + \phi_e^\alpha)/2], -\operatorname{tanh}[\beta(\varepsilon - \phi_e^\alpha)/2], -\operatorname{tanh}[\beta(\varepsilon - \phi_e^\alpha)/2])$, where $\beta$ is inverse temperature.
Taking the trace and integrating over energies, we get
\begin{equation}
  \frac{N_0^\alpha e}{8}\int_{-a}^a \dd{\varepsilon} \Tr(\hat g_s^{K,\alpha}) = \frac{N_0^\alpha e}{8} 16 \phi_e^\alpha = 2N_0^\alpha e \phi_e^\alpha.
\end{equation}
The electrochemical potential is gauge-dependent, so the second term in \cref{eq:obs_charge} is required to cancel the gauge-dependent contribution from $\hat g_s^{K,\alpha}$ in this case.

For the spin density in direction $i$, $\tilde M = \sigma_i/8$.
The projection of spin Pauli matrices onto the conduction band is trivial for the $z$-direction since it commutes with the $S^\alpha$ matrix. 
That is, $(S_c^\alpha)^T \sigma_z S_c^\alpha= \sigma_z$.
However, $(S_c^\alpha)^T \sigma_{x/y} S_c^\alpha= \sqrt{1 - (J^\alpha/\eta^\alpha)^2}\sigma_{x/y}$, so for the directions orthogonal to the Néel vector we get an additional factor $\sqrt{1 - (J^\alpha/\eta^\alpha)^2}$.
If the initial Hamiltonian in material $\alpha$, given by \cref{eq:material_hamiltonian}, has a Zeeman spin-splitting field $\vec h^\alpha$, this gives rise to a self-energy term equal to $\hat\Sigma_{Z}^{R,\alpha} = \vec h^\alpha \cdot \vec \sigma \tau_z$ before projection onto the conduction band.
Inserting this into \cref{eq:observables}, we get that the spin densities are given by
\begin{subequations}
  \begin{align}
    s_x^\alpha = &\sqrt{1 - \left(\frac{J^\alpha}{\eta^\alpha}\right)^2}\frac{N_0^\alpha}{16}\int_{-a}^a \dd{\varepsilon} \Tr(\sigma_x\tau_z\hat g_s^{K,\alpha}) \nonumber\\
    & \qquad\qquad\qquad\qquad\quad - \left[1 - \left(\frac{J^\alpha}{\eta^\alpha}\right)^2\right]N_0^\alpha h_x^\alpha,\\
    s_y^\alpha = &\sqrt{1 - \left(\frac{J^\alpha}{\eta^\alpha}\right)^2}\frac{N_0^\alpha}{16}\int_{-a}^a \dd{\varepsilon} \Tr(\sigma_y\tau_z\hat g_s^{K,\alpha}) \nonumber\\
    & \qquad\qquad\qquad\qquad\quad - \left[1 - \left(\frac{J^\alpha}{\eta^\alpha}\right)^2\right]N_0^\alpha h_y^\alpha,\\
    s_z^\alpha = &\frac{N_0^\alpha}{16}\int_{-a}^a \dd{\varepsilon} \Tr(\sigma_y\tau_z\hat g_s^{K,\alpha}) - N_0^\alpha h_y^\alpha,
  \end{align}
\end{subequations}
where we again dropped the constant.
The extra factor of $\sqrt{1 - (J^\alpha/\eta^\alpha)^2}$ comes from the fact itinerant electrons become more polarized in the direction of the Néel vector as $J^\alpha/\eta^\alpha$ increases, as discussed above.
This polarization comes in through two different aspects.
First, the Zeeman spin-splitting felt by the itinerant electrons is reduced by a factor $\sqrt{1 - (J^\alpha/\eta^\alpha)^2}$.
Second, the $\sigma_x$ and $\sigma_y$ components of the Green's function do not correspond to spin in the same sense as in a normal metal.
In the limit of very strong exchange coupling $J^\alpha$, the itinerant electrons become fully polarized, and $s_x^\alpha = s_y^\alpha = 0$.

To compute the sublattice-resolved charge densities, one can use \cref{eq:projectionOfProjection} together with $\tilde M = e\tau_z\rho_{A/B}/4$, which gives
\begin{equation}
  n_{A/B}^\alpha = \frac 1 2 n_e^\alpha \pm \frac{eJ^\alpha}{\eta^\alpha} s_z^\alpha.
\end{equation}
One can similarly use \cref{eq:observables} to compute energy and spin-energy densities~\cite{bergeret2018} and all associated current.
Another way to derive expressions for currents is to use the expressions for densities together with \cref{eq:usadel} to obtain conservation-laws of the form $\partial n/\partial t + \nabla\cdot \vec j = S$, where $n$ is the density, $\vec j$ can be identified as the current and $S$ is a source-term.
For instance, multiplying \cref{eq:usadel} with $-ieN_0\tau_z/8$, taking the trace, integrating over energy and adding $-2N_0 e\partial\phi_e^\alpha/\partial t$ to both sides of the equality sign, one obtains $\partial n_e^\alpha/\partial t + \nabla\cdot \vec j_e^\alpha = S_e^\alpha$, where the electric current density can be identified as
\begin{equation}
  \vec j_e^\alpha = \frac{N_0^\alpha e}{8}\int_{-a}^a \dd{\varepsilon} \Tr(\tau_z\hat{\vec j}^{K,\alpha}).
\end{equation}

\section{Conclusion}%
\label{sec:conclusion}
We have derived quasiclassical equations of motion which are valid for mesoscopic heterostructures with antiferromagnetic order, superconductivity, impurity scattering, external electric or magnetic fields, spin-orbit coupling, temporally or spatially inhomogeneous Néel vector, or, in principle any other effect that can be modeled using a quadratic Hamiltonian.
These are valid when the distance between the Fermi level and the edges of the conduction band, $\Delta E^\alpha$, is larger than all other energy scales except possibly the exchange energy which couples the itinerant electrons to the localized, antiferromagnetically ordered spins.
The ratio between the exchange energy and the chemical potential relative to the center of the two energy bands, $J^\alpha/\eta^\alpha$, can take any value between 0 and 1.
Our main results are the quasiclassical equation in the dirty regime, which are valid when the elastic impurity scattering rate is high compared to other energies, except for $\Delta E^\alpha$ and possibly $J^\alpha$, and when the isotropic part of the quasiclassical Green's function dominates.
The latter is true when the matrix current is small, which happens for instance when the system varies slowly on the scale of the mean free path, or when the proximity effect is small.
In the limit of very strong exchange coupling, such that $(J^\alpha/\eta^\alpha)^2 \to 1$, the short-ranged correlations can vary on the scale of the mean free path.
However, these correlations become vanishingly small in the diffusive limit.
Therefore, one can solve the equations by projecting the Green's functions onto the set of long-range components.
Being based on Keldysh theory, the equations can be used to study non-equilibrium situations, such as externally driven currents or spin injection.
Additionally, they can also be solved to study time-dependent phenomena, as there are ways to evaluate the circle products~\cite{houzet2008,virtanen2010,fyhn2021_time,fyhn2022}.
In the absence of antiferromagnetism, the equations reduce to the Eilenberger equation~\cite{eilenberger1968} and Usadel equation~\cite{usadel1970} for normal metals, as expected.
However, with antiferromagnetism, there are a few important differences.
First, all self-energy terms must be projected onto the conduction band.
Second, even nonmagnetic impurities behave magnetically because of the coupling between spin and sublattice.
Finally, this also changes the equation for the matrix current in the dirty regime.
We discuss the physical origin and implications of these effects in ref.~\cite{prl_submission}.

We have also derived boundary conditions that are valid in the diffusive regime. 
These are valid as long as the tunneling amplitudes are small, such that the matrix current is small compared to the Fermi velocity.
They take into account both tunneling and reflection and allow for both compensated and uncompensated interfaces, meaning that the coupling can be asymmetric in sublattice.
Additionally, the boundary conditions allow for spin-active boundaries and isolating, spin-active boundaries can be obtained by setting the tunneling matrices to zero.
In the absence of antiferromagnetism, the boundary conditions reduce to the generalized Kupriyanov-Lukichev boundary conditions for spin-active boundaries~\cite{KL1988,eschrig2015}.

Finally, we have derived an expression that can be used to compute observables from the quasiclassical Green's function.
This expression also includes the contribution from energies which are not captured by the quasiclassical Green's function.
As we saw in the example of charge density, the high-energy contribution is needed to make the observables gauge-invariant.

\appendix
\section{Numerical solution algorithm}
\label{sec:solAlg}
As an example, we illustrate how one can solve \cref{eq:usadel,eq:gen_curr,eq:bc_final} in a time-independent one-dimensional system at thermal equilibrium.
The components of the Green's function are not independent because of the normalization condition, so it is necessary to use a parametrization scheme.
For instance, one can use the Ricatti-parametrization~\cite{eschrig2000,konstandin2005} or the $\theta$-parametrization~\cite{ivanov2006}.
In order to solve \cref{eq:usadel,eq:gen_curr,eq:bc_final} numerically, one must first define a set of algebraic equations.
These equations can then be solved for the unknown parameters.
For simplicity, assume we need only solve \cref{eq:usadel,eq:gen_curr,eq:bc_final} in one material because the solution is known in all neighboring materials.
For this reason, we remove the superscript $\alpha$.
Let there be $N$ discretization points, and denote by $u_n^j$ the $j$th parameter at discretization point $n\in\{1,\dotsc,N\}$.
The spherically symmetric part of the quasiclassical Green's function at point $n$ is a function of the $M$ parameters.
Depending on the problem, the number of different parameters needed to characterize the system will vary.
At most, $M = 8$ in thermal equilibrium since it is only necessary to compute the retarded Green's function.
Let $\check g^R_{s,n}$ be the spherically symmetric part of the retarded Green's function in position $x = (n-1)\Delta x$, where $\Delta x$ is the distance between discretization points.
Then,
\begin{equation}
  \hat g^R_{s,n} = \hat g^R_{s,n}(u_n^1, u_n^2, \dotsc, u_n^M)
\end{equation}
is a function of only the local parameters $(u_n^1, u_n^2, \dotsc, u_n^M)$.

In order to solve \cref{eq:gen_curr,eq:usadel}, we need not only the Green's function but also its spatial derivative.
Let the derivative at point $n$ be $(\partial_x \check g_s^R)_n$.
This can be obtained from the gradients of the parameters,
\begin{equation}
  (\partial_x \check g_s^R)_n = \sum_{j=1}^M \frac{\partial \hat g^R_{s,n}}{\partial u_n^j}\frac{\partial u_n^j}{\partial x}. 
\end{equation}
Thus, we have $2M$ unknown parameters at each point: $(u_n^1, \dotsc, u_n^M, \partial_x u_n^1, \dotsc, \partial_x u_n^M)$.
The circle products reduce to normal matrix products in a static system, so, if $\hat j^R_n$ is the retarded matrix current at point $n$, we get from \cref{eq:gen_curr} that
\begin{multline}
  \hat j^R_n = -D\hat g^R_{s,n}(\partial_x\hat g^R_{s})_n + iD\hat g^R_{s,n}[A_x^R,\,\hat g^R_{s,n}] \\
  - \hat g^R_{s,n}\left[\frac{J^2}{2\eta^2}\sigma_z\tau_z\hat g_{s,n}^R\sigma_z\tau_z,\,\hat{j}^R_n\right].
  \label{eq:numCurr}
\end{multline}

\begin{algorithm}
  \caption{Numerical scheme for solving \cref{eq:usadel,eq:gen_curr,eq:bc_final}.}
    \label{alg:numAlg}
  \begin{algorithmic}[1]
    \Require{$(n_i, m_i)$ for $i\in\{1,\dotsc,N\}$ are $N$ different intervals and $(w^i_1,\dots,w^i_{m_i-n_i})$ are corresponding numerical weights.}
    \Statex
    \Function{R}{$\{u_n^j\}, \{\partial_x u_n^j\}$} %\Comment{$n \in (1,\dotsc,N)$, $j \in (1,\dotsc,M)$}
      \For{$i \gets 1 \textrm{ to } N$}
      \State {$\hat g^R_{s,i} \gets \hat g^R_{s,i}(u_i^1,\dotsc,u_i^M)$}
      \State {$(\partial_x\hat g^R_{s})_i \gets (\partial_x\hat g^R_{s})_i(u_i^1,\dotsc,u_i^M,\partial_xu_i^1,\dotsc,\partial_xu_i^M)$}
      \If {$i = 1$ or $i = N$} 
      \State {$\hat j^R_{i} \gets \hat j^R_{i}(\hat g^R_{s,i}, (\partial_x\hat g^R_{s})_i)$} \Comment{\cref{eq:numBC}}
      \Else
      \State {$\hat j^R_{i,0} \gets 0$}
      \State{$\hat j^R_{i} \gets \hat j^R_{i}(\hat g^R_{s,i}, (\partial_x\hat g^R_{s})_i, \hat j^R_{i,0})$} \Comment{\cref{eq:numCurr}}
      \While{$\lvert \hat j^R_{i} - \hat j^R_{i,0}\rvert > \text{tolerance}$}
      \State {$\hat j^R_{i,0} \gets \hat j^R_{i}$}
      \State{$\hat j^R_{i} \gets \hat j^R_{i}(\hat g^R_{s,i}, (\partial_x\hat g^R_{s})_i, \hat j^R_{i,0})$} \Comment{\cref{eq:numCurr}}
      \EndWhile
      \EndIf
      \EndFor
      \For{$i \gets 1 \textrm{ to } N$}
      \State{$r_1^i \gets \hat j^R_{m_i} - \hat j^R_{n_i} + \sum_{k=1}^j w^i_k F_{n_i+k}$} \Comment{\cref{eq:algebraic1}}
      \State{$r_2^i \gets \left\{u_{m_i}^p - u_{n_i}^p - \sum_{k=1}^j w_k^i \partial_x u_{n_i+k}^p\right\}_p$}
      \EndFor
      \State \Return{$\{r_1^i\}$, $\{r_2\}$}
    \EndFunction
    \State \textbf{Solve} $R(\{u_n^j\}, \{\partial_x u_n^j\}) = 0$
  \end{algorithmic}
\end{algorithm}

The boundary conditions, given by \cref{eq:bc_final}, is in this case
\begin{subequations}
  \begin{align}
  \hat j^R_1 = -\left[\hat T_L \hat g^R_{s,L} \hat T_L^\dagger + i \hat R_L,\, \hat g^R_{s,1}\right],
  \\
  \hat j^R_N = \left[\hat T_R \hat g^R_{s,R} \hat T_R^\dagger + i \hat R_R,\, \hat g^R_{s,N}\right],
  \end{align}
  \label{eq:numBC}
\end{subequations}
where $\hat T_L$ and $\hat T_R$ are the tunneling matrices, $\hat R_L$ and $\hat R_R$ are the reflection matrices and $\hat g^R_{s,L}$ and $\hat g^R_{s,R}$ are the quasiclassical Green's functions on the left ($x = 0$) and right ($x = [N-1]\Delta x$) side, respectively.
If a boundary is insulating, then the corresponding tunneling matrix is zero.
A magnetic insulator will have nonzero magnetic components in the reflection matrix, so that $\hat R = r_0 + \vec m \cdot \vec \sigma$ for some scalar $r_0$ and some vector $\vec m$.

We have $2NM$ unknown parameters, so we need $2NM$ algebraic equations.
These can be obtained by integrating \cref{eq:usadel} in space.
\Cref{eq:usadel} can in this case be written
\begin{equation}
  \frac{\partial \hat j^R}{\partial x} + F = 0,
  \label{eq:numerics_ex}
\end{equation}
where
\begin{equation}
  F = -i\left[\tau_z\varepsilon - \hat V_s^R + \frac{iJ^2}{2\tau_\text{imp}\eta^2}\sigma_z\tau_z\hat g_s^R\sigma_z\tau_z,\,\hat g_s^R\right] - i\left[\hat A_x^R, \hat j^R\right].
  \label{eq:Fnum}
\end{equation}
To obtain algebraic equations, we can integrate \cref{eq:numerics_ex} between two discretization points and use a numerical integration scheme to approximate the integral of $F$.
Integrating between $(i-1)\Delta x$ and $(i+j-1)\Delta x$, we get
\begin{equation}
  \hat j^R_{i+j} - \hat j^R_{i} + \sum_{k=1}^j w_k F_{i+k} = 0,
  \label{eq:algebraic1}
\end{equation}
where $(w_1, \dotsc, w_j)$ is the set of weights defined by the numerical integration scheme and $F_{n}$ is \cref{eq:Fnum} evaluated with $\hat g^R_s = \hat g^R_{s,n}$ and $\hat j^R_s = \hat j^R_{s,n}$.
\Cref{eq:algebraic1} is a matrix-valued equation from which one can take $M$ independent scalar equations.
For instance, in the most general case with $M=8$, one can take the upper right and lower left $2\times 2$ blocks of \cref{eq:algebraic1}.
Another $M$ algebraic equations can be found from the same interval by integrating $\partial_x u^p$ for $p\in{1,\dotsc,M}$,
\begin{equation}
  u_{i+j}^p - u_i^p - \sum_{k=1}^j w_k \partial_x u_{i+k}^p = 0.
  \label{eq:algebraic2}
\end{equation}

To obtain $2NM$ algebraic equations, one can choose $N$ different subintervals, each of which defines $2M$ algebraic equations through \cref{eq:algebraic1,eq:algebraic2}.
These can be solved using Newton's method, and one can use for instance forward-mode automatic differentiation or finite differences to determine the Jacobian. 
The algorithm for solving \cref{eq:usadel,eq:gen_curr,eq:bc_final} for arbitrary values of $J/\eta$ in one dimension is summarized in \cref{alg:numAlg}.
Having found the retarded Green's function, one can determine the advanced and Keldysh Green's functions, and thereby compute observables, through
\begin{subequations}
 \begin{align}
   \hat g^A_{s} &= -\tau_z (\hat g^R_{s})^\dagger \tau_z, \label{eq:gaFromgr}\\
   \hat g^K_{s} &= (\hat g^R_{s} - \hat g^A_s)\tanh(\beta\varepsilon/2), \label{eq:gkFromgagr}
 \end{align}
\end{subequations}
where $\beta$ is inverse temperature.
\Cref{eq:gaFromgr} follows from the definition of the advanced and retarded Green's function while \cref{eq:gkFromgagr} follows from the fluctuation-dissipation theorem.

% Fakesection: Acknowledgements
\begin{acknowledgments}
This work was supported by the Research Council of Norway through grant 323766, and its Centres of Excellence funding scheme grant 262633 ``\emph{QuSpin}''.  
J.L. also acknowledges computational resources provided by Sigma2 - the National Infrastructure for High-Performance Computing and 
Data Storage in Norway from project no. NN9577K.
\end{acknowledgments}

% Fakesection: Bibliography
% \clearpage
\bibliography{bibliography}

\end{document}